\documentclass[12pt,a4paper]{article}
\usepackage[utf8]{inputenc}
\usepackage[T1]{fontenc}
\usepackage{lmodern}
\usepackage{amsmath}
\usepackage{amssymb}
\usepackage{graphicx}
\usepackage{xspace}
\usepackage{color}
\usepackage[dvipsnames]{xcolor}
\usepackage{setspace}
\usepackage{cite}
\usepackage{hyperref}
\hypersetup{colorlinks=true,citecolor=blue,linkcolor=blue}

\usepackage{footmisc}

\usepackage{comment}

\usepackage{lineno}

\newcommand\as{\alpha_s}
\newcommand\lamQCD{\Lambda_{\rm QCD}}
\newcommand\dcut{d_{\rm cut}}

\newcommand\rhocut{\rho_{\rm cut}}

\newcommand\aNLO{{\sc\small MG5\_aMC}}

\newcommand\sss{\scriptscriptstyle}
\newcommand\NC{N_{\sss c}}

\newcommand\CF{C_F}
\newcommand\CA{C_A}

\newcommand{\ptt}{p_{t}}

\newcommand{\Ralt}{\mathcal{R}}

\newcommand\NOARXIV[1]{}

\newcommand\code[1]{\texttt{#1}}

\begin{document}
\begin{titlepage}

\vspace*{0.4cm}

\begin{center}
  {\Large \bf Jets at electron-positron colliders}
\end{center}

\par \vspace{2mm}
\begin{center}
{\bf Giovanni Stagnitto}

\vspace{5mm}

Universit\`a degli Studi di Milano-Bicocca \& INFN Sezione di Milano-Bicocca,\\
Piazza della Scienza 3, Milano 20126, Italy

\vspace{5mm}

\end{center}

\par \vspace{2mm}
\begin{center} {\large \bf Abstract}

\end{center}
\begin{quote}
  \pretolerance 10000
  We provide a pedagogical introduction to the physics of hadronic jets and event shapes at electron-positron colliders.
We present some of the main jet definitions and event shape observables studied at lepton colliders and discuss how to produce theoretical predictions in perturbative quantum chromodynamics (QCD), both at fixed order and with resummation or parton showers.
We further introduce important topics in jet substructure that have seen developments in a lepton collider environment, such as the Lund jet plane, Soft Drop, and quark-gluon jet discrimination.
Finally, we briefly elaborate on selected topics, such as flavoured jets, hadronic decays of the Higgs boson, and non-perturbative effects.

\end{quote}

\vspace*{\fill}

\begin{center}
  {\em Invited contribution to Encyclopedia of Particle Physics}
\end{center}


\end{titlepage}

\tableofcontents

\section{Introduction}\label{intro}
The study of hadronic final states at electron-positron colliders has played a
pivotal role in confirming quantum chromodynamics (QCD) as the fundamental
theory of strong interactions.

\begin{figure}[t]
  \centering
  \includegraphics[width=0.35\textwidth]{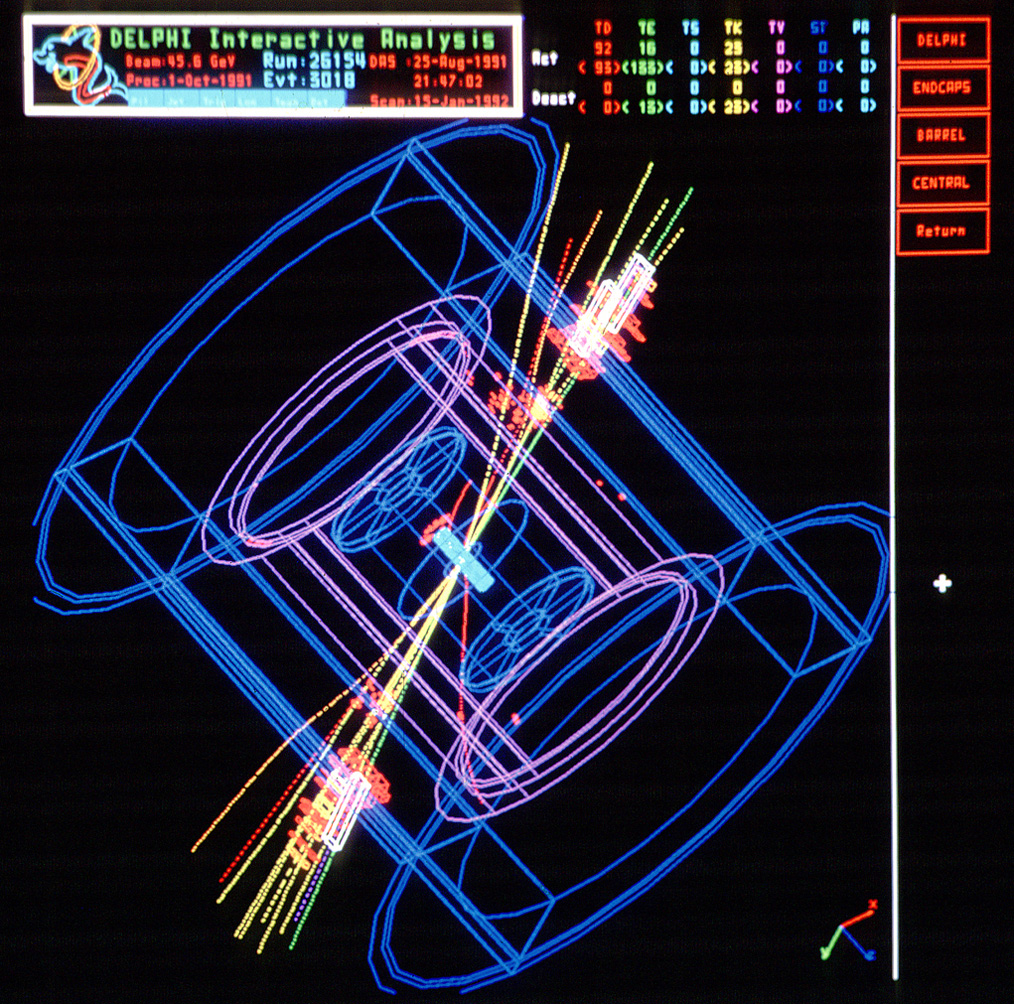}
  \includegraphics[width=0.35\textwidth]{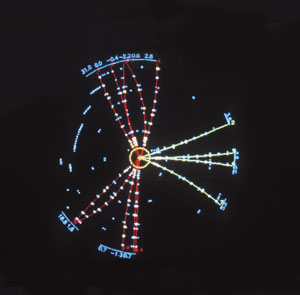}
  \caption{Two-jet event display at LEP DELPHI (left, from
    \texttt{https://cds.cern.ch/record/39449}) and three-jet event display at PETRA
    TASSO (right, from \texttt{https://cds.cern.ch/record/2678861}).}
  \label{fig:evdisp}
\end{figure}

Early experiments in the 1970s at SLAC provided first evidence for collimated
sprays or {\em jets} of high-energy hadrons. Quoting~\cite{Hanson:1975fe}:
\begin{quote}
{\em In quark-parton constituent models of elementary particles, hadron production in
  $e^+e^-$ annihilation reactions proceeds through the annihilation of the $e^+$
  and $e^-$ into a virtual photon which subsequently produces a quark-parton pair,
  each member of which decays into hadrons. At sufficiently high energy the
  limited transverse momentum distribution of the hadrons with respect to the
  original parton production direction, characteristic of all strong interactions,
  results in oppositely-directed jets of hadrons.}
\end{quote}
A typical two-jet event display is shown in Fig.~\ref{fig:evdisp} on the left.
If we identify the jet directions with the quark ones, the angular distributions
of jets will be given by the elementary cross section
\begin{equation}\label{eq:costhLO}
  \frac{\mathrm{d}\sigma(e^+e^- \to \gamma^* \to q\bar{q})}{\mathrm{d}\cos\vartheta}
  = \frac{3}{8} \sigma_0 (1 + \cos^2\vartheta)\,,
\end{equation}
with $\sigma_0$ the leading order total cross section for hadron production,
\begin{equation}\label{eq:sig0qqbar}
  \sigma_0 = \frac{4\pi\alpha^2}{3s} \NC \sum_q e_q^2\,,
\end{equation}
with $\NC = 3$ the number of quark colours, $e_q^2$ the quark charge squared, $s$
the center-of-mass energy squared and $\alpha$ the fine structure constant.
And indeed the $1 + \cos^2\vartheta$ behaviour predicted by~\eqref{eq:costhLO} was
observed in two-jet events~\cite{Hanson:1975fe}, confirming the the spin-1/2
nature of quarks.

Soon after, it was recognized that events exhibiting three distinct jets could
serve as evidence from the gluon, the mediator of the strong force. Such events
were predicted to arise from {\em hard-gluon bremsstrahlung}~\cite{Ellis:1976uc}
from the quark-antiquark pair (the $e^+e^- \to q\bar{q}g$ elementary
scattering).
Experiments at the PETRA electron-positron collider, which operated at DESY from
1979 to 1986 with center-of-mass energies between 12 and 46.7 GeV, provided the
first experimental evidence of three-jet
events~\cite{TASSO:1979zyf,Barber:1979yr,PLUTO:1979dxn,JADE:1979rke}
(see~\cite{Soding:2010zz,Ellis:2014rma} for historical perspectives on the
discovery of the gluon). A three-jet event display from PETRA is shown in
Fig.~\ref{fig:evdisp} on the right.

Quantitative tests of QCD constituted a central component of the physics program
at the Large Electron-Positron Collider (LEP) at CERN, operated during the 1990s
at $\sqrt{s} = 91$~GeV (close to the $Z$-boson resonance) and at $\sqrt{s}$
between 130 and 209 GeV (to scan the $W$-boson pair production cross section).
Measurements of hadronic final states by the four LEP experimental
collaborations (ALEPH~\cite{ALEPH:1990ndp}, DELPHI~\cite{DELPHI:1990cdc},
OPAL~\cite{OPAL:1990yff}, L3~\cite{L3:1989aa}) were instrumental in rigorously
testing perturbative QCD calculations, extracting the value of the strong
coupling constant $\as$, gaining deeper insight into the mechanism of hadron
formation.

The current focus of the high-energy physics community is largely directly
towards hadron colliders, primarily due to the prominence of the Large Hadron
Collider (LHC) at CERN, which has been operational since 2007 and is expected to
remain active for at least another 10–15 years.
It is therefore unsurprising that most recent advancements in
QCD theoretical predictions are tailored to the hadron collider environment.
However, as we will discuss, many of the conceptual developments and
computational techniques devised for the LHC are readily transferable to
electron–positron colliders.

Looking ahead, the future of particle physics may well involve a high-energy
electron-positron collider. The Future Circular Collider
(FCC-ee)~\cite{FCC:2025lpp} or the Circular Electron Positron Collider
(CEPC)~\cite{An:2018dwb}, both currently under discussion, are planned to
operate at various centre-of-mass energies and are expected to produce a vast
amount of hadronic final states from the decay of vector or Higgs bosons.

The purpose of this document is to review in a pedagogical manner the physics of
jets, by focussing on electron-positron colliders.
It is not intended to cover every detail comprehensively, as the study of QCD
final states at colliders is a broad topic with a lot of history.
We will introduce basic concepts---some of them of relevance also for LHC
physics---and the minimum number of equations necessary to drive the physics
message.
Overall, the level of technical detail is kept to a minimum to ease the
discussion.
However, we will also touch on some more advanced topic, to illustrate the
recent developments in the field.

This Chapter is organised as follows.
In Section~\ref{sec:jets} we will introduce the concept of hadronic jet and will
discuss the main jet definitions adopted at $e^+e^-$ colliders.
Section~\ref{sec:evshapes} is devoted to event shapes: in particular, we will
focus on two event shapes, the thrust and the energy-energy correlators.
Section~\ref{sec:highord} will present fixed-order calculations in perturbative
QCD for jets and event shapes, whereas Section~\ref{sec:allorder} will discuss
how to obtain all-order predictions for observables at $e^+e^-$ colliders.
Section~\ref{sec:jetsub} is about the study of the substructure of jets: in this
context, we will briefly introduce the Lund jet plane and quark/gluon jet
discrimination.
Finally, Section~\ref{sec:seltopics} will present a selection of more advanced
topics, or topics where there has been recent development: flavoured jets,
hadronic Higgs decay, non-perturbative corrections.

The content of this Chapter is mainly based on textbooks (the ``pink
book''~\cite{Ellis:1996mzs}, although nearly 30 years old, remains one of the
best introductions to QCD applied to collider physics; a recent book about jet
substructure~\cite{Marzani:2019hun}), on lecture notes (by Stefano
Catani~\cite{Catanilectures}, Michelangelo Mangano~\cite{Mangano:1998fk}, Paolo
Nason~\cite{Nason:1997zu} and Gavin Salam~\cite{Salam:2010zt}) or review papers
(e.g.\ on resummation~\cite{Luisoni:2015xha} or jets~\cite{Salam:2009jx}).

Among the topics not discussed in the Chapter, there is the deep interplay
between modern machine learning techniques and jet physics. For this, we refer
to the lectures in~\cite{Larkoski:2024uoc}.

\section{On the definition of jets}\label{sec:jets}

Naively, a {\em jet} can be seen as a bunch of energetic collimated particles
distributed around a determinate direction.
However, how to properly define a jet in an unambiguous way?
Before answering this question, it may be useful to provide a description of the
physics picture behind the formation of hadronic final states at $e^+e^-$
machines.

\subsection{Jets as macroscopic manifestation of QCD emission dynamics}

\begin{figure}[t]
  \centering
  \includegraphics[width=0.25\textwidth]{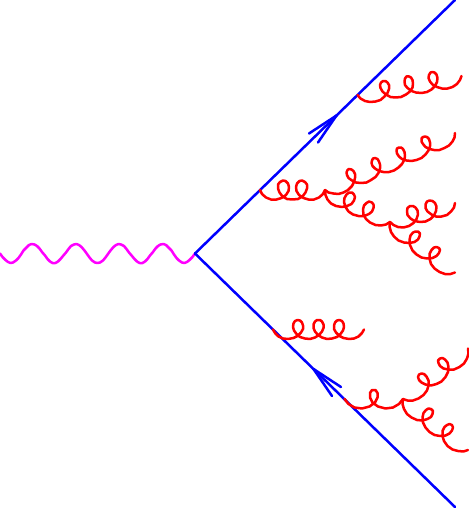}
  \caption{QCD emissions in the final state. From~\cite{Ellis:1996mzs}.}
  \label{fig:QCDps}
\end{figure}

We have seen that the elementary scattering process $e^+e^- \to q\bar{q}$
through the exchange of a virtual $\gamma$/$Z$ boson leads to the creation of a
highly energetic quark-antiquark pair (kinematics fixes the energy of $q$ and
$\bar{q}$ to be equal to $\sqrt{s}/2$).
At this point, the hard partons start progressively loosing energy, by radiating
gluons. Emitted gluons carry colour charge, and they further emit, generating a
cascade of QCD particles, as depicted in Fig.~\ref{fig:QCDps}.

\begin{figure}[t]
  \centering
  \includegraphics[height=2cm]{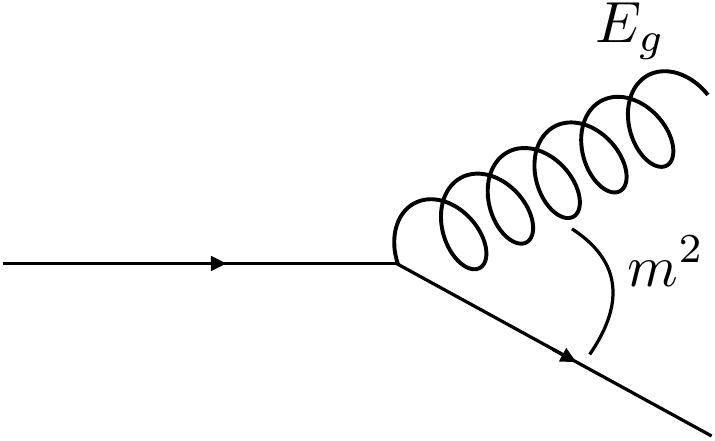}
  \includegraphics[height=2cm]{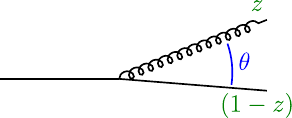}
  \caption{Gluon splitting off a quark, with different choice of variables.}
  \label{fig:q2qg}
\end{figure}

At high energies the coupling of QCD, $\as$, is small, so the usage of
perturbation theory is justified and we can describe the emission process in
terms of an expansion in $\as$.
Let us consider a quark with energy $E$ splitting into a quark-gluon pair $q \to
qg$, see Fig.~\ref{fig:q2qg}. The phase space for the emission depends on two
independent variables, that we can choose to be the energy of the emitted gluon
$E_g$ and the invariant mass squared of quark-gluon system $m^2$,
\begin{equation}\label{eq:msq}
  m^2 = 2 p_q \cdot p_g = 2 E_q E_g (1 - \cos\vartheta) =
  2 E^2 z (1-z) (1 - \cos\vartheta)\,,
\end{equation}
where we have further introduced the angle between the quark and the gluon
$\vartheta$ and the energy fraction $z=E_g/E$ for later convenience.

The dominant behaviour of QCD radiation can be derived by first-principle
considerations, according to the argument proposed in~\cite{Larkoski:2017fip}.
Under the assumption that QCD has no intrinsic scale, so it is a scale-invariant
quantum field theory (which is true if we neglect quark masses), the probability
of a quark emitting a gluon must be invariant under a $\lambda$ rescaling:
\begin{equation}\label{eq:QCDsinv}
P(\lambda E_g, \lambda^2 m^2) \,\mathrm{d}(\lambda E_g) \, \mathrm{d}(\lambda^2 m^2)
= P(E_g, m^2) \,\mathrm{d}E_g \, \mathrm{d}m^2\,.
\end{equation}
The simplest form compatible with~\eqref{eq:QCDsinv} turns out to be the correct
one:
\begin{equation}\label{eq:PqgEm2}
P(E_g, m^2) \, \mathrm{d}E_g \, \mathrm{d}m^2 =
\frac{\alpha_s \CF}{\pi} \, \frac{\mathrm{d}E_g}{E_g} \, \frac{\mathrm{d}m^2}{m^2}\,,
\end{equation}
where we have introduced the strong coupling $\as$ and a colour factor $\CF$
equal to $\CF = (\NC^2-1)/2/\NC = 4/3$ for $\NC = 3$.
We can rewrite~\eqref{eq:PqgEm2} in terms of dimensionless $\vartheta$ and $z$ to
disentangle energy and angular variables,
\begin{equation}\label{eq:Pqgzth}
  P(z, \vartheta^2) \, \mathrm{d}z \, \mathrm{d}\vartheta^2
  = \frac{\alpha_s \CF}{\pi} \, \frac{\mathrm{d}z}{z}\,
  \frac{\mathrm{d}\cos\vartheta}{1 - \cos\vartheta}
  \xrightarrow{\vartheta \ll 1}
  \frac{\alpha_s \CF}{\pi} \, \frac{\mathrm{d}z}{z} \, \frac{\mathrm{d}\vartheta^2}{\vartheta^2}\,,
\end{equation}
where we have additionally taken the small-angle limit.  Hence, QCD dynamics
encourages the production of {\em soft} ($z\to 0$) and/or {\em collinear}
($\vartheta \to 0$) particles.
What we observe in experiments is then a macroscopic manifestation
of~\eqref{eq:Pqgzth}: the final state features energetic particles distributed
around the directions of the original hard partons i.e.\ {\em jets}, possibly
with uniform emission of (undetected) soft particles.
Jets can be then considered as {\em proxies} for the hard partons in the
elementary scattering process.
As a consequence, we would expect most of the events to appear as two-jet
events, with the two jets back-to-back, to reflect the hard $q\bar{q}$ pair
production, and this is what is observed in experiments.

So far, we ignored the fact that in real experiments one would not observe quark
or gluons, but colour-neutral {\em hadrons} (pions, kaons, protons, etc.).
Indeed, when the energy scale involved in the radiation process is~$\lesssim
1$~GeV, we enter in the non-perturbative regime. Quarks and gluons start
recombining and forming bound states in the process of {\em hadronization}.
The main property of hadronization relevant for our discussion is that the
transition from partons to hadrons is {\em local} in phase space. Or in other
words, colour propagation in the QCD shower is such that at non-perturbative
scales we have the formation of colour-neutral and low-mass parton clusters.
Otherwise, at low energy scales, the strong force between partons would be
extremely large, and it would spoil the perturbative approach we adopt at higher
energy scales.
This mechanism is called Local Parton Hadron Duality, and we refer the reader
to~\cite{Dokshitzer:1991wu} for an introduction.

\subsection{Sterman-Weinberg jets}

The first attempt to formalize a jet definition dates back to 1977, with the
seminal paper by Sterman and Weinberg~\cite{Sterman:1977wj}.
It was based on the idea of jet as a {\em cone} containing a determinate
fraction of the total energy present in the final state of the event.
More precisely, by introducing an energy parameter $\epsilon$ and an angular
parameter $\delta$, an event is defined as a {\em two-jet event} if it is
possible to find two cones of half-angle $\delta$ containing $1-\epsilon$ of the
total energy in the event.

\begin{figure}
  \centering
  \includegraphics[width=0.45\textwidth]{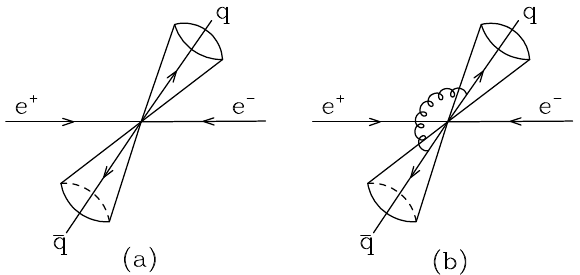}\,
  \includegraphics[width=0.45\textwidth]{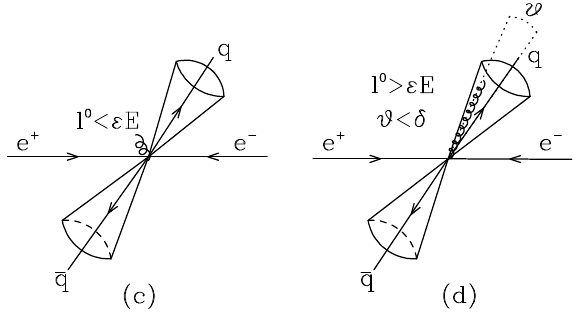}
  \caption{Configurations contributing to Sterman-Weinberg jets up to
    $\mathcal{O}(\as)$. From~\cite{Nason:1997zu}.}\label{fig:SW}
\end{figure}

At leading order in perturbation theory, all events are two-jet events, see
Fig.~\ref{fig:SW}(a). Hence, the jet definition is effective only when real
emission is involved, by limiting the available phase space.
Let us consider the emission of a real gluon. A gluon with energy $l^0 <
\epsilon E$ (with $E$ the total energy in the event) will always contribute to
the two-jet cross section, even if it is emitted outside of the two cones, see
Fig.~\ref{fig:SW}(c). Instead, a gluon with energy $l^0 > \epsilon E$, can only
contribute to the two-jet cross section if emitted inside a cone i.e.\ with an
angle $\vartheta_{q/\bar{q}} < \delta$ with respect to the $q$ or $\bar{q}$
direction, respectively, see Fig.~\ref{fig:SW}(d). In terms of Heaviside step
functions:
\begin{equation}\label{eq:SWps}
  \Theta(l^0 < \epsilon E) + \Theta(l^0 > \epsilon E)
        [\Theta(\vartheta_q < \delta) + \Theta(\vartheta_{\bar{q}} < \delta)]
        = 1 - \Theta(l^0 > \epsilon E) \Theta(\vartheta_q > \delta)
        \Theta(\vartheta_{\bar{q}} > \delta) \,.
\end{equation}
In other words, to obtain the two-jet cross section we discard those events that
would be considered as three-jet events i.e.\ with a ``hard'' gluon ($l^0 >
\epsilon E$) emitted outside of the two cones.
To compute the two-jet rate, we now assume the validity of~\eqref{eq:Pqgzth}
i.e.\ the emitted gluon is soft and collinear to either the $q$ or the
$\bar{q}$. This approximation is enough to derive the dominant behaviour of the
two-jet rate in the limit $\epsilon \ll 1$ and $\delta \ll 1$. By taking into
account the phase-space restrictions of~\eqref{eq:SWps}, the real contribution
is given by
\begin{equation}\label{eq:SWr}
  \sigma_{\rm R} = \sigma_0 \frac{\alpha_s \CF}{\pi} \int_0^1
  \frac{\mathrm{d}z}{z} \int_0^1 \frac{\mathrm{d}\vartheta^2}{\vartheta^2}
  \left[ 1 - 2 \Theta(z > \epsilon) \Theta(\vartheta > \delta) \right]\,,
\end{equation}
with $\sigma_0$ given in~\eqref{eq:sig0qqbar} and with the factor of 2 due to
the presence of the two cones.
Note that~\eqref{eq:SWr} is still divergent in the limits $z \to 0$ and
$\vartheta \to 0$, due to the ``1'' term inside the square brackets.
This term cancels against the virtual contribution, which is divergent as well:
indeed, virtual diagrams feature Born kinematics without any phase space
restriction, see Fig.~\ref{fig:SW}(b), and with an overall minus sign to cancel
the divergent behaviour of the real contribution.\footnote{It is possible to
explicitly show that virtual contributions involving loop integrals in the
divergent limits behave as the real ones, but with an overall minus sign, see
e.g.\ Sec.~2.2 of~\cite{Luisoni:2015xha}. This way of taking into account
virtual effects with a ``-1'' is usually referred to as {\em unitarity} in the
resummation community. This is enough to predict logarithmically enhanced terms
in final results (with constant terms beyond control). Methods based on
unitarity to generate approximations to loop contributions have been
devised~\cite{Rubin:2010xp}. \label{foot:unit}}.
Hence, the two-jet rate up to $\mathcal{O}(\as)$ is given by
\begin{equation}\label{eq:SWxs2jet}
  \sigma_{\rm two-jet} = \sigma_0 \left( 1 - \frac{\alpha_s \CF}{\pi} 2
  \int_\epsilon^1 \frac{\mathrm{d}z}{z}
  \int_\delta^1 \frac{\mathrm{d}\vartheta^2}{\vartheta^2} \right)
  = \sigma_0 \left( 1 - \frac{4 \alpha_s \CF}{\pi} \log(1/\epsilon) \log(1/\delta) \right)
\end{equation}

The interpretation of~\eqref{eq:SWxs2jet} warrants a detailed discussion.

As long as $\epsilon$ and $\delta$ are kept different from zero, the
Sterman-Weinberg two-jet rate is finite.
In the unresolved regions of phase space (i.e.\ for energy below $\epsilon$ and
angle below $\delta$), the singularities in real and virtual contributions are
allowed to cancel, leaving a logarithmic dependence on $\epsilon$ and
$\delta$. In other words, the infinities have been replaced by logarithms of the
parameters that regulate the divergences.
This is a typical feature of all calculations in QCD with phase space
restrictions to the final state.

Moreover, as long as the parameters $\epsilon$ and $\delta$ are not too small,
the coefficient of $\as$ in~\eqref{eq:SWxs2jet} is $\mathcal{O}(1)$, hence the
perturbative expansion is reliable, with the perturbative correction to the
two-jet rate providing a small correction to the Born
result~\footnote{Eq.~\eqref{eq:SWxs2jet} has been derived in the soft and
collinear limit. The exact result for the Sterman-Weinberg two-jet rate is
``rather unwieldy''~\cite{Ellis:1996mzs}.}.
In particular, due to the running of $\as$, increasing the collider
center-of-mass energy results in a smaller coupling, leading to a higher
proportion of two-jet events.

Instead, when $\epsilon \ll 1$ and/or $\delta \ll 1$, the logarithmic terms grow
large, and the validity of the perturbative series becomes questionable.
However, it is sometimes possible to predict the structure of the
logarithmically enhanced terms to {\em all} orders in perturbation theory and
take them into account ({\em resum} them) in a systematic way to regain
predictive power.
We will elaborate on this topic in Sec.~\ref{sec:allorder}.

The definition of jets by Sterman and Weinberg is of historical significance.
However, its formulation renders it suboptimal for the analysis of multi-jet
final states.
Moreover, in the presence of many particles in the final state, identifying the
cones may be computationally intensive.

Modern jet definitions are based on {\em sequential recombination algorithms},
as detailed in the following.
Before we continue, let us state an important property that any jet definition
should satisfy.

\subsection{Infrared and collinear (IRC) safety}\label{sec:IRCsafe}

Given a generic observable $O$, {\em infrared and collinear} (IRC) safety may be
formulated as follows.  Provided a final state with $n$+1 QCD particles, in
the limit in which one particle becomes soft or two particles become collinear,
the observable should scale as
\begin{align}
  O_{n+1}(k_1, \dots, k_i, \dots, k_n) \xrightarrow{k_i \to 0}
  O_{n}(k_1, \dots, k_{i-1},k_{i+1}, \dots, k_n)\,, \label{eq:IRCsafe1} \\
  O_{n+1}(k_1, \dots, k_i,k_j, \dots, k_n) \xrightarrow{k_i \parallel k_j}
  O_{n}(k_1, \dots, k_i+k_j, \dots, k_n)\,, \label{eq:IRCsafe2}
\end{align}
where $O_{n+1}$ and $O_{n}$ is the value of the observable evaluated on a set of
$n$ or $n$+1 particles, respectively.
IRC safety guarantees the cancellation of soft and collinear divergences,
because in the soft and collinear limit $O_{n+1}$ reduces to $O_{n}$ with the
right mapping of momenta to permit the cancellation with the virtual
contributions.
Examples of IRC unsafe observables are the number of particles in the final
state (because e.g.\ the emission of a soft gluon increases by one the number of
particles) or the energy of hardest particle (because e.g.\ a collinear
splitting may change what is the most energetic parton in the event).

IRC safety is a crucial property to be able to perform perturbative
calculations. However, in nature infinities do not appear, because the divergent
behaviour of soft and/or collinear emissions is regulated by non-perturbative
effects, which become relevant at low energy scales around \mbox{$\lamQCD \simeq 0.3$~GeV}.
IRC unsafe observables are then characterised by the presence of large
$\log(Q/\lamQCD)$ terms, with $Q$ the hard scale of the process. Given the
one-loop expression for the running of $\as$, $\as \simeq 1/\log(Q/\lamQCD)$,
then all the terms in the perturbative series become of $\mathcal{O}(1)$,
invalidating the reliability of purely perturbative predictions\footnote{It is
possible to consider IRC unsafe observables, but they require some
non-perturbative input or some empirical model for hadronization effects.}.
Hence, jets or more in general IRC safe observables constitute a way of looking
at collider events designed to be insensitive (or retain little sensitivity) to
all effects beyond perturbative control.

\subsection{Sequential recombination algorithms}

The modern way of defining jets relies on {\em sequential recombination} or {\em
  clustering} algorithms applied to a set of final-state objects. They can be
partons (in fixed-order or resummed calculations), hadrons (in Monte Carlo event
generators, see Sec.~\ref{sec:psevgen}) or detector-level objects in
experimental analysis.

\begin{figure}
  \centering
  \includegraphics[width=0.8\textwidth]{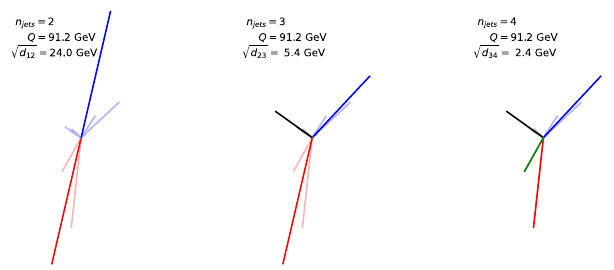}
  \caption{Number of jets as a function of the jet resolution variable. Courtesy of Gregory Soyez.}
  \label{fig:ktjets}
\end{figure}

Sequential recombination algorithms work as follow: for every pair of objects
$i$ and $j$, we define a distance $d_{ij}$. We iteratively find the smallest
$d_{ij}$ and recombine $p_{i+j} = p_i + p_j$ i.e.\ we sum the four-momenta of
the two objects being recombined\footnote{This is referred to as recombination
$E$-scheme. Other recombination schemes have been explored in the
past~\cite{Salam:2009jx}.}.
We can stop the recombination either when we reach a specified number of jets or
when $d_{ij} > \dcut$, with $\dcut$ a given termination condition\footnote{This
is referred to as {\em exclusive} running mode. For algorithms running in {\em
  inclusive} mode, see Sec.~\ref{sec:genkt}.}.
As an example, in Fig.~\ref{fig:ktjets} we show the action of a clustering
algorithm (the $k_t$ algorithm in this case, see Sec.~\ref{sec:ktalg}) as applied to
the final state of an event with $Q = 91.2$~GeV.
In Fig.~\ref{fig:ktjets}, we further report the transition values between the
$n$-jet and $n$+1-jet regions $d_{n,n+1}$ e.g.\ $d_{23}$ is such that if $\dcut
> d_{23}$ the event is considered as an event with two jets (or less), otherwise
as an event with three jets (or more).
As we decrease $\dcut$, the number of reconstructed jets increases.

In the following, we discuss two clustering algorithms adopted at $e^+e^-$
colliders (JADE and $k_t$ algorithm) and we then introduce the generalized $k_t$
family of algorithms.
We refer the reader to~\cite{Moretti:1998qx} for a review of jet algorithms in
$e^+e^-$ collisions, and to~\cite{Salam:2009jx} for a review of jet algorithms
in general.
All of the described algorithms are implemented in the public package {\sc
  FastJet}~\cite{Cacciari:2011ma}.

\subsubsection{JADE algorithm}\label{sec:jadealg}

The JADE algorithm has been first introduced by the JADE collaboration at PETRA
collider~\cite{JADE:1986kta}, with distance
\begin{equation}\label{eq:dijJADE}
  d_{ij} = 2 E_i E_j (1-\cos\vartheta_{ij})\,.
\end{equation}
Measurements of production rates of three-jet events with the JADE algorithm
provided first evidences of the energy dependence of the strong coupling
strength~\cite{JADE:1988xlj}.

\begin{figure}[t]
  \centering
  \includegraphics[width=0.4\textwidth]{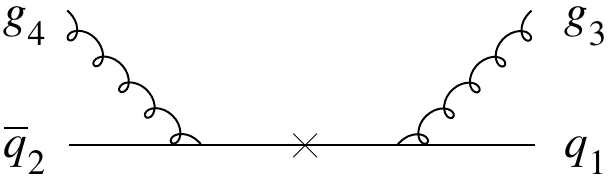}
  \caption{Problematic configurations for JADE algorithm. From~\cite{Dokshitzer:1997in}.}
  \label{fig:confJADECAM}
\end{figure}

Note that for massless particles, $d_{ij}$ in~\eqref{eq:dijJADE} is identical to
the invariant mass squared of the pair of particles, see~\eqref{eq:msq}.
This property renders JADE suitable for higher-order perturbative
calculations\footnote{Note that infrared divergences are sometime called {\em
  mass} divergences because they originate from propagators going on-shell. A
cut on $m^2 > 0$ is then enough to screen both soft and collinear
singularities. Matrix element squared are naturally expressed in terms of
invariants like $s_{ij} = (p_i + p_j)^2$.}.
However, from the experimental point of view, JADE features irregular jets, due
to an enhanced recombination of soft particles. Indeed, by looking
at~\eqref{eq:dijJADE}, pairs of soft particles tend to be recombined in the
early stages of the clustering e.g. a $q\bar{q}gg$ final state, with soft
gluons emitted at large angle, will likely be reconstructed as a three-jet
event, with an unphysical third jet whose direction points halfway between the
gluons' directions, see Fig.~\ref{fig:confJADECAM}. 

Analogously to the Sterman-Weinberg jet rate~\eqref{eq:SWxs2jet}, jet rates with
JADE (or any recombination algorithm) suffer from large logarithms of $\dcut$
when $\dcut$ is small.
The presence of the soft large-angle gluon clusters render jet rates with JADE
difficult to resum, as first observed in~\cite{Brown:1990nm}.

\subsubsection{$k_t$ or Durham algorithm}\label{sec:ktalg}

The $k_t$ or Durham algorithm has been first introduced in~\cite{Catani:1991hj},
with distance
\begin{equation}\label{eq:dijKT}
  d_{ij} = 2\min(E_i^2,E_j^2) (1-\cos\vartheta_{ij})\,.
\end{equation}
At variance with~\eqref{eq:dijJADE}, the presence of the minimum
in~\eqref{eq:dijKT} ensures that a e.g.\ hard quark close in angle with a soft
gluon is recombined before a pair of large-angle soft gluons.
Indeed, the $k_t$ distance~\eqref{eq:dijKT} is equal to the relative transverse
momentum between the two particles $k_t \simeq \min(E_i,E_j) \vartheta_{ij}$
(hence the name $k_t$ algorithm), mimicking the structure of QCD
divergences. The clustering sequence then resembles the physical sequence of QCD
emissions.

The nice physical properties of the $k_t$ algorithm make it suitable for both
fixed-order and resummed calculations. Indeed, in the same paper when it has
been introduced~\cite{Catani:1991hj}, also the resummation of jet rates has been
performed, see Sec.~\ref{sec:resjets}.
One of the disadvantages of the $k_t$ algorithm is that it tends to follow the
aggregation of soft particles, thus resulting in jets with irregular borders.

\subsubsection{Generalized $k_t$ family of algorithms}\label{sec:genkt}

In modern jet terminology, JADE or $k_t$ are run in {\em exclusive} mode,
meaning that the jet algorithm halts based on a specific resolution parameter or
a fixed number of jets. Additionally, all final-state particles are usually
clustered into reconstructed jets.

Jet algorithms can also be run in an {\em inclusive} mode, as usually done in
the hadron collider context.
For instance, the generalised $k_t$ (gen-$k_t$) family of
algorithms~\cite{Cacciari:2008gp} introduces {\rm two} sets of distances, which
in spherical coordinates\footnote{Distances at hadron colliders are usually
expressed in terms of transverse momenta $p_t$ and rapidity differences $\Delta
y$. Both quantities are invariant under longitudinal boosts and this is a
desirable property, because the total momentum component along the beam is not
known. At lepton colliders, it is more natural to express distances in terms of
energies and polar angles.} read~\cite{Cacciari:2011ma}:
\begin{equation}\label{eq:distgenkt}
  d_{ij} = 2 \min\left(E_{i}^{2p},E_{j}^{2p}\right) (1 - \cos\theta_{ij})\,,
  \quad d_{iB} = 2 E_i^{2p} (1 - \cos R)
\end{equation}
where the parameter $R$ is usually referred to as jet radius.
One finds the smallest between $d_{ij}$ and $d_{iB}$: if it is a $d_{ij}$, the
two particles are recombined; instead, if it is a $d_{iB}$, $p_i$ is declared as
jet, and removed from the list.
The algorithm stops when there are no particles left to recombine.
At the end, only jets above an energy cut $E_{\rm cut}$ are retained (without
this final requirement, the algorithm is not soft safe).
Note that we are not requiring a termination condition based on some $d_{\rm
  cut}$ or number of jets. Moreover, at variance with JADE or $k_t$, some
particles in the event may not be part of any final state jet, because they can
be discarded by the energy cut.

The behaviour of the gen-$k_t$ algorithm is dictated by the value of the
exponent $p$ in~\eqref{eq:distgenkt}.
The $p=1$ exponent returns a distance similar to the $k_t$ algorithm of
Sec.~\ref{sec:ktalg}, whereas the version with the $p=0$ exponent is similar to
a purely angular algorithm, like the Cambridge
algorithm~\cite{Dokshitzer:1997in}.
The $p=-1$ exponent is instead referred to as {\em anti-$k_t$} $e^+e^-$
algorithm~\cite{Cacciari:2008gp}.  With this seemingly pathological choice, hard
particles are clustered in the first steps of the algorithm. If the distance
between two hard particles is greater than $R$, each hard particle will tend to
accumulate soft particles, providing perfect conical shapes.
This peculiarity of the anti-$k_t$ algorithm renders it an IRC safe version of a
{\em cone} algorithm. This is one of the reasons behind its adoption as default
jet algorithm in hadron collider environments such as the LHC.

\begin{figure}[t]
  \centering
  \includegraphics[width=0.45\textwidth]{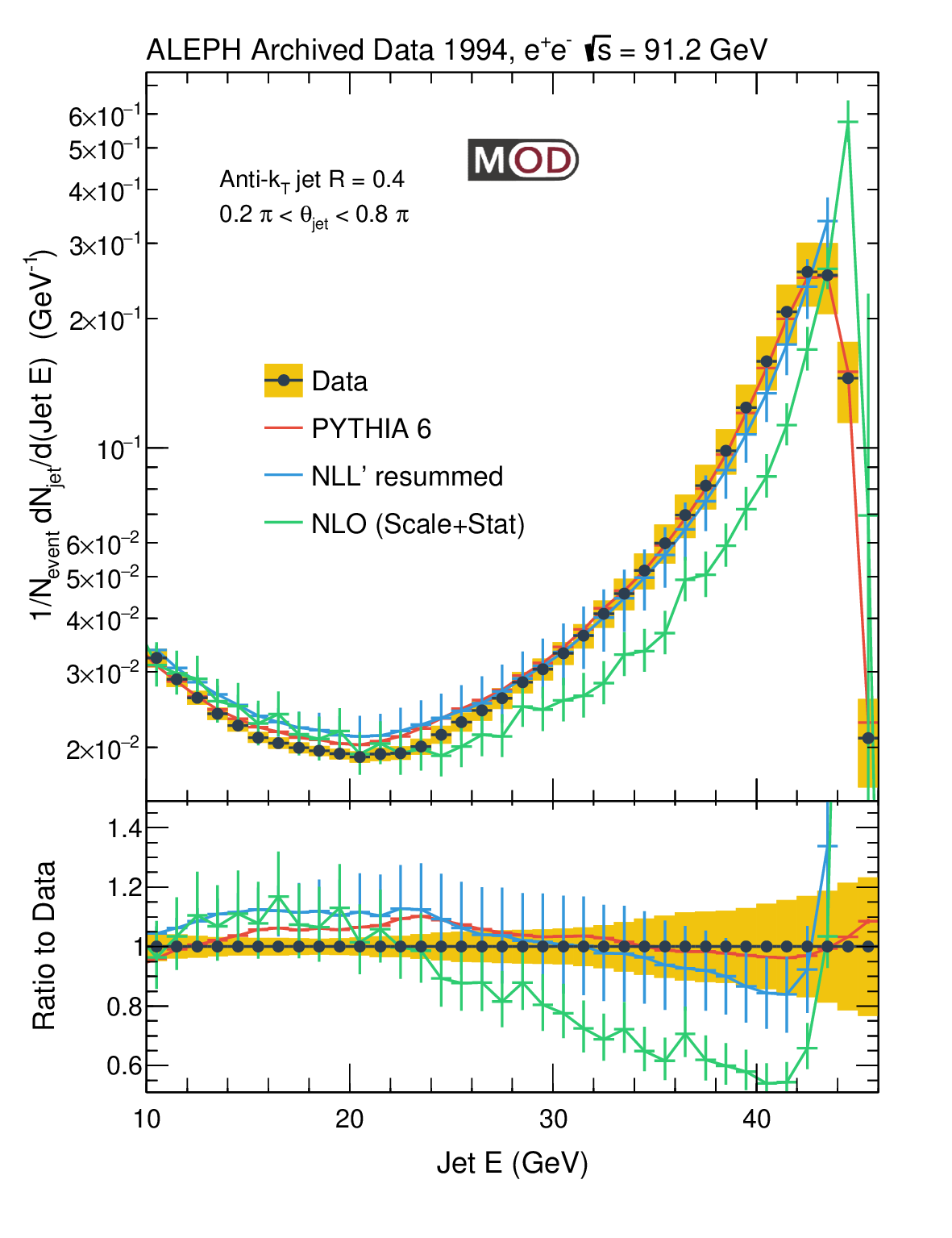}\,
  \includegraphics[width=0.45\textwidth]{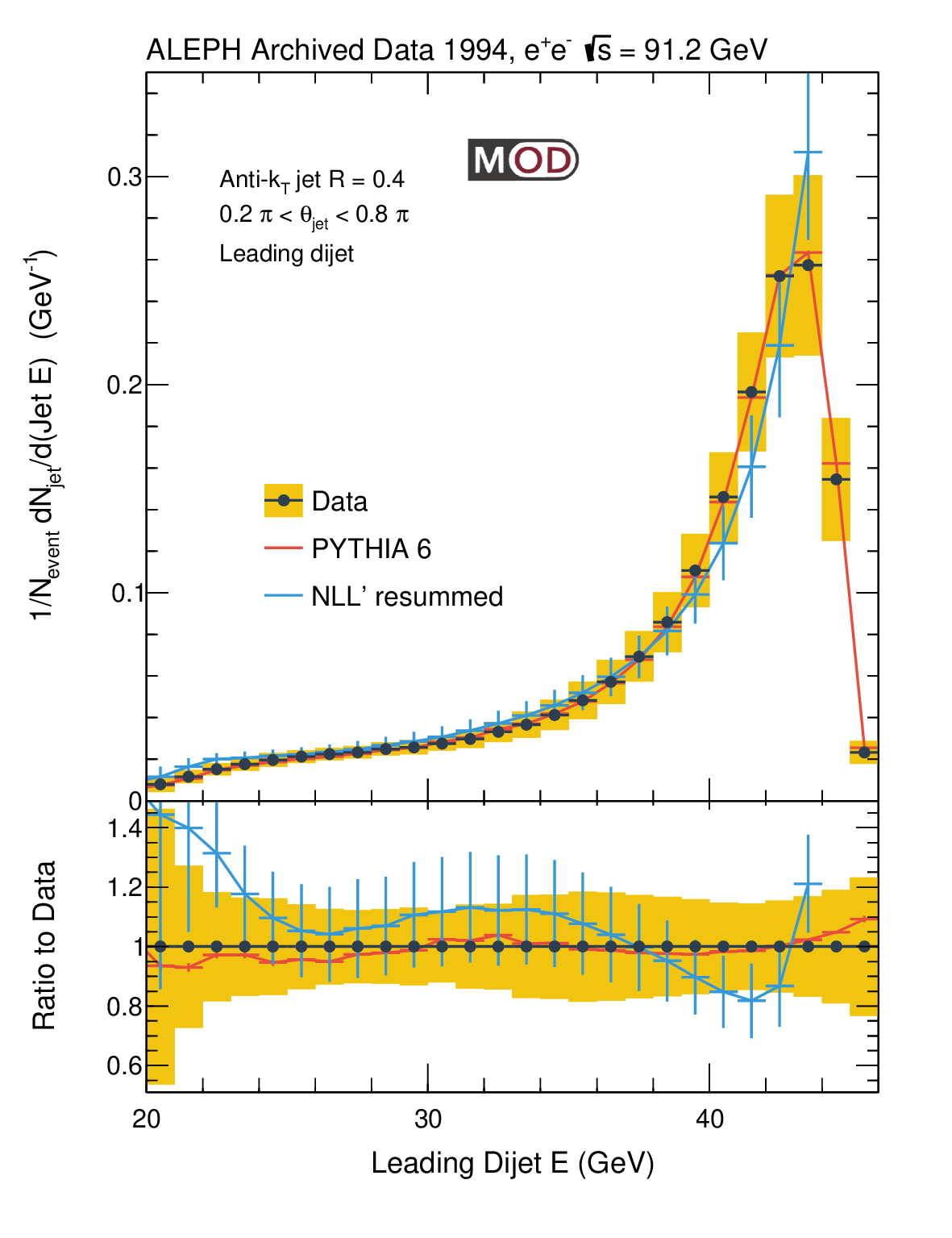}
  \caption{Measurement of energy spectrum (left) and leading dijet energy
    (right) of anti-$k_t$ jets in ALEPH archived data. Experimental data are
    compared to Monte Carlo simulation (red), NLO fixed-order prediction (green)
    and resummed prediction from~\cite{Neill:2021std}
    (blue). From~\cite{Chen:2021uws}.}
  \label{fig:alephEJ}
\end{figure}

The anti-$k_t$ algorithm, or more in general the gen-$k_t$ family of algorithms,
has been designed after LEP times, so it has not been used for LEP analyses.
However, in the recent years, archived ALEPH data (which became publicly
available in the Open Data format) have been reanalysed with modern jet
clustering techniques.
As an example, in Fig.~\ref{fig:alephEJ} we show two measurements of anti-$k_t$
jets (with $R = 0.4$) in $e^+e^-$ collisions~\cite{Chen:2021uws}: the
distribution of the energy of the jets, of all the jets in the event or of the
two most energetic jets in the event.%
We further show theory predictions with a fixed-order calculation, a resummed
calculation and a Monte Carlo simulation.
The following sections will provide a detailed explanation of how this type of
predictions are obtained.

In view of future colliders, recent studies tried to understand how the
gen-$k_t$ algorithm would behave at FCC-ee~\cite{genktFCC}, in the context of
QCD studies at the $Z$-pole, as done at LEP. Indeed, $Z$ production at 91~GeV
has a branching ratio to jets of 70\%, so the vast majority of events involve
QCD in the final state.
However, FCC-ee will involve more energy scales than LEP: in particular, the
processes $e^+e^- \to W^+W^-$ and $e^+e^- \to H Z$ will be accessible at higher
centre-of-mass energies, see Sec.~\ref{sec:hdecay} and Fig.~\ref{fig:higgsxs}.
So it becomes crucial to investigate the potential of using inclusive jet
algorithms to reconstruct the invariant mass peak from colour-singlet decays, to
separate these events from the QCD background.

\section{Event shapes}\label{sec:evshapes}

As alternative to jets, {\em event shapes} provide a way of looking at a
collider event in its entirety.
They are functions of the momenta of the final-state objects (partons, hadrons,
detector inputs), designed to describe the geometry of an event.
Historically, they have played a pivotal role in QCD precision studies at
previous high-energy lepton colliders.
Many event shapes have been proposed, e.g.\ thrust~\cite{Farhi:1977sg},
energy-energy correlators (EEC)~\cite{Basham:1978bw}, heavy jet
mass~\cite{Clavelli:1979md}, C-parameter~\cite{Parisi:1978eg} or
spherocity~\cite{Georgi:1977sf}.

As you may notice from the years of publication of the seminal papers (between
1977 and 1979) event shapes have a long tradition, and they have been studied in
depth over the years.
For the purpose of this Chapter, we limit ourself to introduce in some detail
two of the most common event shapes: thrust and EEC.

Before proceeding, it is worth noting that the design of new event shapes for
studying collider events remains an active field of development, especially at
hadron colliders.
For instance, one can define the notion of a metric in the space of collider
events, and then measure the distance between events or between events and
reference geometries~\cite{Cesarotti:2024tdh,Cai:2024xnt}. An example is the
event isotropy~\cite{Cesarotti:2020hwb}, quantifying the distance between an
event and a spherically-symmetric event, useful to disentangle signals with
quasi-isotropic signatures.

\subsection{Thrust}\label{sec:thrust}

\begin{figure}
  \centering
  \includegraphics[width=0.55\textwidth]{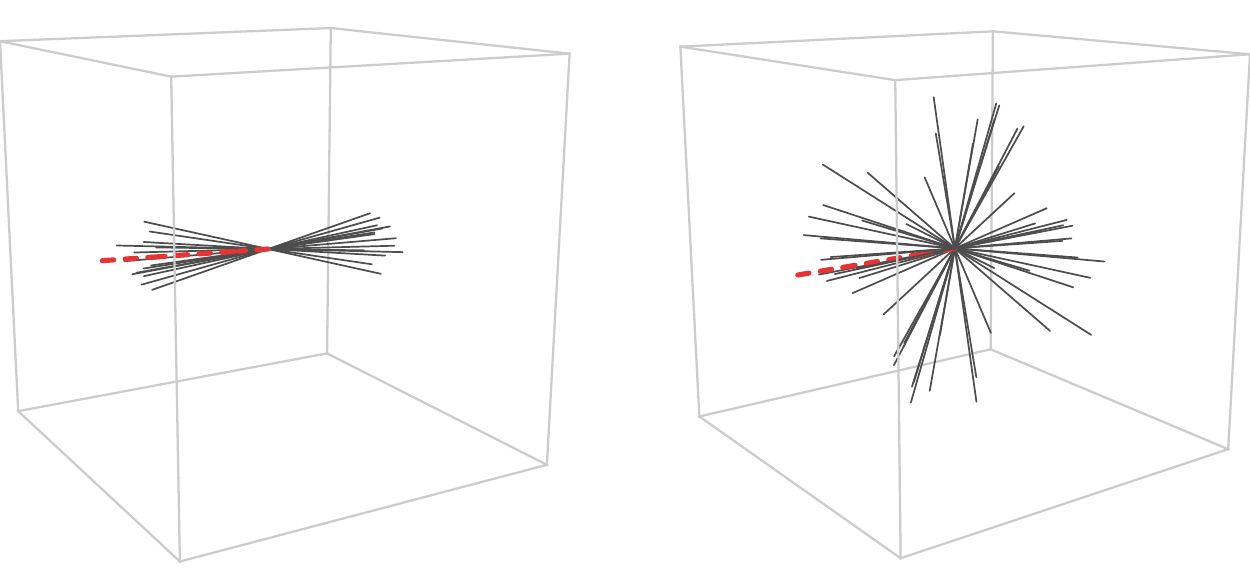}\,
  \includegraphics[width=0.4\textwidth]{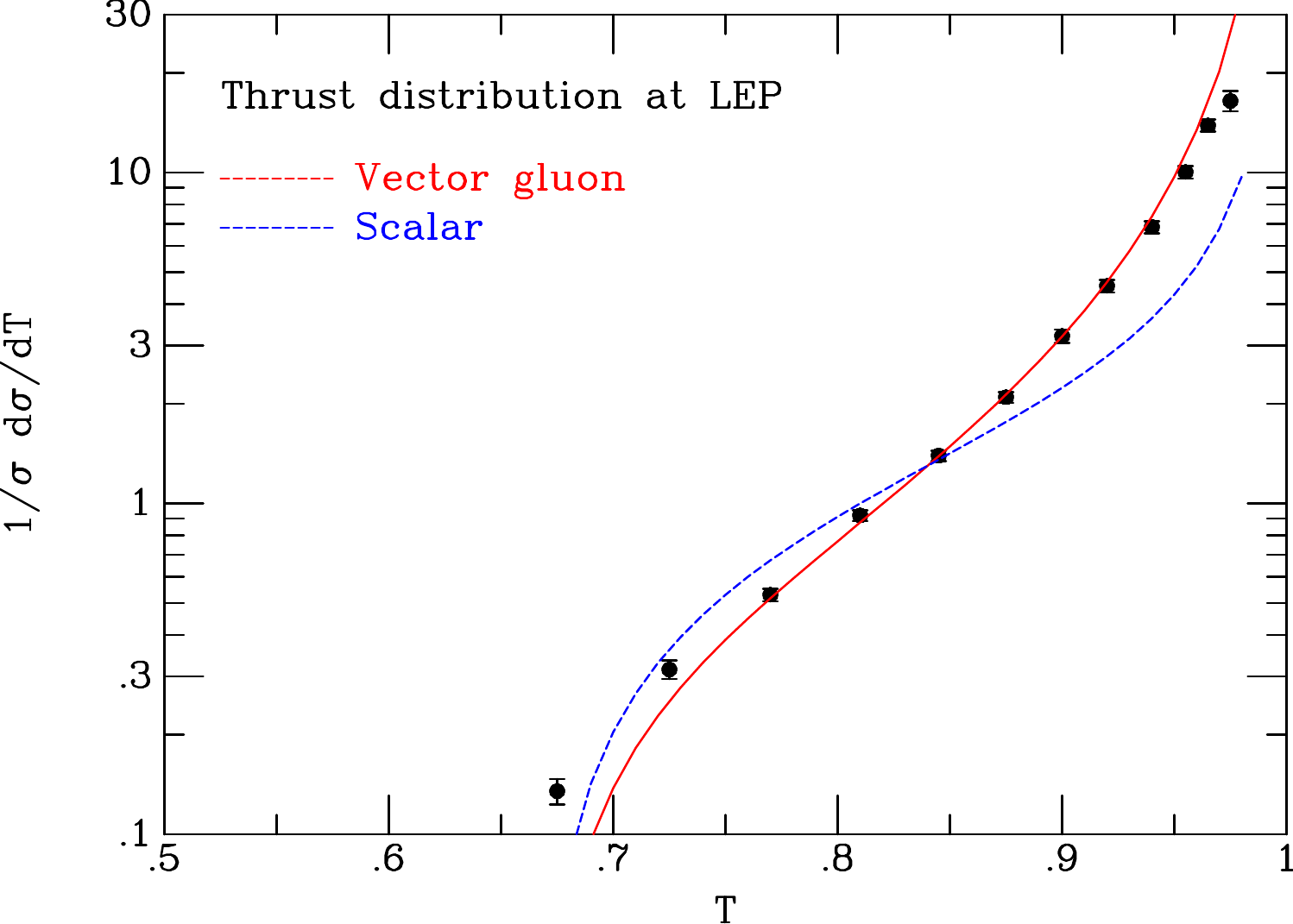}
  \caption{Pencil-like event with large $T \sim 0.998$ and almost spherical
    event with small $T \sim 0.65$, with the dashed red line displaying the
    thrust direction (left, from~\cite{Becher:2018gno}). Predictions for the
    thrust with scalar (blue) or vector gluon (red) compared to data from LEP
    (right, from~\cite{Ellis:1996mzs}).}
  \label{fig:thrust}
\end{figure}

For a final state with $m$ particles with three-momentum $\vec{p}_i$, the thrust
is defined as:
\begin{equation}\label{eq:thrust}
  T_m = \max_{\vec{n}} \frac{\sum_{i=1}^m |\vec{p}_i \cdot
    \vec{n}|}{\sum_{i=1}^m |\vec{p}_i|}\,,
\end{equation}
where the maximum is over all three-momenta $\vec{n}$ of unit norm.
The three-momentum maximizing~\eqref{eq:thrust} is called the {\em thrust axis},
$\vec{n}_T$, and represents the direction along which the longitudinal momentum
is maximized.
The allowed kinematical range for the thrust is between $1/2$ (perfect isotropic
event) and $1$ (perfect 2-jet event), see Fig.~\ref{fig:thrust} on the left.
Note that the value of $1/2$ is approached only in the limit of $n \to \infty$
particles in the final state.
For instance, in the case of a 3-particle final-state, the thrust axis is the
direction of the highest-energy parton, and an explicit expression for the
thrust is
\begin{equation}
  T_3 = \max_{i} \frac{2 E_i}{\sqrt{s}}\,.
\end{equation}
Note that in this case the minimum value of $T_3$ is $2/3 > 1/2$.

We can explicitly check the IRC safety of the thrust: when a momentum $p_k$ is
soft, it drops out from both numerator and denominator, hence $T_{m+1} \to
T_{m}$; when two momenta $p_i$ and $p_j$ are collinear i.e.\ $\vec{p}_i = z
\vec{p}$ and $\vec{p}_j = (1-z) \vec{p}$, with a parent vector $\vec{p} =
\vec{p}_i + \vec{p}_j$, in the numerator we have $|\vec{p}_i \cdot \vec{n}| +
|\vec{p}_j \cdot \vec{n}| = |\vec{p} \cdot \vec{n}|$ and analogously for the
denominator, hence $T_{m+1} \to T_m$. The key property in proving IRC safety for
many event shapes is the linearity in particle momenta e.g.\ event shapes with a
quadratic dependence in particle momenta (like the sphericity, a variable used
in early QCD studies~\cite{Ellis:1976uc}) are collinear unsafe.

Measurement of the thrust distributions have been instrumental in determining
the vectorial nature of the gluon. In Fig.~\ref{fig:thrust} on the right, we
show the predictions at $\mathcal{O}(\as)$ for a scalar and a vector gluon,
compared to data from LEP.
A scalar gluon does not feature a divergence in the soft limit, but only in the
collinear one, so the distribution is less peaked in the two-jet
limit~\cite{Salam:2010zt}.
Data clearly favour the vectorial gluon. The inclusion of higher order effects
improve the agreement at small $T$, as we will discuss in Sec.~\ref{sec:hopred},
whereas resummed predictions improve the agreement at large $T$, as we will
discuss in Sec.~\ref{sec:resjets}.

\subsection{Energy-Energy Correlators (EEC)}\label{sec:EEC}

\begin{figure}
  \centering
  \includegraphics[width=0.25\textwidth]{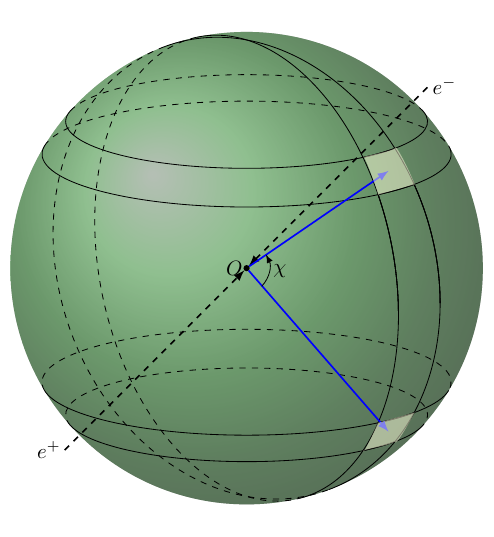}\quad
  \includegraphics[width=0.45\textwidth]{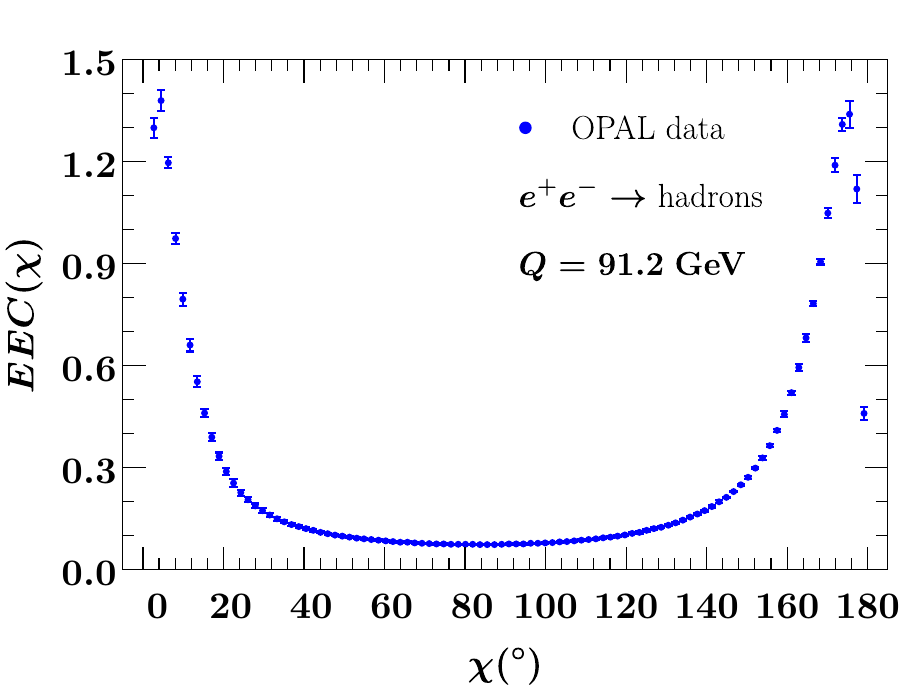}
  \caption{Pictorial representation of EEC (left, from~\cite{Moult:2018jzp}).
    LEP data for the EEC (right, from~\cite{Neill:2022lqx}).}
  \label{fig:EEC}
\end{figure}

Energy-Energy Correlators (EEC) are energy-weighted differential distributions
in the angle $\chi$ between pair of hadrons:
\begin{equation}\label{eq:EECdef}
  {\rm EEC}({\chi}) \equiv \frac{1}{\sigma_{\rm tot}} \frac{\mathrm{d}\Sigma}{\mathrm{d}\cos\chi}
  = \sum_{i,j=1}^n \int \frac{\mathrm{d}\sigma}{\sigma_{\rm tot}}\,
  \frac{E_i}{Q} \frac{E_j}{Q}\,\delta(\cos\chi - \cos\chi_{ij})\,,
\end{equation}
where sum runs over all hadron pairs $(i,j)$ and $\chi_{ij}$ is the opening angle between $i$ and $j$.
The definition is such that the EEC is normalised to 1, because of
\begin{equation}
  \int_{-1}^{+1} \mathrm{d}\cos\chi\,\frac{\mathrm{d}\Sigma}{\mathrm{d}\cos\chi} = \sigma_{\rm tot}\,.
\end{equation}
A pictorial description of EEC is provided in Fig.~\ref{fig:EEC} on the left.
The energy weights $E_i E_j / Q$ in~\eqref{eq:EECdef} are necessary to ensure
IRC safety of the EEC: soft particles trivially do not contribute to EEC;
collinear splittings e.g.\ a hadron with energy $E_i$ into two hadrons with
energies $x E_i$ and $(1-x) E_i$, do not change the overall contribution to the
EEC, because $x E_i + (1-x) E_i = E_i$.

At variance with the thrust, the EEC features two singular regions: the {\em
  back-to-back} region, when $\chi \to \pi$, and the {\em collinear region},
when $\chi \to 0$. In both regions, the cross section is enhanced, as it can be
noticed from Fig.~\ref{fig:EEC} on the right: LEP data clearly show two distinct
peaks at the two kinematical extreme.
The proper theoretical description of EEC in these regions require the all-order
treatment of soft and/or collinear effects through resummation, see
Sec.~\ref{sec:allorder}.

Given the simple definition of EEC in terms of individual particles in the final
state, they are easy to measure. They are relatively insensitive to
hadronization thanks to energy weighting.
Recently, new measurements of EEC appeared thanks to analyses of archived ALEPH
data~\cite{Bossi:2024qeu,Bossi:2025xsi}. These new developments provide also
measurements of EEC by using only information from tracking detectors, which
provide precise angular resolution.
On the theory side, predictions for EEC on tracks require the introduction of
{\em track functions}, non-perturbative functions describing the fragmentation
of quarks and gluons into charged hadrons~\cite{Jaarsma:2023ell}.

From field-theoretic point of view, the EEC or energy correlators in general can
be formulated in terms of matrix elements of energy flow
operators~\cite{Hofman:2008ar}.
The study of formal properties of energy correlators has seen much development
in the recent years.
We refer the reader to~\cite{Moult:2025nhu} for a recent comprehensive review on
EEC.

\section{Jets and event shapes at higher orders}\label{sec:highord}

In this Section, we are going to discuss higher order corrections to jet
production and event shapes.
In Sec.~\ref{sec:QCDpred}, we begin with a recap on the structure of fixed-order
results and renormalization scale dependence of cross section.
In Sec.~\ref{sec:hopred} we then present predictions for jet processes and event
shapes up to high order in QCD.

\NOARXIV{We refer to the ``Perturbative QCD'' chapter for a more complete
  discussion about how to actually compute fixed order predictions (included
  treatment of infrared divergences and multi-loop calculations).}

\subsection{Structure of QCD predictions}\label{sec:QCDpred}

Perturbative series are expressed as an expansion in terms of the strong
coupling constant $\as(\mu_R)$, evaluated at the renormalization scale
$\mu_R$.
At high energies, $\as(\mu_R)$ is small (for instance, at the $Z$-boson mass
$M_Z$, $\as(M_Z) \sim 0.118$) and the perturbative expansion is justified.
We define the N$^k$LO cross section $\sigma_k(Q,\mu_R)$ as:
\begin{equation}\label{eq:sigNkLO}
  \sigma_k(Q,\mu_R) = \sigma_0(Q,\mu_R)
  \left[ 1 + \sum_{n=1}^{k} \left(\frac{\as(\mu_R)}{2\pi}\right)^n \Delta\sigma^{(n)}(Q,\mu_R) \right] \,,
\end{equation}
where we have factored out the leading order (LO) cross section $\sigma_0(Q,\mu_R)$.
We refer to $\Delta\sigma^{(1)}(Q,\mu_R)$ as the next-to-leading order (NLO) coefficient,
$\Delta\sigma^{(2)}(Q,\mu_R)$ as the next-to-next-to-leading order (NNLO) coefficient,
or more generally to $\Delta\sigma^{(k)}(Q,\mu_R)$ as the N$^k$LO coefficient.
$\sigma_k$ may denote a total cross section, possibly with fiducial cuts, or a
cross section differential in a specific observable.

In order to obtain higher order corrections to a specific observable, one would
need to integrate the relevant squared matrix elements over the respective phase
spaces.
In general, given a matrix element squared $|M_l^\ell|^2$ with $l$ final-state
legs and $\ell$ loop integrations, we would need to evaluate\footnote{Note that
$\sigma_l^\ell$ is usually not infrared finite: in real configurations,
singularities arise when performing the phase space integration in the
unresolved (soft and collinear) regions of additional emissions; in virtual
configurations singularities are already present inside $|M_l^\ell|^2$, when
performing the loop integration in the region of small loop momentum. However,
their sum is guaranteed to be finite, thanks to the KLN
theorem~\cite{Kinoshita:1962ur,Lee:1964is} and the IRC safety of the
observable. We assume that a suitable regularization of divergences have taken
place in the intermediate steps of the calculation.\NOARXIV{See Chapter BLABLA
  for a detailed discussion about QCD infrared divergences and their treatment
  in actual calculations.}}
\begin{equation}
  \sigma_l^\ell = \int \mathrm{d}\Phi_l |M_l^\ell|^2 \,V_l(\Phi_l)\,.
\end{equation}
The function $V$ is a measurement function, specifying possible phase space cuts
(through Heaviside step functions) and, in case of differential distributions,
the observable definition (through Dirac delta functions).
For hadronic final states at $e^+e^-$ colliders, it is easy to realize that
$\sigma_l^\ell$ contributes at $\mathcal{O}(\as^{l-2+\ell})$.

\begin{figure}
  \centering
  \includegraphics[width=0.55\textwidth]{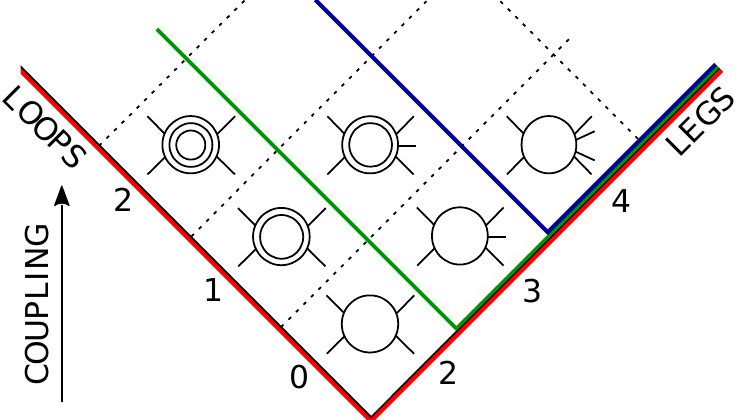}
  \caption{Structure of higher order contributions in perturbative QCD
    calculations. Additional loops are denoted by concentric circles. Each
    horizontal line contains all $\sigma_l^\ell$s contributing to the same power
    of $\as$.}
  \label{fig:loops-legs}
\end{figure}

An important point we would like to make is that what we define as LO, NLO,
etc. depends on the observable we consider.
For instance, the two-jet cross section starts at order $\as^0$, with NLO
corrections of $\mathcal{O}(\as)$, NNLO corrections of $\mathcal{O}(\as^2)$
etc., whereas the three-jet cross section starts at order $\as^1$, with NLO
corrections of $\mathcal{O}(\as^2)$, etc.
However, the NLO of two-jet will feature contributions from both $\sigma_2^1$
and $\sigma_3^0$, whereas the LO of three-jet only from $\sigma_3^0$ (it is
required to have at least three partons to reconstruct three jets).
The situation is depicted in Fig.~\ref{fig:loops-legs}.
More loops in virtual diagrams or more legs in real diagrams increase the power
of the strong coupling constant.
Each tree-level process is at the tip of an inverted pyramid, and it will
receive perturbative corrections from the matrix elements above it inside the
pyramid e.g.\ NLO corrections to two-jet (three-jet) production correspond to
the first line of the red (green) inverted pyramid.

Another important thing to notice is that the N$^k$LO cross section
$\sigma_k(Q,\mu_R)$ in~\eqref{eq:sigNkLO} depends on the value of the
renormalization scale $\mu_R$.
The all-order result (i.e.\ $k \to \infty$) would be independent of $\mu_R$, but
truncated cross sections retain a residual dependence on $\mu_R$.
The dependence on $\mu_R$ in the $\Delta\sigma^{(n+1)}$ coefficient is a
function of lower order terms only, see explicit expressions in
e.g.~\cite{Cacciari:2011ze}.

In order to calculate the cross section we then need to make a choice for
$\mu_R$. What is a good choice for $\mu_R$?
Renormalization scale always appears in fractions like $\mu/Q$ inside a
logarithmic term, where $Q$ is a characteristic hard scale of the process.
Thus, choosing a scale $\mu$ radically different from $Q$ generates large
logarithms which deteriorates the convergence of the perturbative series.  The
simple choice $\mu = Q$ could be a good choice for processes in which only one
characteristic energy appears e.g.\ usually the inclusive ones, such as $e^+e^-
\to$ hadrons\footnote{Actually there is no reason to have $\mu$ \emph{exactly}
equal to $Q$. If we choose for example $\mu = 2Q$ or $\mu = Q/2$ for a fixed
order prediction up to order $p$, there are small logarithms which appear at
order $\mathcal{O}(\as^{p+1})$, which are anyway summed to an unknown
coefficient.}.

The situation is quite different for processes which naturally involve different
physical scales, for example processes with a more exclusive final state.
For instance, in the case of the predictions for thrust, a physical scale
$\mu^2$ much smaller than $Q^2$ seems to be required to describe data in the
region $T \to 1$. However, this is an artifact of the lack of resummation in
that region: with proper resummed predictions, data no longer show a preference
for $\mu^2 \ll Q^2$, see e.g.~discussion in~\cite{Catanilectures}. In other
words, a smaller renormalization scale $\mu^2 \sim (1-T) Q^2$ is mimicking the
effect of resummation in that region.

In any case, while there could be scale choices which are {\em better} than
others, there is never a {\em best} choice, as fixing a scale does not remove
the theoretical error on the prediction: for a quantity up to order
$\mathcal{O}(\as^{p})$, there will always be an error of order
$\mathcal{O}(\as^{p+1})$.
This is referred to as {\em missing higher order uncertainty}.
The residual scale dependence in fixed-order predictions provides a prescription
to estimate these higher order uncertainties: the usual recipe consists in
varying $\mu_R$ within the range $[Q/2,2Q]$ and then taking the minimum and
maximum of the resulting envelope ({\em scale variation} procedure).
Note that a priori there is no reason why scale variation should represent a
sensible estimate for missing higher orders, because the former depends on the
known coefficient, whereas the latter by definition depends on the {\em unknown}
coefficients.
However, it can be shown that the interval given by scale variation and the
remainder of the series are comparable under the assumption that all the
coefficient in the series share the same magnitude~\cite{Cacciari:2011ze}, and
thus the use of scale variation is in some sense justified.

\subsection{Predictions for jet rates and event shapes}\label{sec:hopred}

In this section, we present fixed-order predictions for jet observables at
lepton colliders. We limit ourselves to reporting results rather than explaining
the methods used to obtain them.

NLO corrections to three-jet observable in $e^+e^-$ annihilation and event
shapes were first calculated in the seminal paper~\cite{Ellis:1980wv}.
Note that this is also the first paper to introduce a numerical method for
calculation of differential NLO QCD corrections.
Over the time, more general and flexible approaches for NLO calculations have
been proposed, such as FKS~\cite{Frixione:1995ms} or
Catani-Seymour~\cite{Catani:1996vz}.
For instance, NLO corrections to two- and three-jet production in $e^+e^-$ are
readily available through the \code{Event2} computer code based on
Catani-Seymour subtraction.
Even though numerical predictions offer more flexibility, there is still
theoretical interest in producing analytical predictions for standard candle
observables.
A recent analytic result for the NLO corrections to the EEC has been presented
in~\cite{Dixon:2018qgp}: the calculation required state-of-the-art techniques,
with the final result turning out to be ``remarkably simple''.

\begin{figure}
  \centering
  \includegraphics[width=0.45\textwidth]{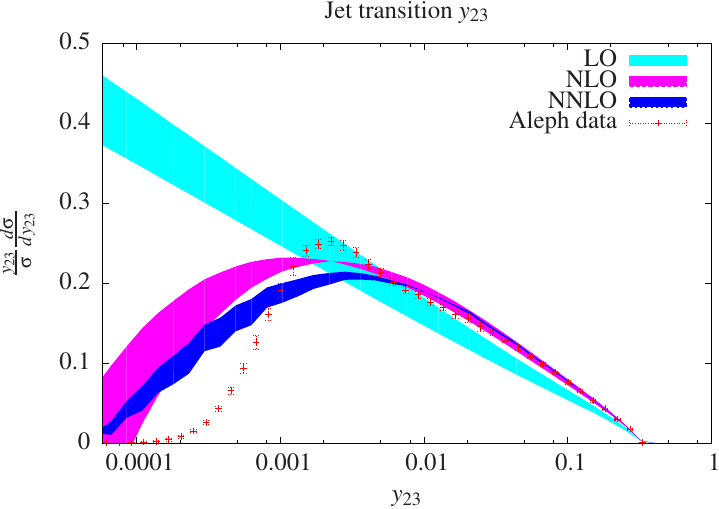}\,
  \includegraphics[width=0.45\textwidth]{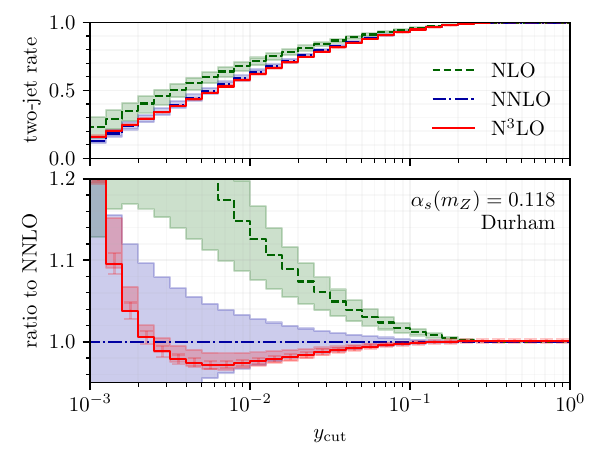}\,
  \caption{Predictions for Durham $y_{23}$ transition variable (left,
    from~\cite{Weinzierl:2009ms}) and for the Durham two-jet rate (right,
    from~\cite{Chen:2025kez}.}
  \label{fig:jetHO}
\end{figure}

\begin{figure}
  \centering
  \includegraphics[width=0.5\textwidth]{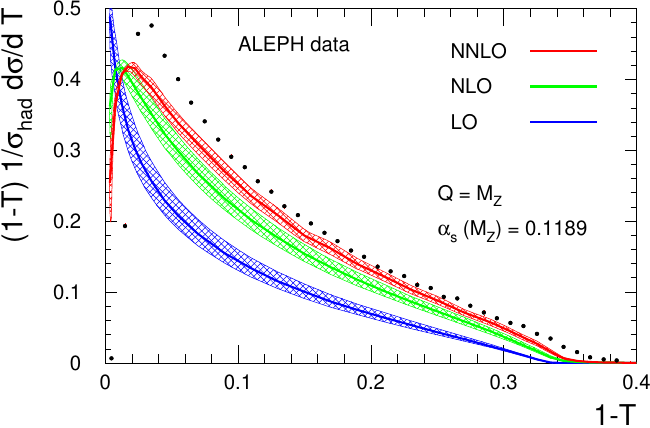}\,
  \includegraphics[width=0.45\textwidth]{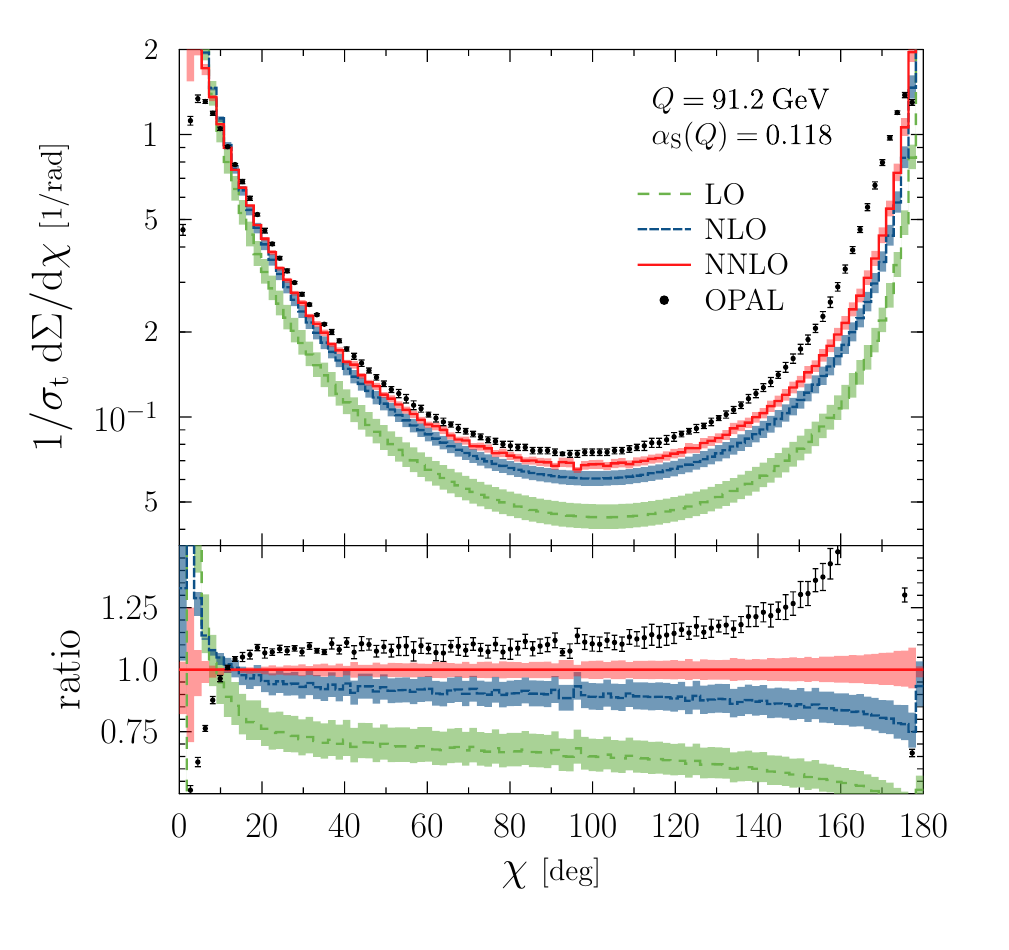}
  \caption{Predictions for thrust (left, from~\cite{Gehrmann-DeRidder:2007vsv})
    and for EEC (right, from~\cite{Tulipant:2017ybb}).}
  \label{fig:evshHO}
\end{figure}

Numerical predictions at NNLO for three-jet production and event shapes first
appeared around 2007.
Results for jet rates~\cite{Gehrmann-DeRidder:2008qsl}, event
shapes~\cite{Gehrmann-DeRidder:2007vsv} and moments of event
shapes~\cite{Gehrmann-DeRidder:2009fgd} at NNLO have been obtained with the
antenna subtraction
method~\cite{Gehrmann-DeRidder:2005btv,Gehrmann-DeRidder:2007foh}.
These results are available through the \code{EERAD3} computer
program~\cite{Gehrmann-DeRidder:2014hxk}.
The same results have also been obtained with an independent implementation of
the antenna subtraction method by another
group~\cite{Weinzierl:2008iv,Weinzierl:2009ms,Weinzierl:2009yz}.
Later, the CoLoRFulNNLO method~\cite{DelDuca:2016csb,DelDuca:2016ily} was used
to validate the results
of~\cite{Gehrmann-DeRidder:2008qsl,Gehrmann-DeRidder:2007vsv,Gehrmann-DeRidder:2009fgd,Weinzierl:2008iv,Weinzierl:2009ms,Weinzierl:2009yz}
and produce new predictions, like the EEC up to NNLO.
As an example, we show fixed-order predictions for the Durham $y_{23}$
transition variable in Fig.~\ref{fig:jetHO} on the left, for the thrust in
Fig.~\ref{fig:evshHO} on the left and for the EEC in Fig.~\ref{fig:evshHO} on
the right. In all plots, LEP data from the OPAL or ALEPH experiment are also
shown.
We first note that when the value of the observable is not close to kinematical
limits (e.g.\ small values of $y_{23}$ or large values of $T$), there is a
convergence of the perturbative series, with NNLO corrections moving theory
predictions towards experimental data.
We can further appreciate how scale variation bands are not really able to
capture the next term in the perturbative series, with the NNLO prediction often
lying outside of the NLO envelope.

Note that predictions for jet rates and event shapes can be obtained from $1 \to
n$ matrix elements for the decay of a virtual boson i.e.\ $\gamma^* \to$
partons\footnote{Neutral current effects with the exchange of a virtual
$Z$-boson can be obtained by simply rescaling the photon-only cross section with
electroweak factors. Singlet diagrams (which spoil this factorised picture) are
usually numerically negligible.}.
Other observables, such as angular correlations between the hadronic final state
and the incoming beams, require the full $2 \to n$ matrix elements for $e^+e^-
\to$ partons e.g.\ five-parton tree-level amplitudes~\cite{Hagiwara:1988pp,
  Berends:1988yn, Falck:1988gc}, four-parton one-loop
amplitudes~\cite{Glover:1996eh,Campbell:1997tv,Bern:1997sc}, three-parton
two-loop amplitudes~\cite{Garland:2001tf,Garland:2002ak}.
First NNLO predictions for event-orientation variables have been calculated
in~\cite{Gehrmann:2017xfb}.
The $e^+e^- \to$ 3 jets process (with related event shapes) is publicly
available within the \code{NNLOJET} code~\cite{NNLOJET:2025rno}.

Finally, very recently, the Durham two-jet rate has been computed as a function
of $y_{\rm cut}$ up to N$^3$LO, see Fig.~\ref{fig:jetHO} on the right.
The two-jet rate at N$^3$LO was already obtained
in~\cite{Gehrmann-DeRidder:2008qsl} by exploiting the three-jet cross section at
NNLO and knowledge of inclusive total cross section up to
N$^3$LO~\cite{Gorishnii:1990vf}: by definition, the two-jet rate is the difference
between the total inclusive cross section and the cross section for the
production of at least three jet.
However, Ref.~\cite{Chen:2025kez} represents the first direct calculation of
two-jet production, by employing a genuine N$^3$LO subtraction method.

\section{Resummation and parton showers}\label{sec:allorder}

In the previous Sections, we have observed the breakdown of perturbation theory
in certain kinematical regions of the phase space.
In these regions, there are logarithmically enhanced contributions that appear
at any perturbative order in $\as$. For instance, in the case of thrust, there
are terms like $\as^n L^m$, with $L = \log(1-T)$, appearing in fixed-order
predictions.
When $T \to 1$, these terms are $O(1)$, compensating the smallness of $\as$ and
then spoiling the convergence of the perturbative series in $\as$.
It is then necessary to go beyond fixed-order perturbation theory and {\em sum}
them to all orders: this is what we mean by {\em resummation}.  Of course, an
{\em exact} all-order treatment is hopeless: the goal is to resum the class of
terms responsible for the large logarithms, associated to soft and/or collinear
emissions%
\footnote{An example of resummation is the running of $\as$ itself: by solving
the QCD $\beta$-function to find $\as(\mu_R)$ as a function of $\as(Q)$, we are
effectively resumming the logarithms $\log(\mu_R/Q)$.}.

As a warmup example, in Sec.~\ref{sec:jetmass} we explain how to get resummed
predictions for the mass of a jet.
In Sec.~\ref{sec:resjets}, we discuss resummed predictions for jet rates, thrust
and EEC, possibly matched to the fixed-order results of Sec.~\ref{sec:hopred}.
Finally, in Sec.~\ref{sec:psevgen}, we introduce the concept of {\em parton
  shower}, which is deeply linked to resummation; parton showers are usually
embedded inside {\em event generators}, important tools for collider
phenomenology.

\subsection{Example of resummation: the jet mass at LL}\label{sec:jetmass}

\begin{figure}[t]
  \centering
  \includegraphics[width=0.25\textwidth]{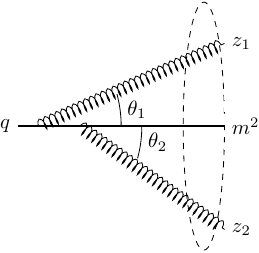}
  \caption{Schematic of a pair of emissions in a jet.}
  \label{fig:jetmassdraw}
\end{figure}

We present a simple calculation of the dominant logarithmic corrections to the
jet mass, based on~\cite{Marzani:2019hun,Larkoski:2017fip}.
Let us consider anti-$k_t$ jets with some radius $R$, obtained with the
algorithm presented in Sec.~\ref{sec:genkt}.
The jet mass is defined as the squared sum of the 4-momenta of all particles in
a jet:
\begin{equation}\label{eq:jetmass}
  m^2 = \left( \sum_{i \in jet} k_i \right)^2\,.
\end{equation}
Note that a jet constituted of a single quark or gluon has zero mass: hence, a
jet acquires mass through radiative emissions. When the emissions are close in
angle, we can rewrite~\eqref{eq:jetmass} as:
\begin{equation}\label{eq:jetmass2}
  m^2 = R^2 E_J^2 \sum_{i \in {\rm jet}} z_i\, \vartheta_i^2\,,
\end{equation}
with $z_i$ the momentum fraction of the emission (rescaled by the jet energy
$E_J$) and $\vartheta_i$ the angular distance from the jet axis (rescaled by the
jet radius $R$), see Fig.~\ref{fig:jetmassdraw}. For later convenience, we
define the normalised jet mass $\rho = m^2 / (R^2 E_J^2)$ and the individual
contributions $\rho_i = z_i \vartheta_i^2$, such that $\rho = \sum_i \rho_i$
according to~\eqref{eq:jetmass2}.

Our aim is to calculate the cumulative distribution $\Sigma(\rho)$, which
corresponds to the probability of measuring a mass smaller than $\rho$'. By
definition, it is given by the integral of the differential distribution up to
$\rho' = \rho$, normalised to total cross section:
\begin{equation}\label{eq:cumxs}
  \Sigma(\rho) = \frac{1}{\sigma} \int_0^{\rho} \mathrm{d}\rho' \frac{\mathrm{d}\sigma}{\mathrm{d}\rho'}
\end{equation}
Calculating the cumulative distribution is often easier than calculating the
differential distribution.
It is always possible to recover the original $\mathrm{d}\sigma/\mathrm{d}\rho$
distribution by differentiating~\eqref{eq:cumxs}.

Let us write an ansatz for $\Sigma(\rho)$, to be justified below:
\begin{equation}\label{eq:sigrho}
\Sigma(\rho) = \sum_{n=0}^{\infty} \frac{1}{n!} \left(
\prod_{i=1}^n
\int \mathrm{d}z_i \mathrm{d}\vartheta_i^2 P(z_i,\vartheta_i^2)
\left[\Theta(\vartheta_i < 1) \Theta\left(\sum_{j=1}^n \rho_j < \rho \right)
  + \Theta(\vartheta_i > 1) - 1\right] \right)\,.
\end{equation}
We are first summing over all possible number of emissions, with each emission
coming with a probability $P(z_i,\vartheta_i^2)$, as given in~\eqref{eq:Pqgzth}.
The real emissions inside jet ($\vartheta_i < 1$) are allowed only if they lead
to a value of the jet mass less than $\rho$ ($\sum_{j=1}^n \rho_j <
\rho$). Instead, real emissions outside jet ($\vartheta_i > 1$) as well as
virtual emissions (``-1'') are always allowed, as they do not affect the value
of jet mass.
Note that for the virtual contributions we have exploited unitarity, see
footnote~\ref{foot:unit}.

\begin{figure}
  \centering
  \includegraphics[width=0.45\textwidth]{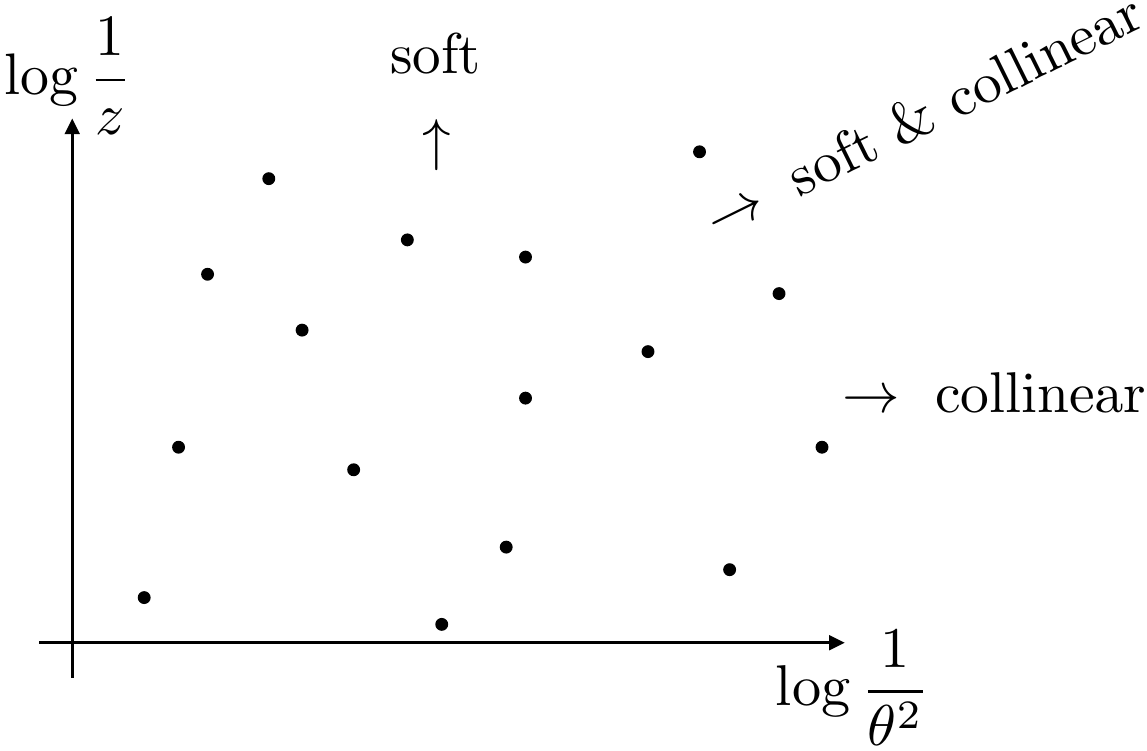}
  \includegraphics[width=0.4\textwidth]{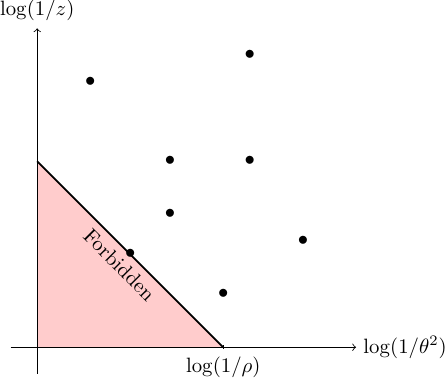}
  \caption{Lund plane (left) and forbidden region for jet mass (right).}
  \label{fig:Lplane}
\end{figure}

For sake of discussion, let us rewrite~\eqref{eq:Pqgzth}
\begin{equation}\label{eq:Pqgzth2}
P(z, \vartheta^2) \mathrm{d}z \mathrm{d}\vartheta^2
= \frac{\alpha_s \CF}{\pi} \frac{\mathrm{d}z}{z}\frac{\mathrm{d}\vartheta^2}{\vartheta^2}
= \frac{\alpha_s \CF}{\pi} \mathrm{d}
\left( \log \frac{1}{z} \right) \mathrm{d} \left(\log \frac{1}{\vartheta^2}\right)\,.
\end{equation}
The crucial observation in~\eqref{eq:Pqgzth2} is that emissions are uniform in
the $(\log 1/\vartheta^2, \log 1/z)$ plane, as depicted in Fig.~\ref{fig:Lplane}
on the left.
Such a plane (or variants thereof) is often referred to as the ``Lund plane''.
Eq.~\eqref{eq:Pqgzth2} implies that emissions are {\em exponentially} apart in
the $(\vartheta^2, z)$ physical plane.
Or in other words, that a single emission dominates the jet mass. Hence, the
theta function can be rewritten in a {\em factorised} form as
\begin{equation}\label{eq:JMfact}
\Theta\left(\sum_{i=1}^n \rho_i < \rho \right)
\simeq \Theta\left(\max_i \rho_i < \rho \right)
= \prod_{i=1}^n \Theta(\rho_i < \rho)\,.
\end{equation}

By exploiting our observations, we obtain {\em exponentiation}
in~\eqref{eq:sigrho}:
\begin{equation}\label{eq:sigrhoLL}
\Sigma(\rho) = \sum_{n=0}^{\infty} \frac{1}{n!} \left( -
\int \mathrm{d}z_i \mathrm{d}\vartheta_i^2 P(z_i,\vartheta_i^2)\,
\Theta(\vartheta_i < 1) \Theta (\rho_i > \rho) \right)^n \equiv \exp\left[-V(\rho)\right]\,,
\end{equation}
with $V(\rho)$ defined as
\begin{equation}\label{eq:Vrho}
V(\rho) = \frac{\alpha_s \CF}{\pi}
\int_0^1 \frac{\mathrm{d}z_i}{z_i} \int_0^1 \frac{\mathrm{d}\vartheta_i^2}{\vartheta_i^2}\,
\Theta (z_i\vartheta_i^2 > \rho)
= \frac{\alpha_s \CF}{\pi} \frac{1}{2} \log^2 \frac{1}{\rho}\,.
\end{equation}
The graphical interpretation in the Lund plane is that we are vetoing all real
contributions that would lead to a value of the mass larger than $\rho$, see
Fig.~\ref{fig:Lplane}. Then only virtual contributions survive in the forbidden
region: $V(\rho)$ is called {\em no-emissions probability} and it is then seen
to be proportional to the area of the forbidden region.
In~\eqref{eq:sigrhoLL}, we are then exponentiating virtual emissions (as can be
understood from the overall minus sign). Such exponential is usually referred to
as {\em Sudakov factor}.
Note that in the limit of $\rho \to 0$, the Sudakov form factor is vanishing,
$\Sigma(\rho) \to 0$ i.e.\ the probability of not having radiation (hence a jet
mass equal to zero) is zero: any scattering process is always accompanied by the
emissions of arbitrarily soft radiation.

The result obtained in~\eqref{eq:sigrhoLL}-\eqref{eq:Vrho} achieves the
resummation of the {\em double-logarithmic} terms, which are of the form $O(\as
L^2)^n$, with $L = \log(1/\rho)$. Note that in our derivation we have ignored
the running of $\as$, as it is formally subleading (even though usually
numerically important). A proper scale for the evaluation of $\as$ is the
relative transverse momentum $k_t$, see Sec.~\ref{sec:LP}.

\subsection{Resummation of jet observables in QCD}\label{sec:resjets}

Generally speaking, resummed predictions for an observable will involve
double-logarithmic terms $(\as^2 L^{2n})$ (DL), then next-to-double-logarithmic
terms $(\as^n L^{2n-1})$ (NDL), next-to-next-to-double-logarithmic terms $(\as^n
L^{2n-2})$ (NNDL), and so on.
In case of cumulative distributions like~\eqref{eq:cumxs}, the structure of the
resummed result for a generic observable $O$ is given by
\begin{equation}\label{eq:resNkDL}
  \Sigma(O < e^L) = h_1(\as L^2) + \sqrt{\as} h_2(\as L^2) + \as h_3(\as L^2) + \ldots\,,
\end{equation}
where we have made manifest the dependence on the large logarithm $L < 0$ we
wish to resum, with $|L| \gg 1$. The function $h_{k+1}$ is responsible for the
resummation of N$^k$DL terms.
However, in case an observable exponentiates\footnote{A notable example of an
observable that does {\em not} exponentiate are jet rates with the JADE
algorithm, see Sec.~\ref{sec:jadealg}.}, it is possible to count logarithms at
the exponent
\begin{equation}\label{eq:resexp}
  \Sigma(O < e^L) = \left(1 + \frac{\as}{2\pi} C_1 + \ldots \right)
  \exp( L g_1(\as L) + g_2(\as L) + \as g_3(\as L) + \ldots )
  \equiv C(\as) e^{G(\as,L)}\,,
\end{equation}
where the function $g_1$ encapsulates the resummation of the leading-logarithmic
terms $\as^n L^{n+1}$ (LL), $g_2$ the next-to-leading-logarithmic terms $\as^n
L^n$ (NLL), $g_3$ the next-to-next-to-leading-logarithmic terms $\as^{n}
L^{n-1}$ (NNLL), and so on.
The constant coefficient $C_1$ can be obtained by matching~\eqref{eq:resexp} to
a fixed-order calculation, and it contributes at NNLL, with $C_2$ contributing
at N$^3$LL, and so on.
By {\em exponentiation} we mean the property that terms like $\as^n L^m$, with
$m > n +1$, are absent from $\ln \Sigma$, whereas they appear in $\Sigma$
itself, as in~\eqref{eq:resNkDL}.
If an observable exponentiates and it is then possible to arrange the
resummation as in~\eqref{eq:resexp}, it is usually better, as one gains in
predictive power, because by organizing the resummation at the exponent we
automatically include terms that would be subleading in~\eqref{eq:resNkDL}.
For instance, for event shapes, NLL resummation implies NDL accuracy, and the
inclusion of $C_1$ is enough to achieve NNDL accuracy~\cite{Catani:1992ua}.

As already mentioned above, in order to obtain a prediction valid for any value
of the observable, we eventually need to perform a matching to fixed-order
calculations. The matched result for $v = e^L$ can be written as
\begin{equation}
  \Sigma_{\rm matched}(v) = C(\as) \exp G(\as,\ln v) + D(\as,v)\,,
\end{equation}
with $D \to 0$ as $v \to 0$. Note that there is no unique way to perform the
matching: the only requirements are that $\Sigma_{\rm matched}$ when expanded in
$\as$ should reproduce the fixed-order result and it should behave
as~\eqref{eq:resexp} in the $v \to 0$ region. So different matching
prescriptions may result in different subleading terms, beyond fixed-order and
logarithmic accuracy. An alternative matching prescription, could be
\begin{equation}
  \Sigma_{\rm matched}(v) = \exp \left[ K(\as) + G(\as,\ln v) + H(\as,v) \right] \,,
\end{equation}
with $H \to 0$ as $v \to 0$. In the literature, this matching is referred to as
``log-R''~\cite{Catani:1992ua}, and it basically amounts to an exponentiation of
non-logarithmic corrections.

It would be ambitious to attempt a comprehensive account of the rich history of
resummation in QCD.
In the following, we restrict ourselves to a brief introduction of the coherent
branching formalism, which underpins many resummed calculations for event shapes
and jet rates. We then outline the main ideas behind the CAESAR approach to
resummation and touch upon the use of effective field theories in this context.
For a more detailed---though still pedagogical—--discussion of resummation
techniques in QCD, we refer the reader to the review~\cite{Luisoni:2015xha}.

\subsubsection{Coherent branching formalism for NLL accuracy}\label{sec:cohbra}

\begin{figure}
  \centering
  \includegraphics[width=0.3\textwidth]{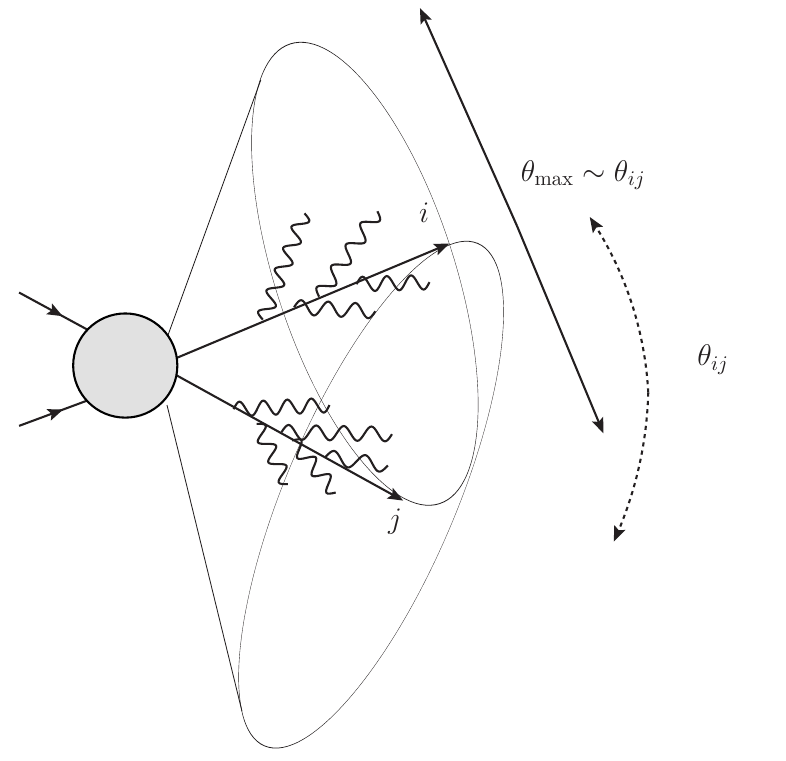}\,
  \includegraphics[width=0.3\textwidth]{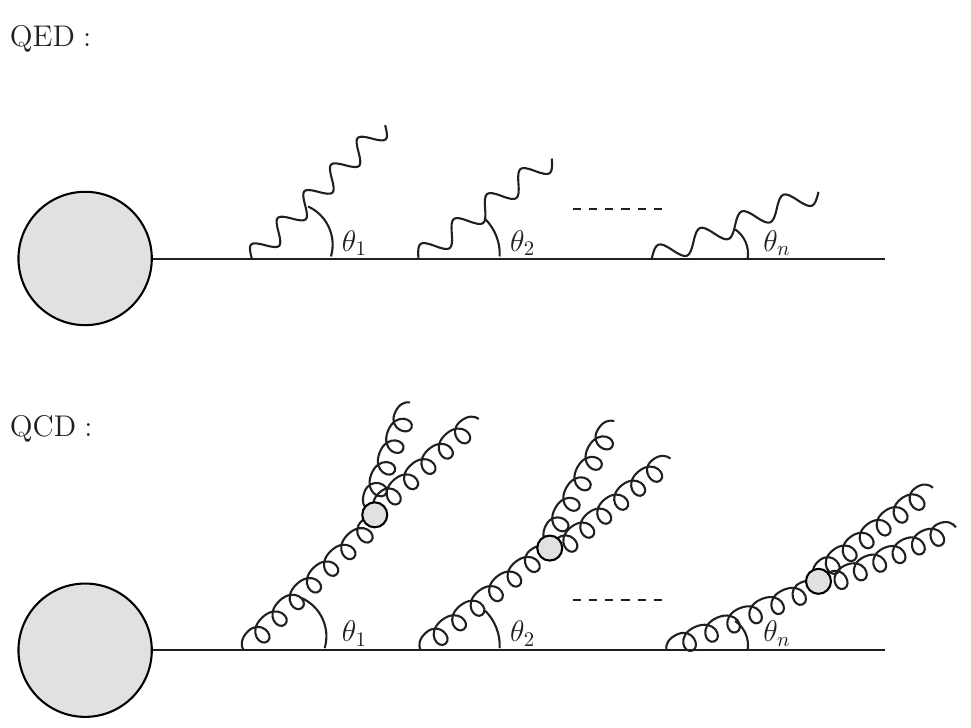}
  \caption{Pictorial representation of coherence and angular ordering constraint
    in QED and QCD. From~\cite{Luisoni:2015xha}.}\label{fig:coher}
\end{figure}

As a preliminary step toward introducing the concept of colour coherence, it is
useful to consider the (simpler) case of photon emissions in QED.
The seminal paper by Yennie-Frautschi-Suura (YFS)~\cite{Yennie:1961ad} in 1961
showed that soft photon emissions to all orders are given by a simple
exponentiation of lowest-order contribution.
The single-photon emission probability can be approximated by a sum of radiation
from {\em independent} emitters with {\em angular ordering} constraints. This is
possible because destructive interference cancels radiation at large
angles. This phenomenon is called {\em coherence}: it arises from quantum
mechanics (waves interfere) and gauge invariance (charge conservation).
See Fig.~\ref{fig:coher} for a pictorial representation.

In QCD, the main difference is that gluons carry colour charge, which
complicates the treatment of multiple emissions.
First of all, colour charge is a matrix; hence, in general, matrix elements are
{\em vectors} in colour space~\cite{Catani:1996vz} and the soft factor is a
matrix\footnote{Note that in the large-$\NC$ limit, the off-diagonal entries of
the matrices vanish, so the treatment of colour is largely simplified.}.
As a result, the single-soft gluon probability is not simply proportional to the
squared matrix element, but includes colour correlations with the hard legs in
the process.
However, in the leading collinear approximation, it is again possible to show
that coherence can be replaced by sum over independent emitters with angular
ordering constraints.
This is the basis of the coherent branching algorithm~\cite{Catani:1990rr},
resulting in an iterative, angular-ordered partonic cascade with a self-similar
structure.
Within the coherent branching formalism, resummed results at NLL accuracy for
global event shapes have been presented~\cite{Catani:1992ua}.

In addition to the angular-ordered constraint to take care of large-angle soft
emissions, two additional ingredients are required to have under control
single-logarithmic terms and claim NLL accuracy.
It is first necessary to promote~\eqref{eq:Pqgzth} to include the correct
$z$-dependence away from the soft limit, as encoded in the splitting functions
$P_{j \leftarrow i}(z)$:
\begin{equation}\label{eq:PSprob}
  \frac{\alpha_s \CF}{\pi} \, \frac{\mathrm{d}z}{z} \, \frac{\mathrm{d}\vartheta^2}{\vartheta^2}
  \longrightarrow
  \frac{\alpha_s}{2\pi}\,P_{j \leftarrow i}(z)\,\mathrm{d}z\,\frac{\mathrm{d}\vartheta^2}{\vartheta^2}
  \equiv \mathcal{P}(\Phi_{\rm rad})\,\mathrm{d}\Phi_{\rm rad}\,.
\end{equation}
As the notation suggests, the splitting function depends on the nature of the
particles $i$ and $j$ involved in the splitting, $i \to j + k$, with $j$ and $k$
with momentum fraction $z$ and $1-z$, respectively.
For instance, the $q \to g + q$ splitting function is given by
\begin{equation}\label{eq:Pqg}
  P_{g \leftarrow q}(z) = \CF \frac{1+(1-z)^2}{z}\,.
\end{equation}
In the gluon soft limit $z \to 0$, it simplifies to $2\CF/z$ and we
recover~\eqref{eq:Pqgzth}. By symmetry, we can obtain the quark-to-quark
splitting function as $P_{q \leftarrow q}(z) = P_{g \leftarrow q}(1-z)$. The
gluon-initiated splitting functions $P_{q \leftarrow g}$ and $P_{g \leftarrow
  g}$ can be read from e.g.~\cite{Buckley:2011ms}.
Note that for later purposes in~\eqref{eq:PSprob} we have introduced the symbols
$\mathcal{P}$ and $\Phi_{\rm rad}$ to generically denote a single-emission
probability and the radiation phase space, respectively.
Regarding the running of $\as$, it has been shown that NLL effects due to soft
and collinear emissions can be accounted for by using a two-loop evolution and
working in the CMW scheme~\cite{Catani:1990rr}, defined as
\begin{equation}
  \as^{\rm CMW}(k_t^2) = \as^{\overline{\text{MS}}}(k_t^2)
  \left(1 + \as^{\overline{\text{MS}}}(k_t^2) \frac{K}{2\pi} \right)\,,
  \quad
  K = \left( \frac{67}{18} - \frac{\pi^2}{6} \right) \CA - \frac{5}{9} N_F\,.
\end{equation}

Finally, an important point to be discussed is the factorization of the
kinematics, as entering in the observable definition. Let us consider the
thrust: in the two-jet region, the thrust $\tau = 1-T$ can be written as a sum
of contributions from individual emissions
\begin{equation}\label{eq:thrustadd}
  \tau =  \sum_i \frac{k_{ti}}{Q} e^{-|\eta_i|} \equiv \sum_i \tau_i \,,
\end{equation}
with $k_{ti}$ and $\eta_i$ transverse momentum and rapidity, as measured with
respect to the $q\bar{q}$ pair.
However, at NLL, we cannot assume that a single emission dominates the value of
the observable, as was done for the jet mass at LL~\eqref{eq:JMfact}.
Often, kinematic factorization can be reached in a {\em conjugate space}, where
kinematic constraint can be diagonalized.
In presence of Heaviside step functions, as in the case of cumulative
distributions, it is customary to exploit the Laplace transform
\begin{equation}
  \Theta(\tau - \sum_i \tau_i) = \frac{1}{2\pi i}
  \int \frac{\mathrm{d}\nu}{\nu} e^{\nu \tau} \prod_i e^{-\nu\tau_i}\,,
\end{equation}
to convert a sum into a product, and then reach a factorised form. At the end,
the Laplace expression is inverted either numerically or analytically (possibly
with a series expansion), to obtain a result in direct space.
Note that it is not necessary to rely on the existence of an integral transform
that factorises the observable, see e.g.\ the CAESAR approach introduced in the
next Section.

\subsubsection{CAESAR: automating NLL resummation}\label{sec:CAESAR}

A general framework for performing resummed calculations at NLL accuracy in a
semi-automatic way for a broad class of observables\footnote{The precise
definition of which observables can be resummed with CAESAR is somewhat
technical; we refer the interested reader to the original
paper~\cite{Banfi:2004yd}.}, including event shapes and jet-resolution
parameters, is provided by the CAESAR approach~\cite{Banfi:2004yd}.
CAESAR is based on the master formula for a cumulative distribution:
\begin{equation}\label{eq:CAESARmeq}
  \Sigma(v) = e^{-R(v)} \mathcal{F}\,,
\end{equation}
i.e.\ a product between the exponential of a single emission and a correction
factor $\mathcal{F}$ to account for the dependence of the observable on multiple
emissions.

Given a $\bar{q}q$ Born configuration, the single emission contribution is
schematically given by
\begin{equation}\label{eq:Rv}
  R(v) = \int [\mathrm{d}k]\,|M^2(k)|\,\Theta(V(\{\tilde{p}\},k) - v)\,,
\end{equation}
where $[\mathrm{d}k]$ is the single-gluon phase space, $|M^2(k)|$ is the matrix
element for the emission of a single gluon soft or collinear to either of the
hard leg and $V(\{\tilde{p}\},k)$ is the value of the observable for a single
emission ($\{\tilde{p}\}$ denotes the hard momenta after recoil from emission
$k$). Note that in the soft and collinear limit the expression of the observable
features the scaling $(k_t/Q)^{a} e^{-b \eta}$, with observable-dependent
coefficient $a$ and $b$; for the thrust, $a=1$ and $b=1$,
see~\eqref{eq:thrustadd}.

Instead, the function $\mathcal{F}$ reads
\begin{align}
  \mathcal{F} = \lim_{\epsilon \to 0}
  \int [\mathrm{d}k_1] & |M^2(k_1)| \exp\left(-R' \ln\frac{v}{\epsilon v_1}\right) \nonumber \\
  & \times \sum_{m=0}^{\infty} \frac{1}{m!} \left(
  \prod_{i=2}^{m+1} \int_{\epsilon v_1}^{v_1} [\mathrm{d}k_i] |M^2(k_i)|
    \right)
    \Theta(v - V(\{\tilde{p}\},k_1,\ldots,k_{m+1}))\,,\label{eq:Fcal}
\end{align}
where $R' = \mathrm{d}R / \mathrm{d} \ln 1/v$ and $v_i = V(\{\tilde{p}\},k_i)$,
while $v_1$ is defined as the largest value among the $k_i$s, with associated
gluon $k_1$.
Behind the derivation of~\eqref{eq:Fcal}, there is the assumption that emissions
widely separated in rapidity are independent, as dictated by colour coherence,
described in~\ref{sec:cohbra}.

The power of~\eqref{eq:CAESARmeq} relies in the clear separation of
contributions.
All double-logarithmic terms are inside $e^{-R(v)}$ and are known
analytically. The function $\mathcal{F}$, usually evaluated numerically by means
of Monte Carlo methods, is single-logarithmic.
Particular care, involving delicate numerical limits, is required to make sure
that spurious subleading contributions do not appear in
$\mathcal{F}$~\cite{Banfi:2001bz}.

Recently, the CAESAR method has been implemented within the SHERPA
framework~\cite{Gerwick:2014gya}.
This allowed to obtain new results e.g.\ resummed predictions for jet-resolution
scales $y_{34}$, $y_{45}$ and $y_{56}$ in multi-jet production with the Durham
algorithm at NLO+NLL'~\cite{Baberuxki:2019ifp}.

A method based on CAESAR to reach NNLL accuracy for some observables has been
proposed in~\cite{Banfi:2014sua}.
The main idea of his method, called ARES, is that NNLL accuracy can be obtained
by starting from a NLL resummation and by modifying a single emission at the
time.
This is naively justified by the fact that to move from NLL $(\as L)^n$ to NNLL
$\as(\as L)^n$, it is sufficient to ``add a power of $\as$''.
With ARES, NNLL predictions for several event shapes~\cite{Banfi:2014sua} and
for the two-jet rate with the Durham and the Cambridge algorithm have been
obtained~\cite{Banfi:2016zlc}.

Two-jet rate predictions have been used to fit the strong coupling constant
$\as$, thanks to their high perturbative accuracy and reduced sensitivity to
hadronization corrections compared to three-jet
observables. In~\cite{Verbytskyi:2019zhh} a determination of $\as$ was carried
out by using N$^3$LO predictions for the two-jet rate (including bottom mass
corrections at NNLO~\cite{Nason:1997nw}) supplemented with the aforementioned
NNLL resummed predictions.

\subsubsection{QCD resummation in effective theories: SCET}\label{sec:SCET}

So we far we discussed techniques based on the study of the all-order behaviour
of QCD emissions by means of suitable approximations to keep only the dominant
terms up to some logarithmic accuracy, usually referred to as {\em direct QCD}
(dQCD) techniques.
As an alternative it is possible to work in an effective theory, where it is the
QCD Lagrangian itself to be modified to retain only the soft and collinear
degrees of freedom, by ``integrating'' out the hard emissions, whose effect
become manifest only through modified couplings.
This theory is called {\em Soft-Collinear Effective Theory}
(SCET)~\cite{Bauer:2001yt,Bauer:2000ew,Bauer:2001ct}. The great advantage of
SCET is that it is possible to build from first principles factorisation
formulae with universal ingredients, where each building block is responsible
for only one type of degree of freedom (hard, soft or collinear).
Each ingredient features a (different) typical scale: resummation is achieved by
evolving via renormalization group equations (RGEs) all ingredients to a common
scale.
Note that the accuracy of each ingredient can be systematically improved: hence,
with a factorization formula at our disposal, it is enough to calculate to
higher orders the relevant ingredients to achieve higher logarithmic
resummation.
For a reader interested in knowing more about SCET, we suggest the
textbook~\cite{Becher:2014oda}.

First works provided results at NLL+NLO accuracy for
thrust~\cite{Schwartz:2007ib} or more in general {\em two-jet event shapes}
i.e.\ those event shapes $e$ whose distribution near $e = 0$ is dominated by
events with two back-to-back collimated jets of particles~\cite{Bauer:2008dt}.
With SCET, it was possible to achieve N$^3$LL accuracy for
thrust~\cite{Becher:2008cf} i.e.\ two logarithmic orders better than the NLL
calculation of~\cite{Catani:1992ua}, which was state-of-the-art at the time.
The first N$^4$LL resummation for an event shape (EEC in the back-to-back
region) has been achieved in~\cite{Duhr:2022yyp}.
For an extensive discussion of the applications of SCET to jet physics and event
shapes, see Sec.~9.3--9.4 of~\cite{Becher:2014oda}.

Finally, one may wonder how dQCD and SCET methods compare. In the case of
two-jet event shapes, in~\cite{Almeida:2014uva} it was shown that they are
equivalent once framed in the same language, provided that the accuracy of
analytical ingredients entering the two formulations is the same.

\subsubsection{Example of resummed results for event shapes}\label{sec:resumres}

\begin{figure}
  \centering
  \includegraphics[width=0.49\textwidth]{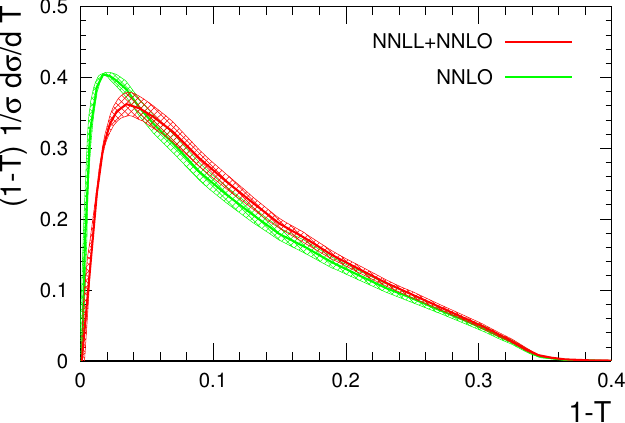}\,
  \includegraphics[width=0.49\textwidth]{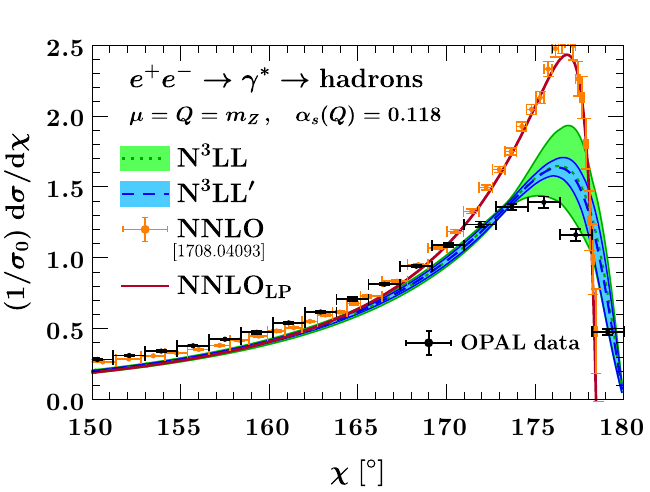}
  \caption{Predictions for thrust at NNLO(+NNLL) (from~\cite{Monni:2011gb},
    left).  Predictions for EEC at NNLO (orange), singular terms of the NNLO in
    the back-to-back region (red), N$^3$LL (green) and N$^3$LL'
    (blue) (from~\cite{Ebert:2020sfi}, right).}
  \label{fig:thrustEEC}
\end{figure}

We start by showing predictions for thrust in Fig.~\ref{fig:thrustEEC} on the
left, where the fixed-order at NNLO is compared to matched NNLL+NNLO
predictions. We first note that the effect of resummation is to suppress events
at small values of $1-T$, in order to let the distribution vanish rather than
diverge in this limit.
The effect of the resummation is visible everywhere, not only in the peak, by
shifting the fixed-order prediction upwards also in the region of $1-T$ between
0.1 and 0.3.

It is important to notice that predictions for thrust (or event shapes in
general) are highly dependent on the value of $\as$.
Indeed, thrust distributions are usually adopted to fit $\as$ e.g.\ see the
recent determinations~\cite{Aglietti:2025jdj,Benitez:2024nav}.
However, in order to properly describe data, the introduction of a
non-perturbative (NP) function, with some free parameters, is required.
Hence, there is some interplay between the NP parameters and the value of $\as$,
and they are often fitted together.
For instance, in~\cite{Gehrmann:2012sc}, predictions for the thrust at NNLL+NNLO
have been used to determine $\as$, with the adoption of a dispersive
model~\cite{Dokshitzer:1995qm,Dokshitzer:1997ew} for NP corrections.
Such a model is based on the introduction of an effective coupling $\alpha_{\rm
  eff}(k^2)$ regularised down to $k^2 \to 0$. It depends on a single parameter
$\alpha_0$, which is then fitted together with $\as$.
We will elaborate more on non-perturbative effects in Sec.~\ref{sec:nonpert}.

About EEC, in the seminal paper~\cite{Collins:1981uk} first predictions at NLL
in the back-to-back region were computed.
For EEC, the proper conjugate space to diagonalize kinematics is the
impact-parameter space or $b$-space, where $b$ is the conjugated variable to
$Q\sqrt{1-z}$.
This prediction was further refined to NNLL in~\cite{deFlorian:2004mp} and to
N$^3$LL in~\cite{Aglietti:2024xwv} (dQCD) and in~\cite{Moult:2018jzp}
(SCET). In~\cite{Ebert:2020sfi}, the missing ingredient to achieve N$^3$LL'
accuracy was completed\footnote{The ``primed'' counting refers to the inclusion
of higher-order constants coefficients $C_i$ in~\eqref{eq:resexp} i.e.\ NLL' is
NLL with $C_1$ included, NNLL' is NNLL with $C_2$ included, and so on.}.
Matched NNLL+NNLO predictions were produced in~\cite{Tulipant:2017ybb}.
Resummation in the collinear region was considered in~\cite{Dixon:2019uzg}.

In Fig.~\ref{fig:thrustEEC} on the right we show state-of-the-art predictions
compared to experimental data from LEP.
As already noticed for the thrust, resummation cures the divergent behaviour of
the fixed-order predictions in the singular limits.
N$^3$LL and N$^3$LL' predictions are closer to data in the peak region, whereas the
NNLO result agrees better with the fixed-order away from the peak.
Finally, the theory uncertainty decreases when moving from N$^3$LL to N$^3$LL'.

\subsection{Event generators and parton showers}\label{sec:psevgen}

In the previous Section, we have touched upon a variety of techniques adopted to
perform resummed calculations. As shown, such techniques often rely on dedicated
calculations that need to be adjusted when changing observable, exploring other
processes or introducing different fiducial cuts. Even the powerful SCET
approach is limited by the requirement of having at disposal a proper
factorization formula in order to achieve resummation.

\begin{figure}[t]
  \centering
  \includegraphics[width=0.45\textwidth]{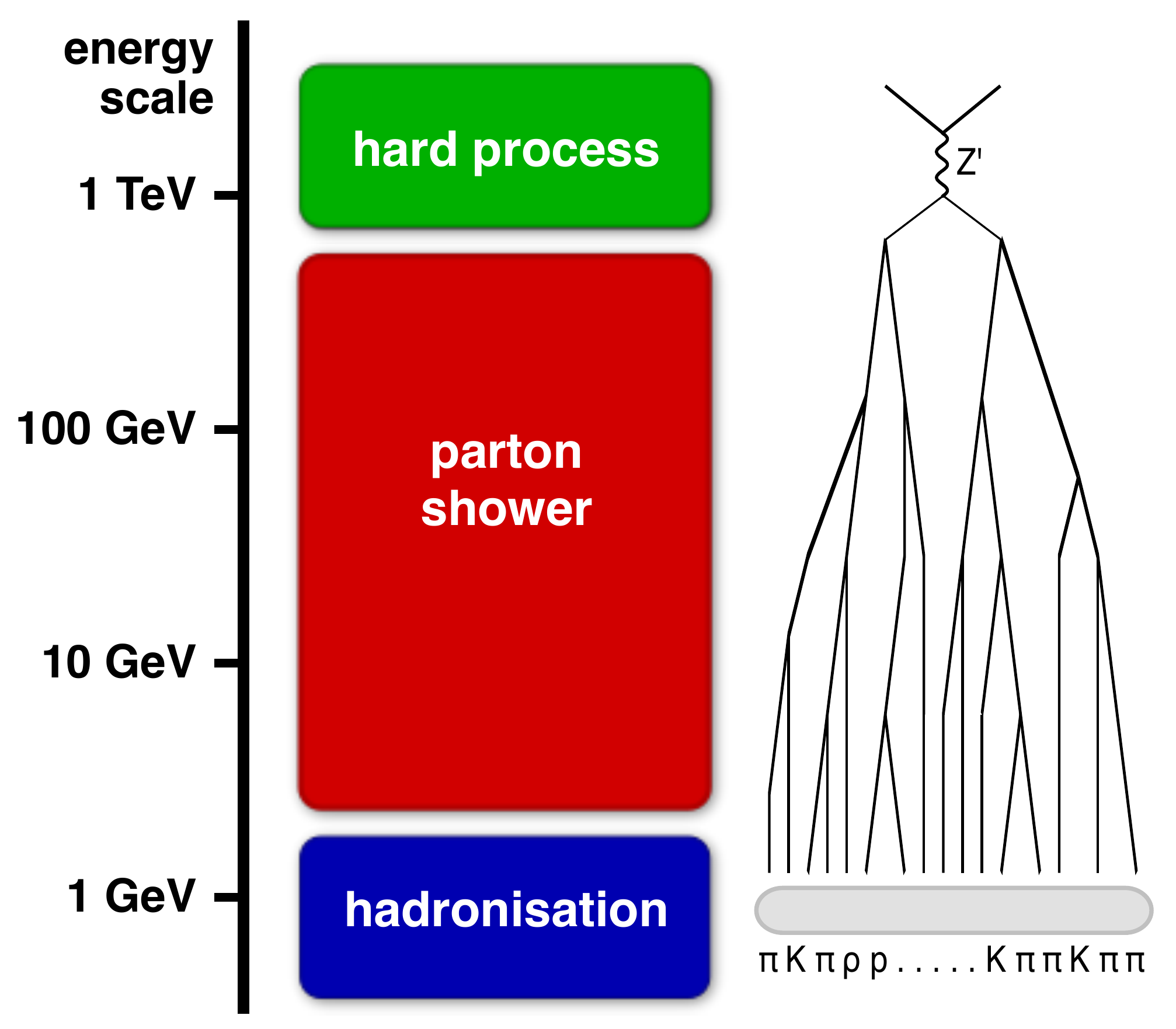}
  \caption{Schematic representation of event generators in $e^+e^-$
    collisions. Courtesy of Gavin Salam.}
  \label{fig:PSsilvia}
\end{figure}

An alternative and more flexible approach is through {\em parton showers}, based
on Monte Carlo methods.
Starting with a handful of hard particles, we can reconstruct the complexity of
a real event through parton showers — albeit approximately.
Parton showers are at the core of more general tools called {\em event
  generators}, that provide a description of collider events in their
entirety\footnote{At hadron colliders, they further include a description of the
hadronic activity in the initial state.}.
The situation is depicted in Fig.~\ref{fig:PSsilvia}: parton showers control the
evolution of the final state from the scale of the hard process down to the
scale of non-perturbative physics; at this stage, some heuristic model to
describe the parton-to-hadron transition is required.
The major event generators are \code{Herwig}~\cite{Bewick:2023tfi},
\code{PYTHIA}~\cite{Bierlich:2022pfr} and \code{Sherpa}~\cite{Sherpa:2024mfk}
(the references correspond to the most recent version of the codes).
We refer to the lectures~\cite{Skands:2012ts} for a more in-depth introduction
to parton showers and event generators in general.
See also a comprehensive review of event generators~\cite{Buckley:2011ms} and
the more recent community paper~\cite{Campbell:2022qmc}.

In the following, we provide a brief description of the basic ideas behind any
parton shower algorithm. In order to populate the phase space with more
emissions, we assume that the cross section for $n+1$ particles can be written
as the cross section for $n$ particles times some emission probability
$\mathcal{P}$ differential in the radiation phase space $\Phi_{\rm rad}$, as
denoted in~\eqref{eq:PSprob},
\begin{equation}\label{eq:PScross}
  \mathrm{d}\sigma_{n+1} \simeq \mathrm{d}\sigma_n\,\mathcal{P}(\Phi_{\rm rad})\,
  \mathrm{d}\Phi_{\rm rad}\,.
\end{equation}
In parton showers emissions are generated by iterating~\eqref{eq:PScross}, with
emissions ordered according to an evolution variable $t$ e.g.\ the virtuality of
the parent parton $q^2 = z(1-z)\vartheta^2 E^2$ or the relative transverse
momentum $k_t^2 = z^2(1-z)^2 \vartheta^2
E^2$~\cite{Gustafson:1987rq,Sjostrand:2006za}.
Assuming $t$ is a quantity with mass dimension, the starting value for the
evolution is $t_{\rm max} \sim Q$, the hard scale of the process, and the final
value is $t_{\rm min} \sim 1$~GeV, a non-perturbative scale.

The main analytical ingredient of a shower is the Sudakov form factor
$\Delta(t_1,t_2)$, which encodes the probability of having {\em no-emission}
between $t_1$ and $t_2$:
\begin{equation}\label{eq:DeltaSud}
  \Delta(t_1,t_2) = \exp\left[ - \int \mathrm{d}\Phi_{\rm rad}
    \mathcal{P}(\Phi_{\rm rad})\,\Theta(t_1 > t(\Phi_{\rm rad}) > t_2) \right]\,.
\end{equation}
At the exponent of~\eqref{eq:DeltaSud}, we are basically integrating over all
possible splittings that would result in a value of $t$ between $t_1$ and
$t_2$. The presence of the minus sign is due to the fact that we are effectively
exponentiating virtual emissions, see the discussion at the end of
Sec.~\ref{sec:jetmass}.
Note that $\Delta$ is bounded between 0 and 1, with $\Delta = 1$ corresponding
to $t_2 = t_1$.

Given the Sudakov form factor~\eqref{eq:DeltaSud}, emissions are generated as
follow. We start by imposing $t = t_{\rm max}$. We then generate a
uniformly-distributed random number $r$, we solve $\Delta(t,t_{\rm rad}) = r$
and we determine $t_{\rm rad}$ by inverting the equation. At this point,
radiation variables $\Phi_{\rm rad}$ are generated\footnote{In this context,
veto techniques are usually adopted, see e.g.\ Appendix A
of~\cite{Frixione:2007vw}.} according to $\mathcal{P}$, under the requirement
that $t_{\rm rad} = t(\Phi_{\rm rad})$. Finally, we set $t = t_{\rm rad}$ and we
iterate until we find $t < t_{\rm min}$.

The Sudakov form factor $\Delta$ is related to the emission probability
$\mathcal{P}$ by the differential equation
\begin{equation}\label{eq:deqDelta}
  \frac{\mathrm{d}\Delta(t_{\rm max},t)}{\mathrm{d}t} =
  \Delta(t_{\rm max},t) \frac{\mathrm{d}\mathcal{P}}{\mathrm{d}t}\,,\quad
  \frac{\mathrm{d}\mathcal{P}}{\mathrm{d}t} = \int \mathrm{d}\Phi_{\rm rad}
    \mathcal{P}(\Phi_{\rm rad})\,\delta(t - t(\Phi_{\rm rad}))\,.
\end{equation}
The definition of $\Delta$ itself~\eqref{eq:DeltaSud} is indeed a solution
of~\eqref{eq:deqDelta}. The relation between $\Delta$ and $\mathcal{P}$ ensures
unitarity i.e. the shower does not affect total production rates. Namely, when
integrating over all possible radiation, we recover the Born cross section.

With dedicated approaches, it is possible to preserve the accuracy of the hard
scattering process even at higher perturbative orders. At NLO, there exist
techniques (e.g.\ MC@NLO~\cite{Frixione:2002ik} or
POWHEG~\cite{Nason:2004rx,Frixione:2007vw}) to match NLO calculations with
parton showers and obtain NLO+PS predictions that preserve NLO accuracy for
integrated distributions and Born-like quantities. In essence, these algorithms
provide a way to exactly describe one hard emission and fill the remaining phase
space with a shower, avoiding possible double counting.
For specific classes of processes, NNLO calculation can be embedded in event
generators through frameworks such as MiNNLO$_{\rm PS}$~\cite{Monni:2019whf} or
GENEVA~\cite{Alioli:2015toa,Alioli:2012fc}.

As stated above, parton showers share many similarities with resummation: during
the evolution, they are effectively resumming large logarithms due to soft
and/or collinear emissions, as encoded in the splitting
probability~\eqref{eq:PSprob}.
In this regard, an issue that has attracted the attention of the community in
recent times is related to the logarithmic accuracy of parton showers.
Parton showers offer the most exclusive description of an event, but it is
difficult to quantify their formal accuracy e.g.\ whether they are able to
capture only the dominant LL effect or more than that.
For instance, it is known that the coherent branching formalism presented in
Sec.~\ref{sec:cohbra} is able to reproduce full-color NLL accuracy for global
observables such as event shapes.
Inspired by it, a class of parton showers adopts angles as evolution
variables~\cite{Marchesini:1983bm,Marchesini:1987cf,Gieseke:2003rz}.
With a careful treatment of the kinematics of the
emissions~\cite{Bewick:2019rbu}, such angular–ordered showers maintain NLL
accuracy accuracy for global observables.

\begin{figure}[t]
  \centering
  \includegraphics[width=0.35\textwidth]{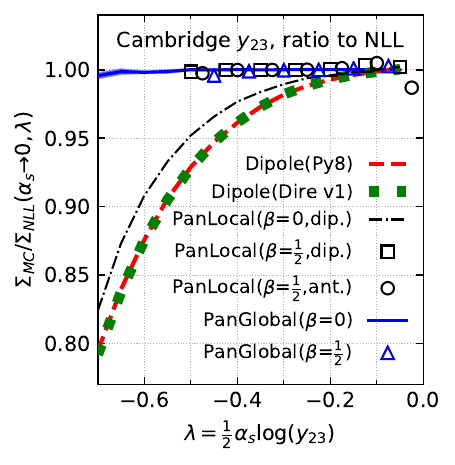}\,
  \includegraphics[width=0.55\textwidth]{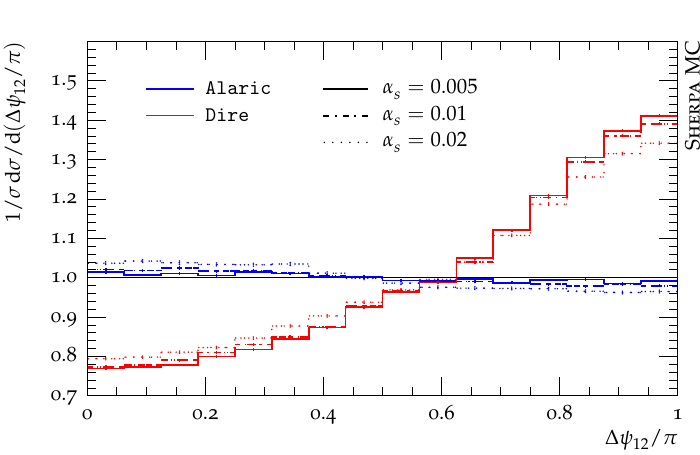}
  \caption{Proof of NLL accuracy for the PanScales family of showers (left,
    from~\cite{Dasgupta:2020fwr}) and for Alaric (right,
    from~\cite{Herren:2022jej}).}
  \label{fig:nllproof}
\end{figure}

A new generation of parton showers with NLL accuracy has been developed in the
past few years~\cite{Dasgupta:2020fwr,Herren:2022jej,Forshaw:2020wrq,Preuss:2024vyu}.
To prove NLL accuracy, it was necessary to design ad-hoc tests, such as the fact
that the parton shower result with fixed $\lambda \propto \as L$ and in the
limit $\as \to 0$ should reproduce the known NLL result\footnote{In this
respect, it is important to stress that these accuracy tests were possible only
thanks to the availability of resummed results obtained with traditional
techniques.}, because higher-order contributions vanish in that limit.
Note that these shower accuracy tests are extremely challenging from a numerical
point of view~\cite{Dasgupta:2020fwr}.
An example of these tests is shown in Fig.~\ref{fig:nllproof} on the left for
the distribution of the transition variable $y_{23}$ of the Cambridge
algorithm~\cite{Dokshitzer:1997in}, for the PanScales family of showers: in the
$\as \to 0$ limit, most of the PanLocal and PanGlobal variants correctly
reproduce the NLL result, whereas predictions obtained with public dipole
showers deviate from it.
In Fig.~\ref{fig:nllproof} on the right, we show a NLL test for the Alaric
parton shower. The variable adopted for this test is $\Delta\psi_{12}$, the
azimuthal angle between the two leading Lund plane declustering (see
Sec.~\ref{sec:LP} and~\cite{Dasgupta:2020fwr} for the proper definition).
The NLL resummation predicts a flat behaviour for this observable, which is
correctly reproduced by Alaric in the limit of $\as \to 0$.

Finally, we point out that a first parton shower with proved NNLL accuracy for
global event shapes has been very recently achieved within the PanScales
framework~\cite{vanBeekveld:2024wws}\footnote{The claim of NNLL accuracy is
still limited to global event shapes at $e^+e^-$ colliders.}.
The interested reader can find in~\cite{FerrarioRavasio:2025soo} a short
pedagogical discussion about the ingredients required to achieve NLL (and NNLL)
accuracy in parton showers.

\section{Jet substructure}\label{sec:jetsub}

In Sec.~\ref{sec:jets}, we demonstrated that jets provide a powerful framework for
classifying hadronic final states at lepton colliders.
However, by clustering radiation into jets, we inevitably lose valuable information
that could contribute to a more detailed description of the event.
In contrast, the event shapes introduced in Sec.~\ref{sec:evshapes} offer a global
characterization of the radiation pattern across the entire event.
Ideally, one would combine both approaches to study the radiation pattern within
individual jets, the {\em substructure} of the jet.

\begin{figure}[t]
  \centering
  \includegraphics[width=0.3\textwidth]{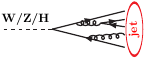}\;\;\;\;\;\;
  \includegraphics[width=0.3\textwidth]{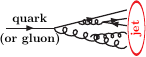}
  \caption{Jets from decay of colour-singlet (left) or QCD radiation (right).}
  \label{fig:WvsQCD}
\end{figure}

At hadron colliders, the interest in studying jet substructure is largely
motivated by the presence of highly boosted jets, i.e.\ jets whose transverse
momentum $\ptt$ is much larger than their invariant mass $m$\footnote{Note that
the boosted scenario is common at the LHC due to the presence of electroweak
particles with mass $m \sim 100$~GeV that can be produced with $p_t \sim
1$~TeV.}.
For instance, consider a $W$ boson decaying hadronically into two jets: a simple
calculation shows that the angle $\theta$ between the decay products is
proportional to $m/\ptt$.
Hence, when $\ptt$ is much larger than $m$, the two jets can no longer be
individually resolved, as they are clustered into a single jet. Thus, in the
boosted regime, the 2-pronged decay structure appears as a QCD jet with the same
mass.
The situation is depicted in Fig.~\ref{fig:WvsQCD}.
To discriminate between the ``signal'' ($W$-boson decay) and the ``background''
(QCD jet), it is necessary to go beyond the monolithic picture of a jet by
studying its substructure: for instance, by identifying the hard prongs within
the jet and quantifying the amount of radiation around them.

The study of the substructure of hadronic jets at the LHC has seen significant
developments over the past 15 years.
For a more in-depth introduction to the field of jet substructure, we refer the
reader to the book~\cite{Marzani:2019hun}.
In the following, we present selected topics that have undergone significant
developments at the LHC and can also be reformulated to suit the
electron-positron environment: the Lund (Jet) plane (Sec.~\ref{sec:LP}), Soft
Drop (Sec.~\ref{sec:SD}), and quark vs.\ gluon discrimination
(Sec.~\ref{sec:QG}).

\subsection{The Lund (Jet) plane}\label{sec:LP}

\begin{figure}
  \centering
  \includegraphics[width=0.45\textwidth]{figs/lund_diag.pdf}
  \includegraphics[width=0.35\textwidth]{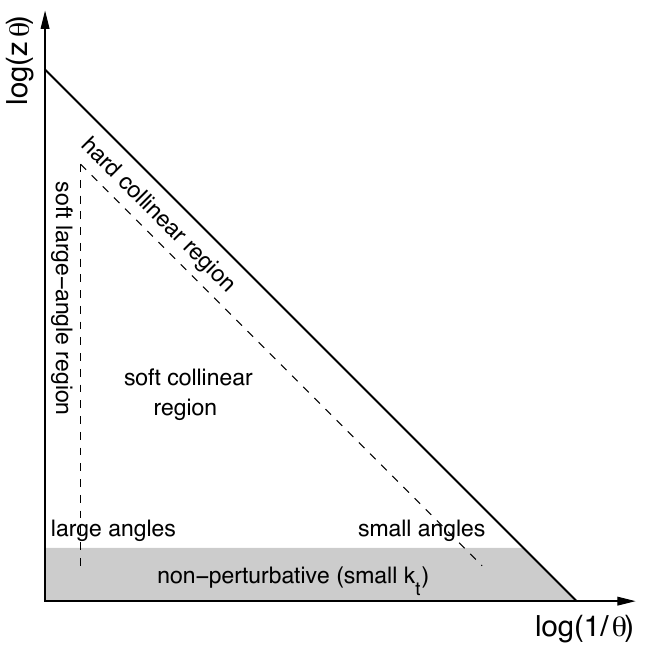}
  \caption{Lund plane in terms of $(\log 1/\vartheta^2, \log 1/z)$ (left) or
    $(\log 1/\vartheta, \log z \theta)$ (right)
    variables. From~\cite{Marzani:2019hun}.}\label{fig:Lplane2}
\end{figure}

We have already introduced the Lund plane in Sec.~\ref{sec:jetmass}, while
discussing the resummation of the jet mass.
The Lund plane, originally proposed in~\cite{Andersson:1988gp}, provides a way
of depicting the pattern of QCD radiation, inside a jet or in a whole event.
We have seen how emissions are uniform in the $(\log 1/\vartheta^2, \log 1/z)$
plane, as depicted in Fig.~\ref{fig:Lplane} on the left, reported for sake of
clarity also in Fig.~\ref{fig:Lplane2} on the left.
In Fig.~\ref{fig:Lplane2} on the right we show an alternative version of the
Lund plane, written in terms of the $(\log 1/\vartheta, \log z \vartheta)$
variables. The product $z \vartheta$ is equal to the relative transverse
momentum between emissions, see~\eqref{sec:ktalg}.
Hence, in this version of the plane, the non-perturbative region is confined to
small $k_t$ values towards the bottom of the plane, and there is a clear
separation between QCD regimes.
Also depicted in Fig.~\ref{fig:Lplane2} on the right is the soft-collinear
region (constituting the bulk of the plane), the soft large-angle region (as
vertical slice on the left) or the hard-collinear region (as diagonal slice on
the top-right).

Other than a graphic tool, the Lund plane may constitute an observable itself.
Indeed, to each high-energy jet, we can associate a kinematic structure on the
Lund plane in the following way~\cite{Dreyer:2018nbf}. Let us first formulate
the Lund jet plane at hadron colliders, by using as kinematic variables $p_{t}$
(transverse momentum), $y$ (rapidity) and $\phi$ (azimuthal angle); we will then
discuss its lepton collider version.

\begin{figure}
  \centering
  \includegraphics[width=0.6\textwidth]{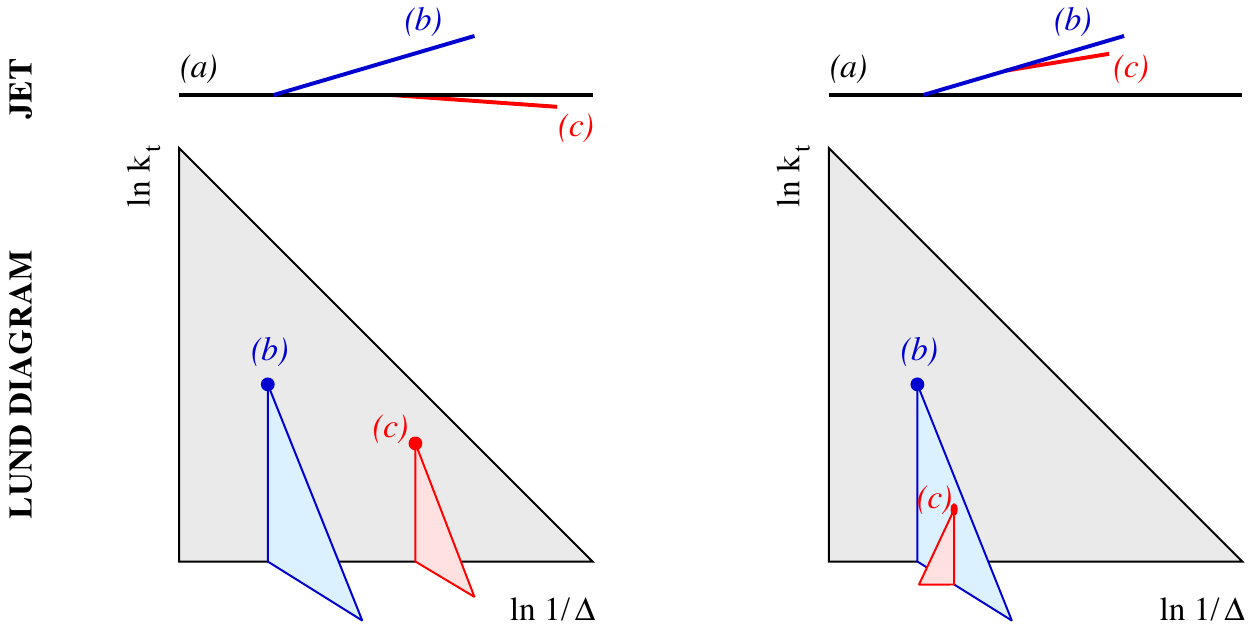}
  \includegraphics[width=0.3\textwidth]{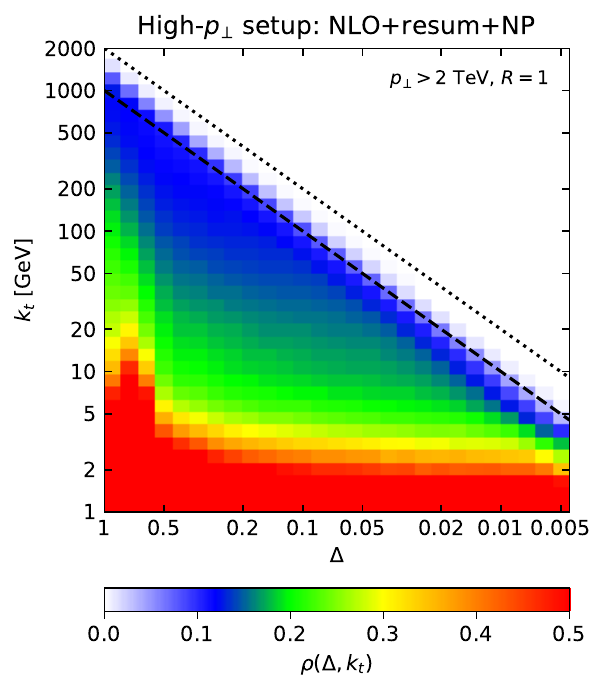}
  \caption{Primary, secondary and tertiary Lund planes (left,
    from~\cite{Dreyer:2018nbf}) and prediction for Lund plane density (right,
    from~\cite{Lifson:2020gua}).}\label{fig:Lplanes}
\end{figure}

Given a jet obtained with any IRC safe algorithm e.g.\ anti-$k_t$, see
Sec.~\ref{sec:genkt}, we first {\em recluster} it with a purely angular distance
among particles. By {\em reclustering}, we mean obtaining a clustering sequence
by using an algorithm different from the one adopted in the first place.
For each step of the declustering, involving pseudo-jets $a$ and $b$ with
$p_{t,a} > p_{t,b}$, we record the variables:
\begin{equation}\label{eq:defs}
  \Delta \equiv \Delta_{ab} = \sqrt{(y_a-y_b)^2+(\phi_a-\phi_b)^2},
  \quad k_t = p_{t,b} \Delta_{ab}, \quad z = \frac{p_{t,b}}{p_{t,a} + p_{t,b}}\,.
\end{equation}
We iterate the collection of variables~\eqref{eq:defs} on both branches of the
declustering tree.
At the end, we can plot the set of points $(\ln 1/\Delta, \ln k_t)$ in a {\em
  primary} Lund plane (if related to an emission off the hardest branch) or in a
{\em secondary}, {\em tertiary}, etc.\ Lund plane, see Fig.~\ref{fig:Lplanes} on the left.

The simplest observable defined on the primary Lund plane is the Lund jet plane
density, counting the number of emissions falling in each ``pixel'' of the
plane, defined as\footnote{The observables is infrared and collinear safe
provided the pixel has area different from zero.}:
\begin{equation}
  \rho(\Delta, k_t) = \frac{1}{N_{\rm jet}}
  \frac{\mathrm{d} n_{\rm emissions}}{\mathrm{d}\ln k_t \mathrm{d}\ln 1/\Delta}\,,
\end{equation}
with $N_{\rm jet}$ the total number of jets.
At leading order, by exploiting the uniformity of emissions in the Lund plane,
we have:
\begin{equation}\label{eq:LPdens}
  \rho_i \simeq \frac{2\alpha_s(k_t) C_i}{\pi}\,,
  \quad \text{with} \quad C_q = \CF\,,\,C_g = \CA\,.
\end{equation}
where the colour factor depends on the nature of the parton originating the jet.
More refined analytical calculations of the Lund plane density have been
performed. In~\cite{Lifson:2020gua}, the logarithmically dominant terms with
structure $\alpha_s^{n+1} \ln^m \Delta \ln^{n-m} z$, $0 \leq m \leq n$, are
resummed to all-orders, and then matched to the fixed-order NLO result.
Non-perturbative effects are estimated through Monte Carlo event generators.
The final prediction for jets with high transverse momentum at the LHC is shown
in Fig.~\ref{fig:Lplanes} on the right. The fact that the density becomes larger
as $k_t$ becomes smaller is in agreement with the naive estimation
in~\eqref{eq:LPdens}, reflecting the running of $\as$ with $k_t$.

A lepton collider version of the Lund plane has been proposed
in~\cite{Medves:2022ccw} in the context of the study of {\em jet
  multiplicities}\footnote{Note that early measurements of multiplicities, such
as the {\em hadronic multiplicity} counting the total number of hadrons in the
final state, have played a pivotal role in the study of singular structure of
QCD.
As observable, the hadronic multiplicity is clearly IRC unsafe, so its
calculation require the introduction of an infrared cutoff $Q_0$ (to be fitted
to data).
Instead, its energy dependence is predictable, being fully perturbative.  See
e.g.~\cite{Ellis:1996mzs,Mangano:1998fk} for more details.}.
A jet multiplicity is defined as the average number of reconstructed jets above
a certain resolution cut. As long as the resolution cut vetoes the
non-perturbative region, the jet multiplicity is an IRC safe quantity.
For instance, the resolution cut can be a transverse momentum cut, $k_{t,\rm
  cut}$, with $k_{t,\rm cut} > \lamQCD$.
It is also possible to study a {\em subjet multiplicity}, with the introduction
of a pair of scales $k_{t0,\rm cut}$ and $k_{t1,\rm cut}$, with $k_{t1,\rm cut}
< k_{t0,\rm cut}$: jets are defined with $k_{t0,\rm cut}$, then for each
individual jet one counts the number of subjets above $k_{t1,\rm cut}$.

The Lund multiplicity, $N^{(Lund)}$, introduced in~\cite{Medves:2022ccw}, is
defined as the (average) number of Lund declusterings (in the full Lund tree)
with $k_t \geq k_{t,{\rm cut}}$.
Its precise definition is as follows. We first produce an angular-ordered
clustering sequence of the full event.
Then, by undoing the last step of clustering, two back-to-back hemisphere jets
are obtained.
For each hemisphere jet, we undo the last step to generate $j_1$ and $j_2$, with
$E_1 > E_2$, and we calculate the relative transverse momentum of the splitting,
which at $e^+e^-$ colliders is defined as
\begin{equation}
  k_t = \min (E_1, E_2) \sin\theta = E_2 \sin\theta\,.
\end{equation}
If $k_t \geq k_{t, \rm cut}$, we increment $N^{(Lund)}$ by one and we iterate
for each of the subjet. Otherwise, we iterate only within the hardest subjet.
Event-wide Lund multiplicity is finally defined as the sum of $N^{(Lund)}$ for
each of the two exclusive jets.

Predictions for jet multiplicities feature the presence of large logarithmic
terms $L = \ln(Q/k_{t,{\rm cut}})$ that require resummation.
The seminal paper~\cite{Catani:1991pm} provided the resummation of Durham jet
multiplicities up to NDL, see~\eqref{eq:resNkDL}.
In~\cite{Medves:2022ccw}, resummed predictions for Lund multiplicity and
Cambridge multiplicity have been pushed to NNDL accuracy, with a novel
resummation technique.
With this method, subleading contributions are identified as originating from
configurations where the subleading part is associated only to a subset of
emissions: these emissions can be computed at fixed-order and dressed with an
arbitrary number of soft-and-collinear emissions.

\begin{figure}[t]
  \centering
  \includegraphics[width=0.45\textwidth]{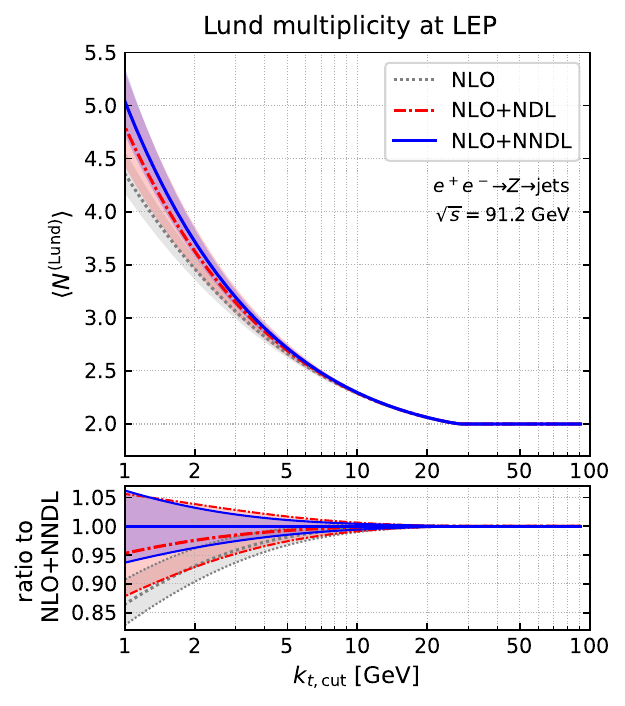}\,
  \includegraphics[width=0.33\textwidth]{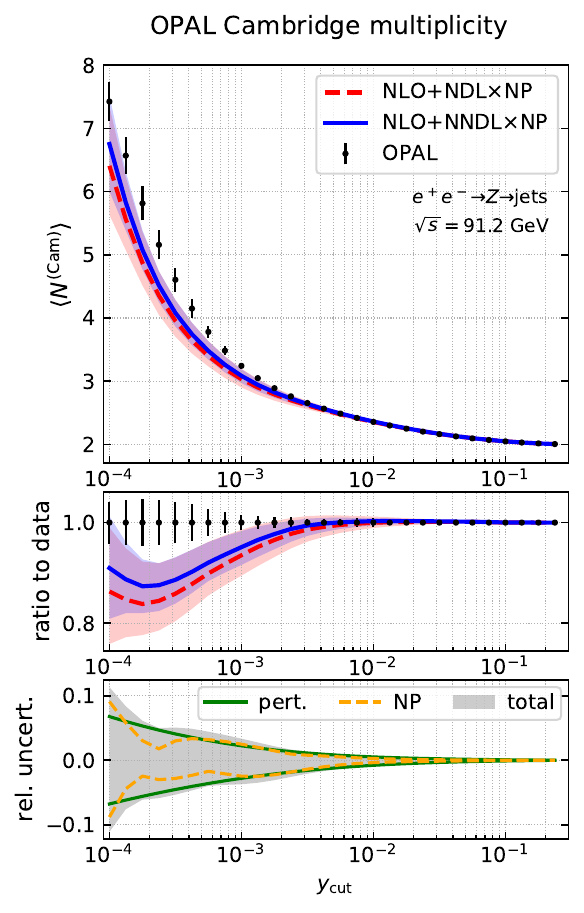}
  \caption{Matched predictions for Lund multiplicity (right) and Cambridge
    multiplicity (left). From~\cite{Medves:2022ccw}.}
  \label{fig:LundLEP}
\end{figure}

In Fig.~\ref{fig:LundLEP}, we show resummed predictions up to NNDL matched to
fixed-order results up to NLO for Lund and Cambridge multiplicities, as a
function of the resolution variables $k_{r,{\rm cut}}$ and $y_{\rm cut} =
k^2_{t,{\rm cut}}/Q^2$, respectively.
In both cases, we see that NNDL improves the NDL prediction towards small values
of the resolution variable, by reducing the theory uncertainty and, in the case
of Cambridge multiplicity, by getting closer to experimental LEP data.
Note that the Cambridge prediction is further supplemented with an estimation of
non-perturbative uncertainties, obtained with Monte Carlo event generators.

\subsection{Soft Drop}\label{sec:SD}

Soft Drop, originally proposed in~\cite{Larkoski:2014wba} is one of the most
used jet substructure techniques at the LHC.
Given a jet, Soft Drop can be used both as {\em tagger} and as {\em groomer}.
By {\em tagging}, we mean finding a particular structure inside the jet
e.g.\ looking for a 2-prong decay (in case of Higgs boson or vector boson decay)
or a 3-prong decay (in case of top quark hadronic decay).
By {\em grooming}, we mean removing background contaminations inside the jet
(e.g.\ soft gluon emissions entering the jet), while keeping the bulk of
perturbative radiation.
As done for the Lund jet plane, we first introduce its hadron collider version,
and then we discuss its lepton collider variant.

To apply Soft Drop to a given jet, we first recluster the jet constituents with
an angular-ordered distance. We then undo the last stage of declustering, $(i+j)
\to i + j$, and we check the Soft Drop condition:
\begin{equation}\label{eq:SDcrit}
  \frac{\min(p_{t,i},p_{t,j})}{p_{t,i} + p_{t,j}}
  > z_{\rm cut} \left( \frac{\Delta R_{ij}}{R} \right)^\beta\,,
\end{equation}
with $\beta$ and $z_{\rm cut}$ two free parameters. If it is satisfied, then we
declare $(i+j)$ as the soft-drop jet; otherwise we iterate on the subjet with
the largest $p_t$.

\begin{figure}
  \centering
  \includegraphics[width=0.45\textwidth]{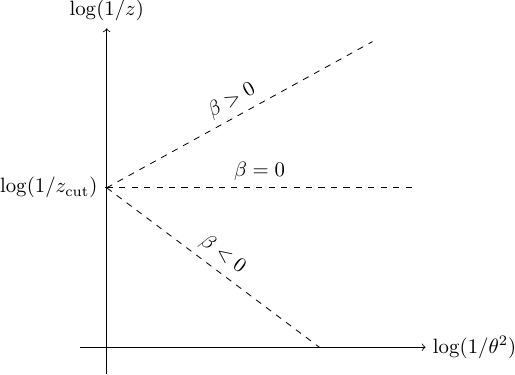}\,
  \includegraphics[width=0.40\textwidth]{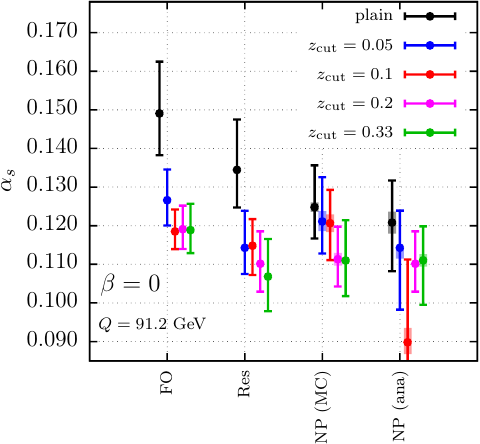}
  \caption{Soft Drop condition in the Lund plane (left) and value of $\as$
    obtained by fitting the Soft Drop thrust (right,
    from~\cite{Marzani:2019evv}).}\label{fig:SDimg}
\end{figure}

The action of Soft Drop can be understood by inspecting the meaning of the
condition~\eqref{eq:SDcrit} in the Lund plane. Note that in the soft limit we
can rewrite~\eqref{eq:SDcrit} as $z > z_{\rm cut} \theta^\beta$, with the usual
$z$ and $\theta$ variables, such that the Soft Drop condition becomes a line in
the Lund plane, as depicted in Fig.~\ref{fig:SDimg} on the left.
Hence, according to the value of $\beta$, we are effectively vetoing part of the
radiation in the Lund plane.
If $\beta > 0$, we remove all soft radiation and some soft-collinear radiation
from the jet. If $\beta = 0$, we remove all soft-collinear radiation (in this
case, eq.~\eqref{eq:SDcrit} reduces to a symmetry condition, also known as
mMDT~\cite{Butterworth:2008iy,Dasgupta:2013ihk}). If $\beta < 0$, we allow for
the removal of hard-collinear radiation.

The power of Soft Drop relies on the fact that it removes soft radiation from
the jet in a dynamical way.
Predictions for the mass of soft-drop jets have reached high accuracy
e.g.~\cite{Frye:2016aiz}.

The lepton collider version of Soft Drop has been introduced
in~\cite{Baron:2018nfz}, in the context of the definition of the {\em Soft Drop
  thrust}, defined as follows.
One first defines two hemisphere jets by using the angular-ordered version of
the $e^+e^-$ gen-$k_t$ algorithm of Sec.~\ref{sec:genkt}.
Then the Soft Drop algorithm is applied on each hemisphere jet by using the
$e^+e^-$ variant of~\eqref{eq:SDcrit}, which reads
\begin{equation}
  \frac{\min(E_i,E_j)}{E_i+E_j} > z_{\rm cut} (1-\cos\vartheta_{ij})^{\beta/2}\,.
\end{equation}
In this way, some radiation is removed from the event in order to obtained a
{\em groomed} version of each hemisphere.
At this point, the thrust is calculated for each groomed hemisphere separately,
by using the groomed hemisphere-jet momenta as reference axes, and finally the
Soft Drop thrust is then assembled.

The reason behind the introduction of the Soft Drop thrust compared to the usual
thrust is that one would expect reduced hadronization effects.
Indeed, it has been proved~\cite{Marzani:2019evv} that fits of $\as$ by using
the Soft Drop thrust are more stable under hadronization effects compared to
fits that exploit the usual thrust event shape.
This is shown in Fig.~\ref{fig:SDimg} on the right: by comparing predictions at
fixed-order (FO), with resummation (Res) and with the inclusion of
non-perturbative effects through Monte Carlo (MC) or analytical models (ana), we
note a larger shift of $\as$ values when adopting the usual ``plain'' thrust for
the fit; such a shift is reduced when using the Soft Drop thrust for any choice
of $z_{\rm cut}$ values.

\subsection{Quark- vs.\ gluon-jet discrimination}\label{sec:QG}

\begin{figure}
  \centering
  \includegraphics[width=0.2\textwidth]{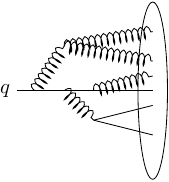}\quad
  \includegraphics[width=0.2\textwidth]{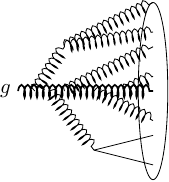}
  \caption{Pictorial representation of fragmentation of quark (left) or gluon
    (right).}\label{fig:qgjet}
\end{figure}

Another typical case scenario where one may be interested in going beyond the
monolithic picture of jet is in the context of {\em quark- versus gluon-jet
  discrimination}. Namely, being able to disentangle jets that can be thought of
as originating by the fragmentation of a high-energy quark from the ones
originating from a gluon, see Fig.~\ref{fig:qgjet}.
This is important for precision QCD studies, such as the determination of $\as$,
but also as a way to isolate specific scattering processes and to search for new
physics.

\begin{figure}[t]
  \centering
  \includegraphics[width=0.8\textwidth]{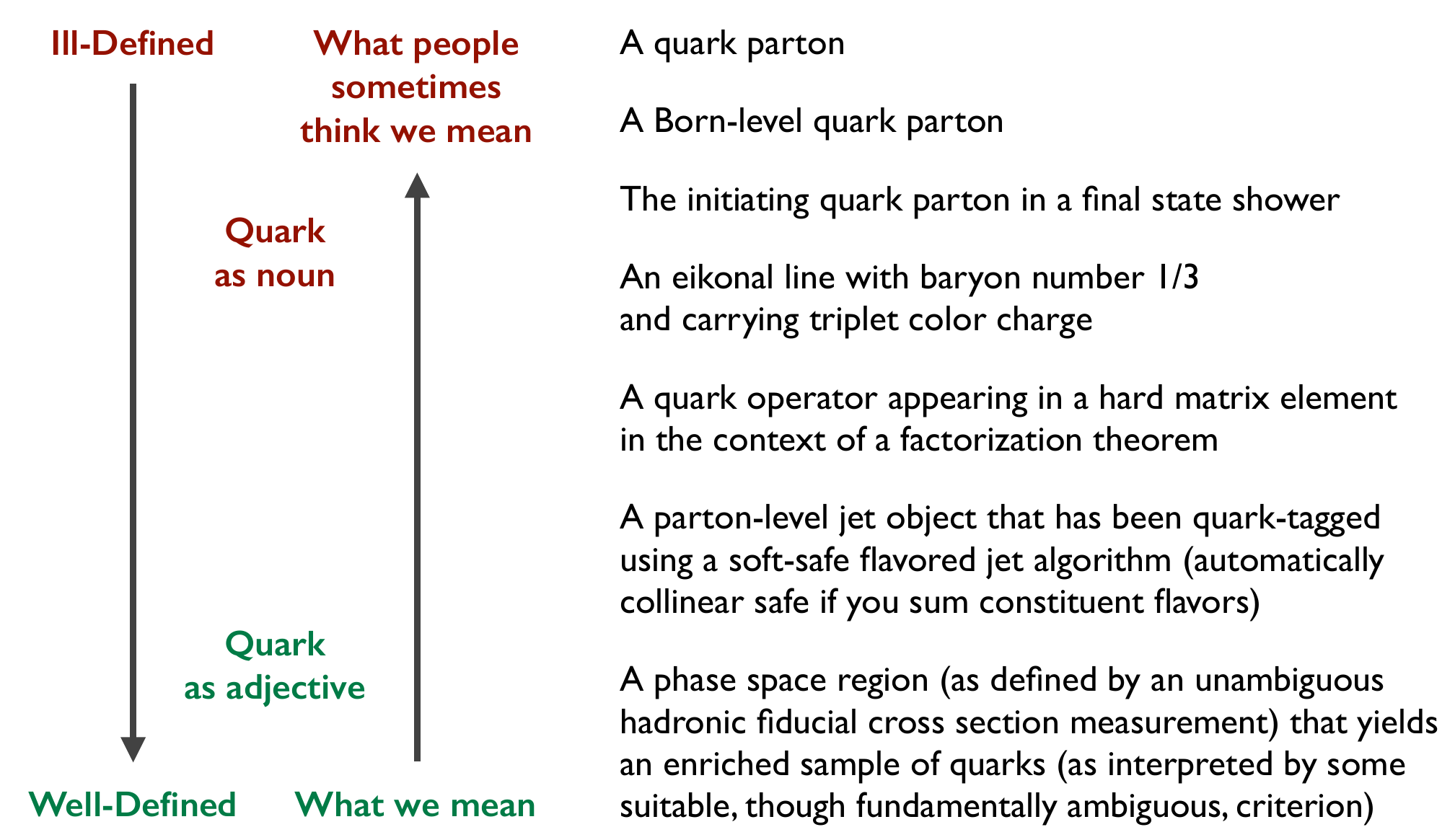}
  \caption{Possible definitions of quark jet. From~\cite{Andersen:2016qtm}.}
  \label{fig:qgdef}
\end{figure}

In a first approximation, the radiation potential for a gluon is greater than
one for a quark, due to the different colour factor. Hence, by quantifying the
amount of radiation around a hard prong we can in principle discriminate between
a quark- and a gluon-jet.
Note that the association of a single parton to a final-state jet is an
intrinsically ambiguous---if not ill-defined---operation, essentially because of
higher-order corrections. It is however possible to employ operational
definitions that associate the value of a measurable quantity to an enriched
sample of quarks or gluons~\cite{Badger:2016bpw,Gras:2017jty}.  See also
Fig.~\ref{fig:qgdef} for possible definitions of a quark jet\footnote{An
alternative way to define a quark- or gluon-jet is through a flavoured jet
algorithm, see Sec.~\ref{sec:flav}.}.

\begin{figure}
  \centering
  \includegraphics[width=0.35\textwidth]{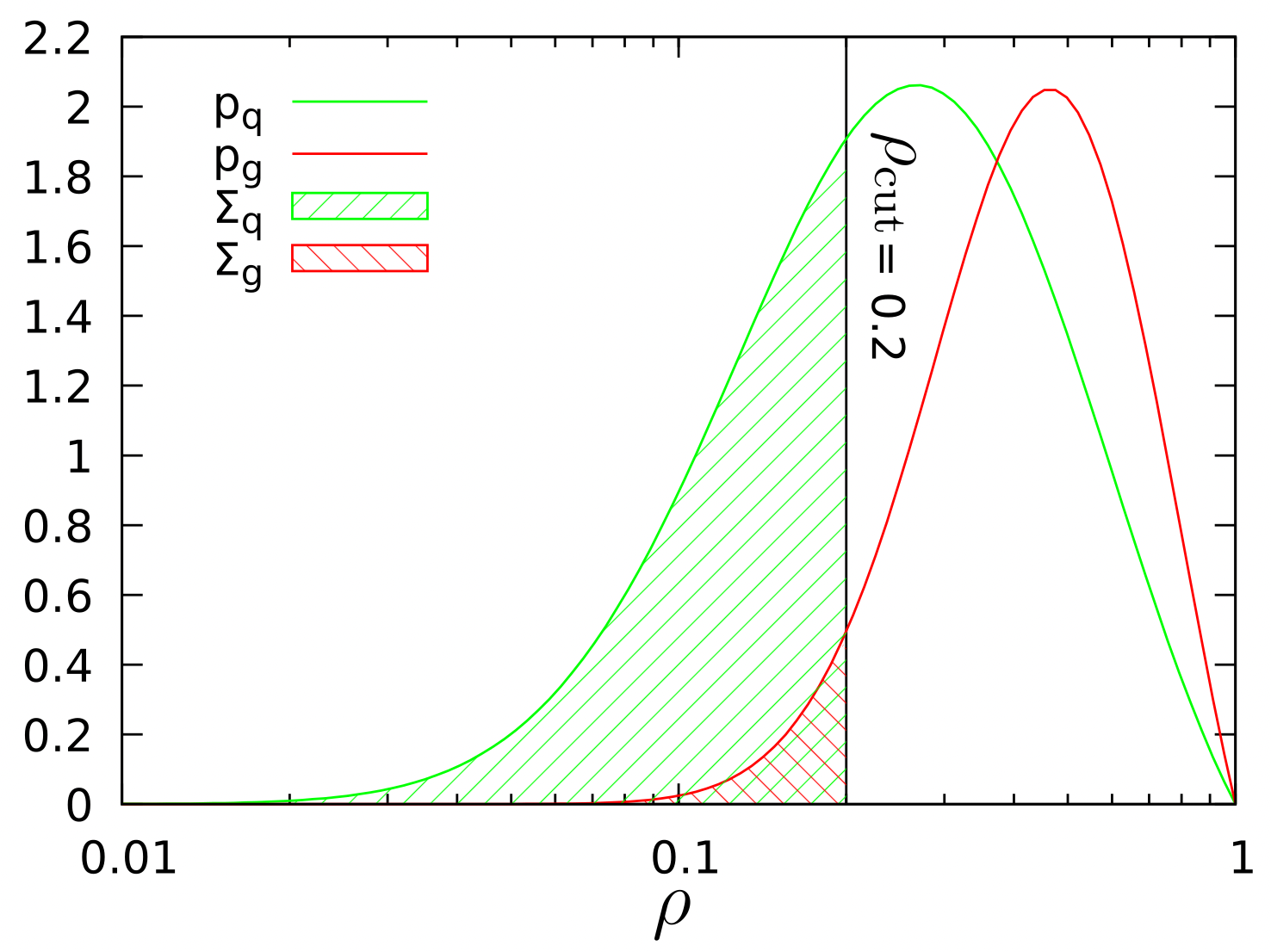}\,
  \includegraphics[width=0.3\textwidth]{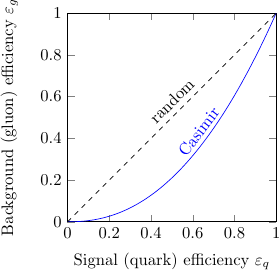}\,
  \includegraphics[width=0.3\textwidth]{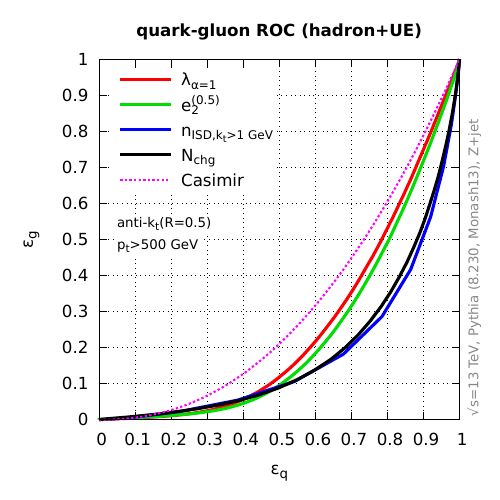}
  \caption{Differential $p_i(\rho)$ and cumulative $\Sigma_i(\rho < \rho_{\rm
      cut})$ with $\rho_{\rm cut} = 0.2$ distributions for quark and gluon jets
    (left). ROC curves for a random classifier and the classifier corresponding
    to a simple cut on $\rho$ (middle). ROC curves for a set of quark-gluon
    taggers, with e.g.\ $\lambda_{\alpha=1}$ the angularity (right,
    from~\cite{Marzani:2019hun}).}\label{fig:qgROC}
\end{figure}

Among the simplest one-dimensional discriminant between quark and gluon jets, we
can consider the jet mass\footnote{Note that the (normalised) jet mass $\rho$ is
equal to the $N$-subjettiness~\cite{Thaler:2010tr} jet substructure variable
$\tau_N$, quantifying the amount of radiation around $N$ prongs within the jet,
with $N=1$ and parameter $\beta = 2$.}, whose LL resummation has been discussed
in Sec.~\ref{sec:jetmass}.
The discussion in Sec.~\ref{sec:jetmass} was focused on a jet originated from
the fragmentation of a high-energy quark.
However, if we repeat the calculation for a high-energy gluon, we realise that
the final result of~\eqref{eq:sigrhoLL}--\eqref{eq:Vrho} only differs by the
colour factor:
\begin{equation}\label{eq:qgcum}
\Sigma_q(\rho) = \exp\left[-\frac{\alpha_s \CF}{\pi} \frac{1}{2} \log^2 \frac{1}{\rho}\right]\,,\quad
\Sigma_g(\rho) = \exp\left[-\frac{\alpha_s \CA}{\pi} \frac{1}{2} \log^2 \frac{1}{\rho}\right]
\end{equation}
We can straightforwardly derive the differential distribution for observing a
specific value of the mass by simply taking the derivative of~\eqref{eq:qgcum}:
\begin{equation}
  p_i(\rho' = \rho) = C_i \frac{\as}{\pi} \frac{\log\left(1/\rho\right)}{\rho}\,
  \exp\left( - C_i \frac{\as}{2\pi} \log^2\left(\frac{1}{\rho}\right) \right) \,,
  \quad \text{with} \quad C_q = \CF\,,\,C_g = \CA\,.
\end{equation}
These differential distributions for quark and gluon jets are shown in
Fig.~\ref{fig:qgROC} on the left. The cumulative distributions for an
illustrative cut $\rhocut = 0.2$ are also shown. It is clear from
Fig.~\ref{fig:qgROC} on the left that the less the two differential distributions
overlap, the more a simple cut $\rhocut$ on $\rho$ is effective when separating
quark jets from gluon jets.

In order to assess the discriminating power of an observable, Receiver Operating
Characteristic (ROC) curves are often considered. These curves show the
background (gluon) efficiency against the signal (quark) efficiency and are
remarkably useful to directly compare the performance of different tools. In
terms of the normalised cumulative distributions for quark and gluon jets,
$\Sigma_q$ and $\Sigma_g$ respectively, the ROC curve is defined as:
\begin{equation}\label{roc-def}
  \text{ROC}(x) = \Sigma_g\left (\Sigma_q^{-1}\left(x \right)\right)\,,
\end{equation}
where $x$ is the signal (quark) efficiency. If the observable is $\rho$, it's
trivial to find an explicit expression for the ROC curve:
\begin{equation}
  \Sigma_g(\rho) = \left[\exp\left( - \CF \Ralt(\rho) \right)\right]^{\CA/\CF}
  \equiv \left[\Sigma_q(\rho)\right]^{\CA/\CF} \longrightarrow
  \text{ROC}(x) = x^{\CA/\CF}\,.
\end{equation}
Such a ROC curve is shown in Fig.~\ref{fig:qgROC} in the middle.
This feature is called Casimir scaling and provides a benchmark for expectations
when using more sophisticated taggers.
The area under the ROC curve (AUC) can be used as a single-number quantifier of
the discrimination power. $\text{AUC}=0$ corresponds to perfect performance,
whereas $\text{AUC}=1/2$ to a random classifier. In this latter case the ROC
curve is just the straight line $y=x$, see Fig.~\ref{fig:qgROC} in the middle.
The AUC for a cut on $\rho$ is given by:
\begin{equation}
  \text{AUC}=\int_0^1 dx\, x^{\CA/\CF} = \frac{\CF}{\CF+\CA} \simeq 0.308\,.
\end{equation}

A class of variables often used in the context of quark/gluon discrimination are
{\em angularities}~\cite{Larkoski:2014pca}. Usually they are defined by
clustering jets with the $e^+e^-$-variant of the anti-$k_t$ algorithm of
Sec.~\ref{sec:genkt}; then for each jet one computes
\begin{equation}\label{eq:angul}
  \lambda^\kappa_\alpha = \sum_{i\in {\rm jet}} z^\kappa_i \theta_i^\alpha\,,
\end{equation}
with energy fractions $z_i$ and angle relative to the jet axis $\theta_i$. Note
that angularities are IRC safe only with $\kappa=1$ i.e.\ linear dependence on
the momentum fractions entering~\eqref{eq:angul}. Because of this, sometimes
$\kappa=1$ is understood and omitted.
At LL accuracy, IRC-safe angularities satisfy Casimir scaling, with
discrimination power independent of $\alpha$. However, beyond LL, Casimir
scaling is typically violated.
In Fig.~\ref{fig:qgROC}, we show the ROC curves for angularity $\lambda^1_1
\equiv \lambda_{\alpha=1}$ and other quark-gluon taggers, as obtained with a
Monte Carlo simulation, and they are compared to Casimir scaling.

\begin{figure}
  \centering
  \includegraphics[width=0.4\textwidth]{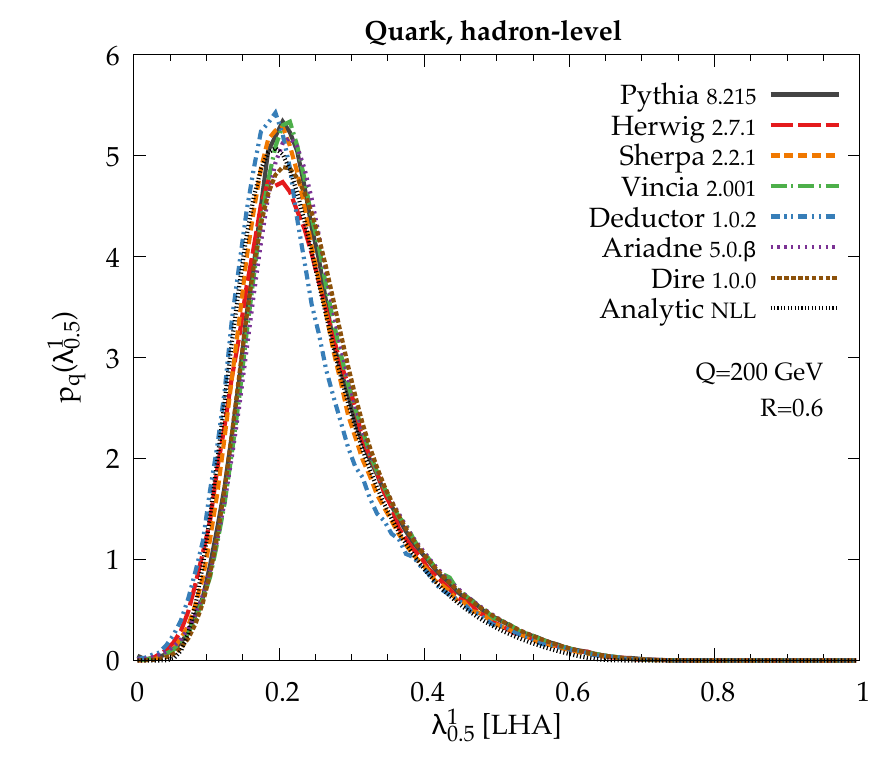}\,
  \includegraphics[width=0.4\textwidth]{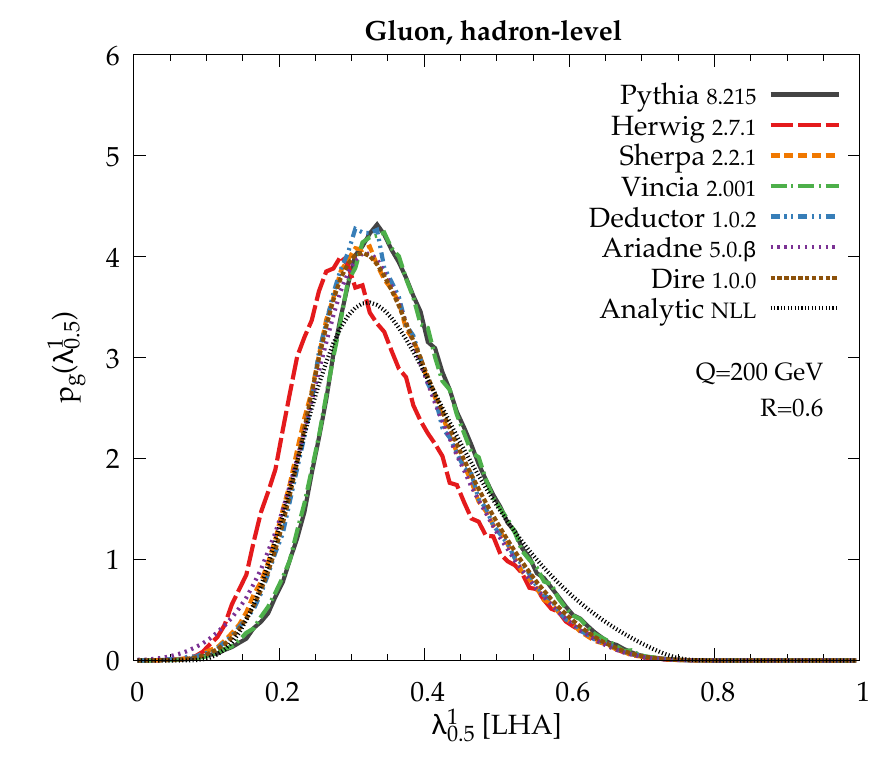}\,
  \caption{Predictions for angularities evaluated on quark and gluon jets,
    obtained with Monte Carlo event generators or analytical
    calculations. From~\cite{Gras:2017jty}.}\label{fig:qgang}
\end{figure}

Resummed results at NLL for angularities, with some modelling of
non-perturbative effects, has been presented in~\cite{Gras:2017jty}.
In Fig.~\ref{fig:qgang} we show the analytical NLL predictions for angularities
on quark and gluon jets. Also shown are predictions obtained with different
Monte Carlo codes.
One thing that it is worth to notice in Fig.~\ref{fig:qgang} is the different
level of agreement between Monte Carlo generators in the case of quark or gluon
jets: while for quark jets all the generators provide similar predictions, in
the gluon case there is an increase spreads between predictions.
This is mostly due to the lack of experimental sets of data enriched in gluon
jets: Monte Carlo event generators are tuned to reproduce LEP data, which
consists of quark jets from $Z \to q\bar{q}$ elementary process, with gluon jets
only appearing at higher QCD orders.
The situation may change at future colliders. First, future colliders will
increase significantly the statistics of three-jet events $Z \to
q\bar{q}g$. Moreover, by looking at Higgs decays to gluons\footnote{It is
customary to refer to the ``$H \to gg$'' elementary process, even though the
Higgs does not couple directly to gluons. Namely, the gluon decay channel
features a loop of top quarks coupling to the Higgs, with gluons emitted from
this loop. However, in the large top mass limit (which provides a very good
approximation), it is possible to work in an effective theory in which the Higgs
boson couples directly to gluons, hence the ``$H \to gg$'' terminology is
justified. See e.g.~\cite{Maltoni:2014iea} for a pedagogical discussion.}, it
may be possible to obtain a large set of gluon jet events, see
Sec.~\ref{sec:hdecay}.

\section{Selected topics}\label{sec:seltopics}

\subsection{Flavoured jets}\label{sec:flav}

Similarly to quark and gluon jets, discussed in Sec.~\ref{sec:QG}, we can also
focus on jets originated by quarks of different {\em flavours}: it is common to
distinguish between light quarks, such as up $u$, down $d$ and strange $s$,
whose masses are much smaller than the proton mass (around 1 GeV), and heavy
quarks, such as charm $c$ and bottom $b$, whose masses are approximately 1.3~GeV
and 4.2~GeV, respectively.
When a heavy quark fragments into a heavy hadron, it leaves distinct signatures
in experimental detectors: for instance, $b$-hadrons feature a lifetime of
around 1.5 ps and they travel a macroscopic distance (a few millimeters) before
decaying, so decay tracks are displaced from the central interaction point.

From a theory point of view, jets consistent with being initiated by a charm or
bottom quark, also dubbed {\em flavoured jets}, are important to pinpoint
specific scattering processes and reject backgrounds, and are useful both for
Standard Model measurements and new physics searches. For instance, the Higgs
boson decays primarily into $b$ quarks, hence processes featuring $b\bar{b}$
pairs play a central role in studies of the Higgs mechanism. See also
Sec.~\ref{sec:hdecay}.

\begin{figure}
  \centering
  \includegraphics[width=0.7\textwidth]{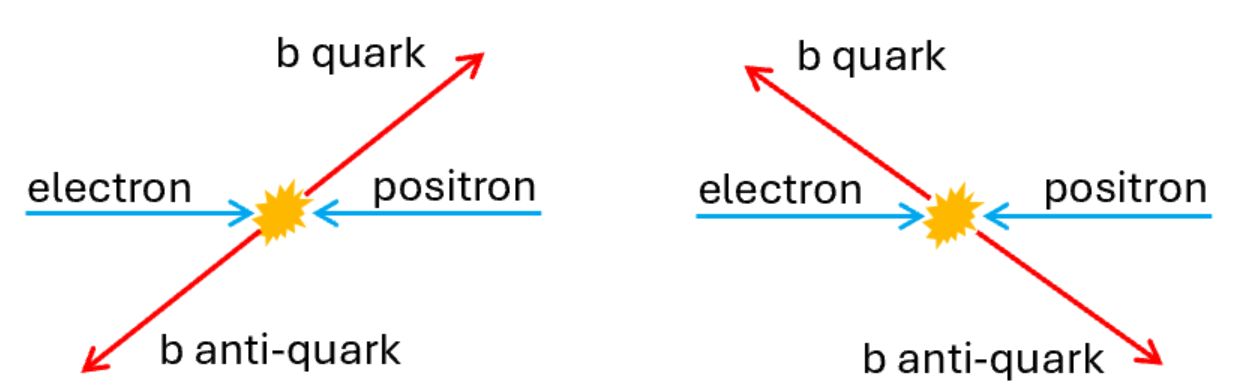}
  \caption{Pictorial representation of a forward (left) and a backward (right)
    event. Courtesy of Matt Strassler.}\label{fig:FAB}
\end{figure}

An important example of a flavour-aware observable at lepton colliders is the
bottom forward-backward asymmetry, defined as
\begin{equation}
  A_{\rm FB} = \frac{\sigma_F - \sigma_B}{\sigma_F + \sigma_B}\,,
\end{equation}
where $\sigma_F$ and $\sigma_B$ are defined as the cross section for the
production of a bottom quark in the {\em forward} or {\em backward} hemisphere,
respectively.
In other words, if we denote by $\theta_b$ the angle between the $b$-quark and
the electron momentum, $\sigma_F$ and $\sigma_B$ are defined as
\begin{equation}
  \sigma_F = \int_0^1 \mathrm{d} \cos\theta_b \frac{\mathrm{d}\sigma}{\mathrm{d}\cos\theta_b}\,,\quad
  \sigma_B = \int_{-1}^0 \mathrm{d} \cos\theta_b \frac{\mathrm{d}\sigma}{\mathrm{d}\cos\theta_b}\,.
\end{equation}
See Fig.~\ref{fig:FAB} for an example of both configurations.
It is possible to adopt different choices for the axis identifying the forward
direction. For instance, another common choice is the oriented thrust axis:
namely, one computes the thrust axis and then determines the direction by
proximity to the $b$-quark direction.
With a simple back-of-the-envelope calculation (see e.g.~\cite{Catani:1999nf}),
it is possible to shown that the leading order prediction for $A_{FB}$ in
unpolarised $e^+e^- \to Z \to b\bar{b}$ events is
\begin{equation}
  A_{\rm FB}^{\rm LO} = \frac{3}{4} \mathcal{A}_e \mathcal{A}_b\,, \quad
  \mathcal{A}_f = \frac{2 v_f a_f}{v_f^2 + a_f^2}\,, \;\;f=e,b\,,
\end{equation}
where $v_f$ and $a_f$ are the vector and axial couplings of fermion $f$ to the
$Z$. Hence $A_{\rm FB}$ provides a direct handle to access the value of the weak
mixing angle $\sin^2 \theta_W$, one of the fundamental parameters of the
electroweak sector of the Standard Model.
Note that $A_{\rm FB}$ can also be extracted from a fit of the angular
distribution
\begin{equation}
  \frac{\mathrm{d}\sigma}{\mathrm{d}\cos\theta_b} = \sigma_{b\bar{b}} \left(
  \frac{3}{8} (1 + \cos^2\theta_b) + A_{\rm FB} \cos\theta_b \right)\,,
\end{equation}
basically extending~\eqref{eq:costhLO} to include the exchange of a $Z$ boson.
QCD theory predictions for the bottom $A_{\rm FB}$ are known at
NLO~\cite{Arbuzov:1991pr,Jersak:1981sp,Djouadi:1994wt,Lampe:1996rt} and NNLO
accuracy~\cite{Altarelli:1992fs,Ravindran:1998jw,Catani:1999nf,Weinzierl:2006yt}.
State-of-the-art predictions include bottom mass effects at
NNLO~\cite{Bernreuther:2016ccf}.

\begin{figure}
  \centering
  \includegraphics[width=0.4\textwidth]{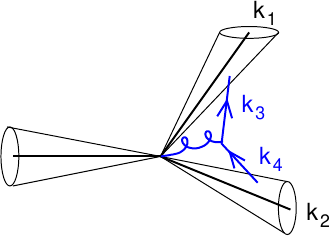}\,
  \includegraphics[width=0.5\textwidth]{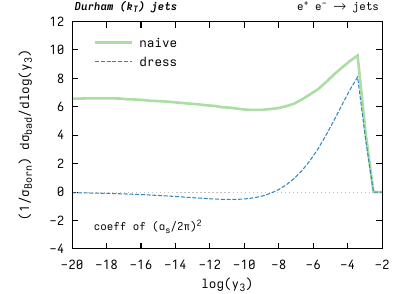}
  \caption{Example of problematic configuration for infrared safety of flavoured
    jet definition (right, from~\cite{Banfi:2006hf}). Test of infrared safety of
    flavour dressing algorithm, by studying the behaviour of the cross section
    for ``bad'' mis-identified events in the unresolved $y_{3} \to 0$ limit
    (left, from~\cite{Gauld:2022lem}).}
  \label{fig:jetflavbad}
\end{figure}

The precise definition of a flavoured jet is not free from ambiguities. A naive
way of assigning flavour to jets would be to declare a jet as flavoured if it
contains at least a heavy (anti-)quark inside it (or a heavy hadron in Monte
Carlo samples or experimental measurements).
However, such a definition is not IRC safe from the point of view of flavour
i.e.\ a soft or collinear emission can alter the value of the observable.
For instance, according to this definition, a gluon splitting to a flavoured
quark pair, $g \to f \bar{f}$, would always lead to a flavoured jet, even in the
collinear limit.
As discussed in Sec.~\ref{sec:IRCsafe}, IRC safety is a crucial property to
perform perturbative calculations and it guarantees reduced sensitivity to
non-perturbative effects.
The problematic collinear configuration can be avoided by modifying our
``counting'' of flavoured particles within the jet: if charge information is
available, and it is then possible to distinguish between quark and anti-quark,
we can assign a flavour number of $+1$ to $f$ and of $-1$ to $\bar{f}$, in such
a way a $f\bar{f}$ system would have net zero flavour; instead, if we cannot
distinguish between flavoured quarks and anti-quarks, we can consider an {\em
  even} number of flavoured particles as a flavourless system.
But there are genuine IRC unsafe configurations, that cannot be solved in a
simple way. An example is shown in Fig.~\ref{fig:jetflavbad} on the left: a
three-jet configuration is dressed with a soft gluon emitted outside of a jet,
splitting into a large-angle $f(k_3)\bar{f}(k_4)$ pair, with the $f(k_3)$ ending
to be close in angle to hard $\bar{f}(k_1)$ and the $\bar{f}(k_4)$ close to
$f(k_2)$. In the limit when the gluon is soft, the event features two
flavourless jets, whereas the original event had two flavoured jets.
Note that this configuration first appears at $\mathcal{O}(\as^2)$.

A first proposal for an IRC safe flavoured jet algorithm was put forward
in~\cite{Banfi:2006hf}, by taking as reference the $k_t$ algorithm of
Sec.~\ref{sec:ktalg} and by modifying its distance~\eqref{eq:dijKT} to render it
flavour-aware:
\begin{equation}\label{eq:dij-flavour}
  d_{ij}^{(F)} = 2(1-\cos\vartheta_{ij}) \times\left\{
    \begin{array}[c]{ll}
      \max(E_i^2, E_j^2)\,, & \quad\mbox{softer of $i,j$ is flavoured,}\\
      \min(E_i^2, E_j^2)\,, & \quad\mbox{softer of $i,j$ is flavourless.}
    \end{array}
  \right.
\end{equation}
Thanks to the introduction of a ``maximum'' instead of a ``minimum'' in the
distance~\eqref{eq:dij-flavour}, pairs of flavoured soft particles are
recombined before anything else (because their distance will be the smallest in
the event), thus solving the configuration in Fig.~\ref{fig:jetflavbad} on the
left.
The resulting jet algorithm of~\cite{Banfi:2006hf} was dubbed {\em
  flavour-$k_t$}. Note that the kinematics of the resulting jets is different
from the one of the original $k_t$ algorithm, as the clustering distance itself
is modified.

The flavour-$k_t$ algorithm was adopted in~\cite{Weinzierl:2006yt} to define the
reference axis for the bottom $A_{\rm FB}$, thus providing an infrared-safe
definition of the asymmetry.
Note that the usage of the $b$-quark direction or the oriented thrust axis as
reference axis are not IRC safe for {\em massless} quarks.
Indeed, the discussion so far was always assuming massless particles, but the
bottom (or charm) quark features a non-negligible mass.
Hence strictly speaking there are no QCD divergences in the $g \to f \bar{f}$
splitting, because the quark mass $m_f$ regulates them, and there is not need
for an IRC safe definition of the observable.
However, a non-zero mass implies the presence of potentially large logarithms
$\log(m^2_f/s)$, with $s \gg m^2_f$, that could invalidate the convergence of
the perturbative series, so even in the presence of heavy quarks it is sometime
preferable to carry out the calculations in their massless limit, where these
logarithms are absent.
Even when the quarks are kept as massive, IRC safety guarantees the absence of
these logarithmic effects, because there are no QCD divergences they can
originate from.

Recently, the problem of an IRC safe definition of flavoured jets has received a
renewed interest in the LHC context. This progress was motivated by the
availability of NNLO QCD calculations to processes with flavoured jets at hadron
colliders~\cite{Gauld:2019yng,Behring:2020uzq,Gauld:2020deh,Czakon:2020coa,Hartanto:2022qhh,Hartanto:2022ypo,Czakon:2022khx,Gauld:2023zlv,Gehrmann-DeRidder:2023gdl}.
Several new flavoured jet algorithms have been
proposed~\cite{Caletti:2022hnc,Czakon:2022wam,Gauld:2022lem,Caola:2023wpj,Behring:2025ilo}
that are IRC safe to all-orders or up to high orders and that preserve in an
exact or approximate way the kinematics of a given underlying flavour-agnostic
jet algorithm, like the anti-$k_t$.
Among the different proposals, there are a couple that can be immediately recast
to the $e^+e^-$ environment: Interleaved Flavour Neutralisation
(IFN)~\cite{Caola:2023wpj} and flavour dressing~\cite{Gauld:2022lem}.

As an example, in Fig.~\ref{fig:jetflavbad} on the right, we show a test of IRC
safety of the flavour dressing algorithm, when used to assign flavour to $k_t$
jets.
We consider the fraction of mis-identified events, defined as those events where
the flavour assignment is different from the one evaluated on a Born level
configuration i.e.\ events with a number of flavoured jets different from two,
whereas the Born process $e^+e^- \to f\bar{f}$ features two flavoured jets.
If the algorithm is IRC safe, in presence of soft or collinear emissions only,
such a fraction should vanish.
In Fig.~\ref{fig:jetflavbad} on the right, we are differential in the Durham
resolution variable $y_3 \equiv y_{23}$ and the limit $y_3 \to 0$ corresponds to
the infrared region.
We notice that the fraction of ``bad'' events with an IRC safe algorithm
vanishes as $y_3 \to 0$, whereas a naive flavour assignment results in a
non-vanishing probability of mis-identification in the unresolved regime.

\subsection{Hadronic Higgs decays}\label{sec:hdecay}

\begin{figure}
  \centering
  \includegraphics[width=0.6\textwidth]{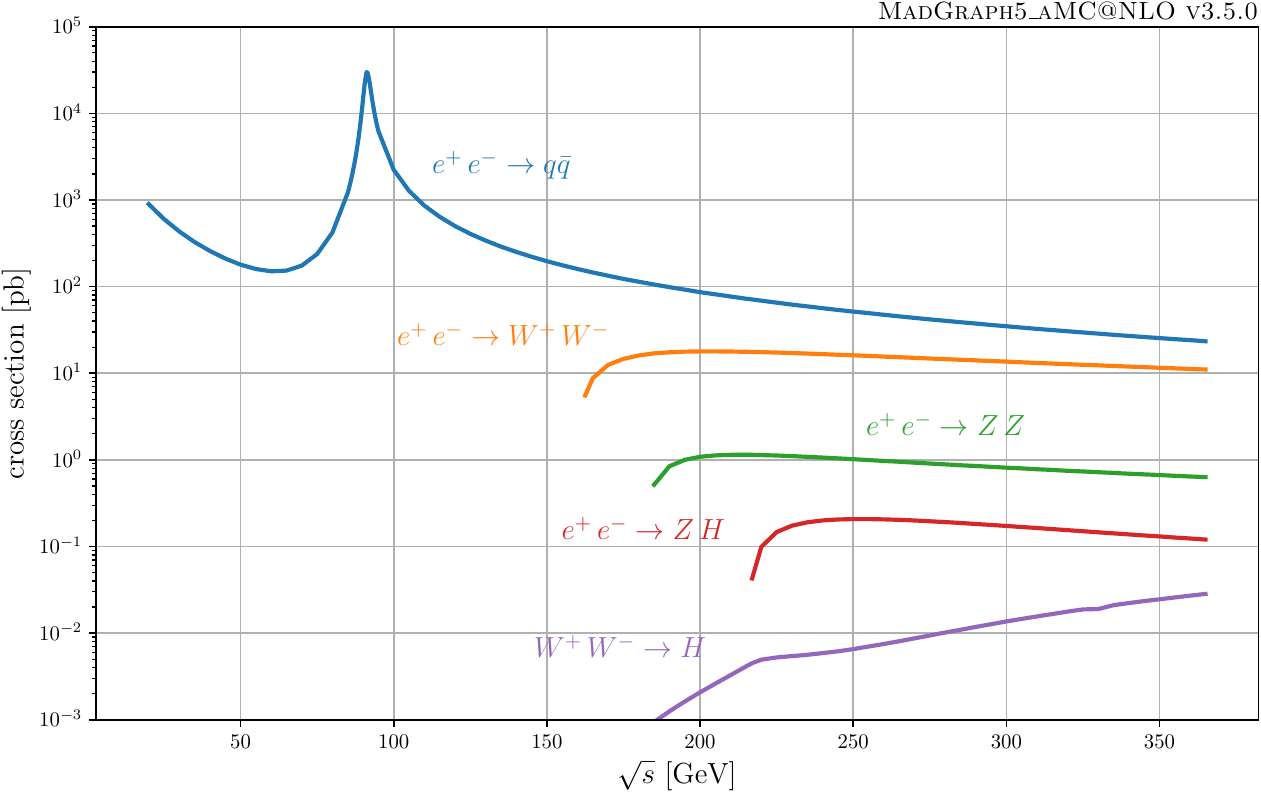}\,
  \includegraphics[width=0.2\textwidth]{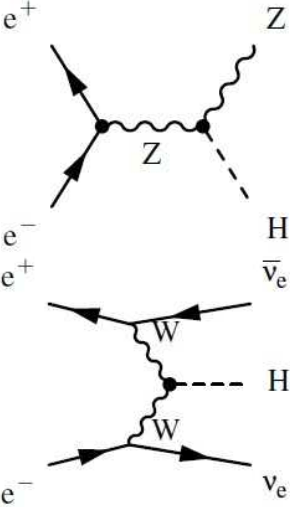}
  \caption{Relevant cross sections at $e^+e^-$ colliders obtained with
    \aNLO~\cite{Alwall:2014hca,Frederix:2018nkq,Bertone:2022ktl,Frixione:2021zdp}
    (left, courtesy of Michele Selvaggi~\cite{selvaggi_2025_kmsg5-47f34}).
    Leading order Feynman diagrams for $e^+e^- \to ZH$ and $WW \to H$ processes
    (right).}
  \label{fig:higgsxs}
\end{figure}

The dominant mechanism for production of jets at electron-positron colliders is
through the elementary process $e^+e^- \to \gamma^*/Z \to q\bar{q}$.
However, at higher energies, other processes start to contribute: for instance
the production of a pair of vector bosons, such as $W^+W^-$ or $ZZ$, that can
subsequently decay either leptonically or hadronically. See
Fig.~\ref{fig:higgsxs} on the left for a summary of cross section values as a
function of the centre of mass energy of the collider.
In particular, for $\sqrt{s} \gtrsim 220$~GeV, it is possible to produce Higgs
bosons in the final state, through the processes $e^+e^- \to Z H$ or $e^+e^- \to
\nu_e \bar{\nu}_e H$, whose leading order Feynman diagrams are depicted in
Fig.~\ref{fig:higgsxs} on the right.
High-energy future lepton colliders, such as the Future Circular Collider
(FCC-ee) or the Circular Electron Positron Collider (CEPC)~\cite{An:2018dwb},
will be able to produce many Higgs bosons, through dedicated runs at $\sqrt{s} =
240$~GeV, with more than 2 millions of Higgs' produced through the $ZH$ process.
To make the most out of those runs, it will be important to study the hadronic
decays of the Higgs boson.
For instance, the Higgs will predominantly decay hadronically, with around 80\%
of the branching ratio in hadronic decays. The dominant decay mode is $H \to
b\bar{b}$ (57.7\%), followed by $H \to WW/ZZ \to 4q$ (11\%), $H \to gg$
(8.6\%), $H \to c\bar{c}$ (2.9\%), $H \to s\bar{s}$ (0.024\%).
The study of such hadronic decays in the clean $e^+e^-$ environment is relevant
for extracting the Yukawa coupling to charm and bottom quarks (and possibly also
to strange), which are difficult to access at hadron colliders.

\begin{figure}
  \centering
  \includegraphics[width=0.5\textwidth]{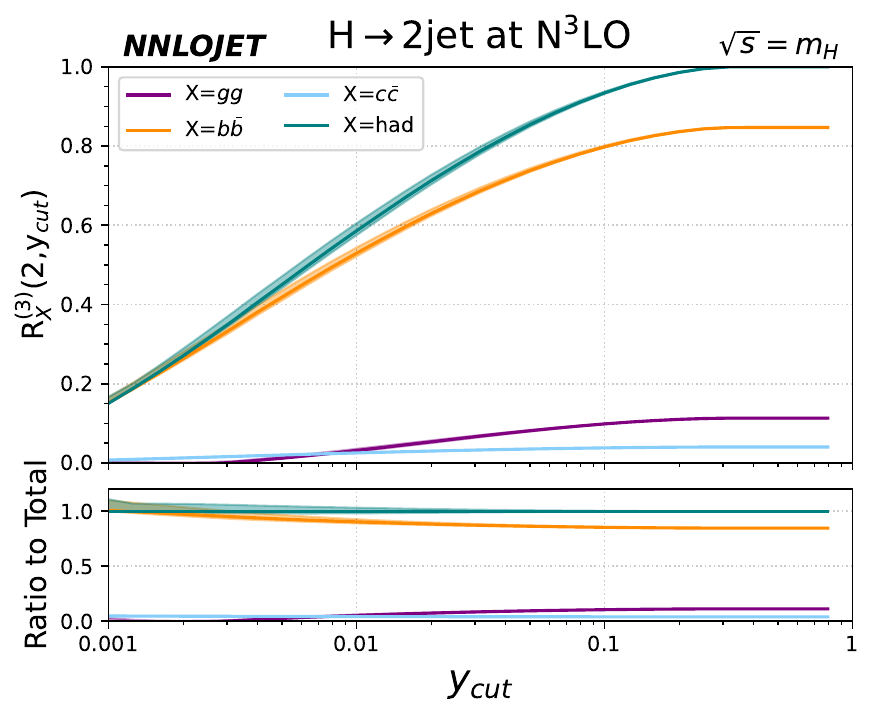}
  \caption{Two-jet rate in hadronic Higgs decays. From~\cite{Fox:2025cuz}.}
  \label{fig:higgs2jet}
\end{figure}

On the theory side, QCD corrections to the decays of the Higgs boson have
reached a high level of accuracy. The inclusive cross section for Higgs decaying
to hadrons is known up to N$^4$LO~\cite{Herzog:2017dtz}.
Fully differential predictions for $H \to b\bar{b}$ are known up to
N$^3$LO~\cite{DelDuca:2015zqa,Mondini:2019gid}.
Very recently, three-jet rates in Higgs decays have been calculated up to NNLO
accuracy~\cite{Fox:2025cuz}.
With the knowledge of the inclusive cross section, it is then possible to infer
the exclusive two-jet rate up to N$^3$LO, shown in Fig.~\ref{fig:higgs2jet} as a
function of the Durham $y_{\rm cut}$.
The relative size of the individual decay channels $H \to b\bar{b}$, $H \to gg$
and $H \to c\bar{c}$ reflect the hierarchy already observed in total branching
ratios reported above.

Predictions for event shapes in Higgs decays are available up to NLO, both in
three-jet~\cite{Coloretti:2022jcl} and four-jet~\cite{Gehrmann-DeRidder:2023uld}
events, and implemented in a new version of the \code{EERAD3}
code~\cite{Aveleira:2025svg}.
Fixed-order predictions for three-jet event shapes have been supplemented with
resummation to reach overall NLO+NLL' accuracy~\cite{Gehrmann-DeRidder:2024avt}.
Finally, IRC safe flavoured jet algorithms (see Sec.~\ref{sec:flav}) have been
applied to the study of hadronic Higgs decay~\cite{CampilloAveleira:2024fll}.

\subsection{Non-perturbative effects}\label{sec:nonpert}

The perturbative description of hadronic final states at $e^+e^-$ collisions is
able to provide a good description of many observables, but it is bound to fail
once we become sufficiently exclusive. In general, a cross section features
non-perturbative (NP) corrections expressed as powers of $\lamQCD/Q$, with
$\lamQCD$ a non-perturbative scale $\mathcal{O}$(1~GeV) and $Q$ the hard scale
of the process $\mathcal{O}$(100~GeV). Hence, for a generic $\sigma$, we can
write\footnote{Note that infrared and collinear safety (see
Sec.~\ref{sec:IRCsafe}) ensures the absence of logarithmic terms in
$\lamQCD/Q$.}
\begin{equation}
  \sigma(Q) = \sigma_{0}(Q) \left[ 1 + \sum_{n=1}^\infty c_n \left( \frac{\as(Q^2)}{2\pi} \right)^n +
  \mathcal{O}\left( \left( \frac{\lamQCD}{Q} \right)^p \right) \right]\,.
\end{equation}
The precise value of the power $p$ depends on the process and on the
observable. For instance, it has been shown that the inclusive cross section for
hadron production exhibits $p=4$~\cite{ParticleDataGroup:2024cfk}, so power
corrections are essentially negligible; but for less inclusive observables, $p$
may be equal to 1 or 2, commonly known as {\em linear} or {\em quadratic} power
correction, respectively.

For instance, the mean value of the thrust $\langle 1-T \rangle$ is known to
suffer from linear power corrections~\cite{Dokshitzer:1995zt}. The perturbative
value at LEP is given by~\cite{Altarelli:1989hv}
\begin{equation}
  \langle 1-T \rangle = 0.335\,\as + 1.02\,\as^2 + \mathcal{O}(\as^3)
  \simeq 0.040 + 0.014\,,
\end{equation}
whereas a linear power correction would result in a NP effect of order
$\lamQCD/Q \simeq 0.01$, hence comparable with the perturbative
$\mathcal{O}(\as^2)$ correction.

One way to introduce non-perturbative corrections in perturbative calculations
is through an effective coupling, as already briefly mentioned in
Sec.~\ref{sec:resumres}.
Another way is through {\em shape functions}~\cite{Korchemsky:1999kt}, based on
the factorization of short-distance perturbative and long-distance
non-perturbative effects. To be concrete, let us consider the differential
distribution in $t = 1-T$. In the $t \to 0$ region, the physical distribution
evaluated at $t$ is written as a convolution between the shape function
$F(\epsilon)$ and the perturbative distribution evaluated at a shifted $\hat{t}
= t - \epsilon/Q$ point
\begin{equation}\label{eq:shapefun}
  \frac{\mathrm{d}\sigma}{\mathrm{d}t}(t) = \int \mathrm{d}\epsilon \,
  F(\epsilon) \frac{\mathrm{d}\sigma}{\mathrm{d}t}\left(t - \frac{\epsilon}{Q}\right)\,.
\end{equation}
The physical interpretation of~\eqref{eq:shapefun} is related to the
identification of the shape function as the contribution coming from the
emissions of many soft gluons in the two-jet region: the overall effect of these
emissions is to change the value of the thrust entering the perturbative
prediction, with a modulation dictated by $F(\epsilon)$.

These historical approaches for the incorporation of NP effects are formally
valid in the two-jet region only (identified as the limit $e \to 0$ for some
three-jet event shape $e$ going to zero).
Understanding NP power corrections away from the two-jet limit has been a
longstanding problem, but recent developments have shed new light on it.
In~\cite{Luisoni:2020efy}, the NP corrections for the C-parameter in the
symmetric three-jet limit has been evaluated, and found to be a factor of two
larger then in the two-jet limit.
By exploiting the renormalon calculus~\cite{Beneke:1998ui}, it was possible to
analytically compute the leading NP corrections to thrust and
C-parameter~\cite{Caola:2021kzt} in the full spectrum. In the case of the
C-parameter, they were found to interpolate between the known two-jet and
symmetric three-jet value, but with a sharp decrease close to the two-jet
region. These analytical models have been later used in fit of
$\as$~\cite{Nason:2023asn,Nason:2025qbx}.

To conclude, let us mention that it is also possible to include NP effects
within effective theories like SCET, that we very briefly introduced in
Sec.~\ref{sec:SCET}: fit of $\as$ by focussing on the thrust distribution in the
two-jet regime have been carried out~\cite{Abbate:2010xh,Benitez:2024nav}.

\section{Conclusions}
\label{sec:conclusions}

In this Chapter, we have provided a pedagogical introduction to the study of
hadronic final states at lepton colliders. As highlighted throughout our
discussion, these studies have played a crucial role in establishing QCD as the
correct theory of the strong interactions. Let us conclude with a few final
considerations regarding specific aspects of our discussion.

From a theoretical perspective, the fact that QCD radiation at electron-positron
colliders is confined to the final state simplifies the computation of
higher-order corrections, both at fixed order and in resummed perturbation
theory. Fixed-order calculations have now reached N$^3$LO
accuracy~\cite{Chen:2025kez}, while event-shape observables have been resummed
up to N$^4$LL~\cite{Duhr:2022yyp}. Moreover, the clean $e^+e^-$ environment
provides an ideal testing ground for the development of novel theoretical tools,
such as the new generation of parton showers with improved logarithmic accuracy,
as discussed in Sec.~\ref{sec:psevgen}.

The legacy data from LEP are so precise that they continue to drive theoretical
developments. A notable example is the determination of $\as$ from event shapes:
as documented in~\ref{sec:nonpert}, achieving a percent-level determination of
$\as$ requires careful control over both perturbative and non-perturbative
effects, and significant progress has been made in this direction in the recent
years. In this context, reanalyzing LEP data with the theoretical and
computational expertise developed during the LHC era could prove highly
impactful in the coming years. An illustrative case was presented in
Sec.~\ref{sec:genkt}.

Looking ahead, the future of high-energy physics will likely involve a
next-generation electron-positron collider, where QCD will continue to play a
central role~\cite{2921876,Verbytskyi:2025sod}. Access to higher center-of-mass
energies will enable, among other things, the production of Higgs bosons.
Studying their hadronic decays as described in Sec.~\ref{sec:hdecay} will offer
a powerful probe of the Yukawa sector of the Standard Model, which remains
largely unconstrained at present.

\noindent {\bf Acknowledgements}.
I am grateful to Leonardo Bonino, Simone Caletti, Xuan Chen, Matteo Marcoli, and
Christian Preuss for their valuable comments on the manuscript.

\clearpage
\bibliographystyle{JHEP}
{\footnotesize
\bibliography{jetsepem}}

\end{document}